%% file: main.tex
\documentclass[twocolumn]{aastex63}

\usepackage{colortbl} 
\usepackage{multirow}
\usepackage{verbatim}
\usepackage{comment} 
\usepackage{amsmath}
\usepackage{array}

\usepackage{rotating}
\usepackage{enumitem}
\usepackage{natbib}
\usepackage{graphics,graphicx}
\usepackage{caption}

\makeatletter
\def\@hex@@Hex#1%
 {\if a#1A\else \if b#1B\else \if c#1C\else \if d#1D\else
  \if e#1E\else \if f#1F\else #1\fi\fi\fi\fi\fi\fi \@hex@Hex}
\makeatother

\definecolor{asparagus}{rgb}{0.53, 0.90, 0.52}
\definecolor{darkseagreen}{rgb}{0.56, 0.74, 0.56}
\definecolor{seagreen}{rgb}{0.18, 0.55, 0.34}
\definecolor{bostonuniversityred}{rgb}{0.8, 0.0, 0.0}
\definecolor{babyblue}{rgb}{0.54, 0.81, 1.}
\definecolor{beaublue}{rgb}{0.74, 0.83, 0.9}
\definecolor{ceruleanblue}{rgb}{0.16, 0.32, 0.75}
\definecolor{smalt(darkpowderblue)}{rgb}{0.0, 0.2, 0.6}

\revised{\today}
\graphicspath{{./}{figures/}}

\begin{document}

\include{definitions}

\title{Learning the Universe: physically-motivated priors for dust attenuation curves}

\author[0000-0002-2906-2200]{L. Sommovigo}
\affiliation{Center for Computational Astrophysics, Flatiron Institute, 162 Fifth Avenue, New York, NY 10010, USA}
\footnote{email:lsommovigo@flatironinstitute.org}

\author[0000-0001-8855-6107]{R. K. Cochrane}
\affiliation{Institute for Astronomy, University of Edinburgh, Royal Observatory, Blackford Hill, Edinburgh, EH9 3HJ, UK}
\affiliation{Department of Astronomy, Columbia University, New York, NY 10027, USA}

\author[0000-0003-2835-8533]{R. S. Somerville}
\affiliation{Center for Computational Astrophysics, Flatiron Institute, 162 Fifth Avenue, New York, NY 10010, USA}

\author[0000-0003-4073-3236]{C. C. Hayward}
\affiliation{Eureka Scientific, Inc., 2452 Delmer Street, Suite 100, Oakland, CA 94602, USA}
\affiliation{Kavli Institute for the Physics and Mathematics of the Universe (WPI), The University of Tokyo Institutes for Advanced Study, The University of Tokyo, Kashiwa, Chiba 277-8583, Japan}
\affiliation{Center for Computational Astrophysics, Flatiron Institute, 162 Fifth Avenue, New York, NY 10010, USA}

\author[0000-0001-7964-5933]{C. C. Lovell}
\affiliation{Institute of Cosmology and Gravitation, University of Portsmouth, Burnaby Road, Portsmouth, PO1 3FX, UK}

\author[0000-0003-2539-8206]{T. Starkenburg}
\affiliation{Center for Interdisciplinary Exploration and Research in Astrophysics (CIERA), Northwestern University, 1800 Sherman Ave, Evanston IL 60201, USA}
\affiliation{Department of Physics and Astronomy, Northwestern University, 2145 Sheridan Rd, Evanston IL 60208, USA}
\affiliation{NSF-Simons AI Institute for the Sky (SkAI), 172 E. Chestnut St., Chicago, IL 60611, USA}

\author[0000-0003-1151-4659]{G. Popping}
\affiliation{European Southern Observatory, Karl-Schwarzschild-Str. 2, D-85748, Garching, Germany}

\author[0000-0001-9298-3523]{K. Iyer}
\affiliation{Department of Astronomy, Columbia University, New York, NY 10027, USA}

\author[0000-0003-4295-3793]{A. Gabrielpillai}
\affiliation{Department of Astrophysics, The Graduate Center, City University of New York, 365 5th Ave, New York, NY 10016, USA}

\author[0000-0003-3207-8868]{M. Ho}
\affiliation{Department of Astronomy, Columbia University, New York, NY 10027, USA}

\author[0000-0001-8867-5026]{U. P. Steinwandel}
\affiliation{Max Planck Institute for Astrophysics, Karl-Schwarzschild-Str. 1, D-85748, Garching, Germany}

\author[0000-0002-8449-1956]{L. A. Perez}
\affiliation{Center for Computational Astrophysics, Flatiron Institute, 162 Fifth Avenue, New York, NY 10010, USA}
\affiliation{Department of Astrophysical Sciences, Princeton University, Princeton, NJ 08544, USA}

\begin{abstract}
Understanding the impact of dust on the spectral energy distributions (SEDs) of galaxies is crucial for inferring their physical properties and for studying the nature and evolution of interstellar dust.
In this study, we analyze dust attenuation curves of $\sim 6400$ galaxies ($M_{\star}\sim10^9-10^{11.5}\,\mathrm{M_{\odot}}$) at $z=0.07$ from the IllustrisTNG50 \& TNG100 simulations. Using radiative transfer post-processing, we generate synthetic attenuation curves and fit them with a versatile parametric model that encompasses both known extinction and attenuation curves (e.g. Calzetti, MW, SMC, LMC) and more exotic forms. 
We present the distributions of the best-fitting parameters — UV slope (\(c_1\)), optical-to-NIR slope (\(c_2\)), FUV slope (\(c_3\)), \(2175\,\angstrom\) bump strength (\(c_4\)), and normalization (\(A_{\rm V}\)) — accounting for scatter from orientation effects. 
Key correlations emerge between $A_{\rm V}$ and the star formation rate surface density $\Sigma_{\rm SFR}$, as well as the UV slope $c_1$. Furthermore, the UV and FUV slopes ($c_1, c_3$) and the visual attenuation and bump strength ($A_{\rm V}, c_4$) exhibit robust internal correlations (anticorrelation in the latter case). The optical-to-NIR slope exhibits minimal variations. 
Using these insights from simulations, we provide a set of scaling relations that predict a galaxy’s median (averaged over line of sight) dust attenuation curve based solely on its $\Sigma_{\rm SFR}$ and/or $A_{\rm V}$. These predictions agree well with observed attenuation curves from the GALEX-SDSS-WISE Legacy Catalog, although there are minor differences in bump strength. 
This study delivers the most comprehensive library of synthetic attenuation curves for local galaxies, and provides a foundation for physically motivated priors for SED fitting and galaxy inference studies, such as those performed as part of the Learning the Universe Collaboration. 
\end{abstract}


\section{Introduction}
Understanding the role of dust in the interstellar medium (ISM) is crucial to interpreting observations and constraining the physical properties of galaxies. The complexity of this topic stems partly from the very wide range of scales involved in and affected by dust-related processes, from microscopic (the scales of individual dust grains), to molecular clouds (star-dust geometry), to galaxy-wide (as interstellar dust impacts the global galaxy emission).

Dust grains primarily form in the late evolutionary stages of massive stars, including Asymptotic Giant Branch (AGB) winds, planetary nebulae, and supernova ejecta \citep{Dwek80,Gehrz89,Schneider23}. Once formed, they evolve through interactions with hot gas, stellar radiation, and cosmic rays, undergoing processes like sputtering, vaporization, and coagulation \citep{Draine79,Jones1996ApJ,Tielens99,Todini00,LiGreenberg03dust,Yan2004,Hirashita20}. 
%
Crucially, for the sake of galaxy studies, dust grains absorb (extinct) and scatter UV and optical starlight, and re-emit it between mid-infared (MIR) and far-infrared (FIR) wavelengths \citep{Draine89,Meurer99,Calzetti00,Weingartner01,Draine03}. The extinction of light by dust substantially alters the spectral energy distribution (SED) of star-forming regions \citep[e.g.;][]{SeonDraine16} and, globally, of galaxies \citep[see][for a review]{Salim20}. Thus, the accurate inference of galaxy physical properties, such as stellar mass and star formation rate, demands a proper accounting for dust. \\
\indent Within our own galaxy, dust extinction has been measured along different lines of sight using the "pair method", whereby the spectra of two stars of the same spectral class, one of which is reddened, are compared \citep[see e.g.; ][]{Cardelli89,Fitzpatrick99}. The mean Milky Way (MW) extinction curve, obtained by averaging over several sightlines, is characterized by a rapid rise in extinction, $A_{\lambda}$, from IR to far UV wavelengths, and a broad absorption feature at $2175\,\angstrom$ \citep{Cardelli89}.

Our knowledge of dust extinction laws in other galaxies is still fairly limited. So far, the pair method has been successfully applied to two satellites of the MW, the Small and Large Magellanic Cloud (SMC, see e.g., \citealt{Prevot84,Bouchet85}, and LMC, see e.g. \citealt{Koornneef81,Nandy81}). The SMC extinction curve has been used as a template to quantify dust attenuation in high-redshift sources, due to its younger stellar ages and lower metallicity compared to the MW ($Z_{\rm SMC} < 0.1\,\mathrm{Z_{\odot}}$, see e.g., \citealt{Draine03}). In the SMC, the extinction curves of most sightlines display a nearly linear ($A_{\lambda} \propto \lambda^{-1}$), steep rise from visual to far-UV (FUV) wavelengths and no $2175\,\angstrom$ bump \citep{Bouchet85,Prevot84}. The LMC extinction curve is characterized by a weak $2175\,\angstrom$ bump and a strong FUV rise \citep{1981ApJ...247..860K,1981MNRAS.196..955N}, and is approximately intermediate between the SMC and MW curve. 
As in our Galaxy, spatial variations are observed in both the SMC and LMC extinction curves.

In galaxies where individual stars cannot be resolved, other approaches can be applied to derive attenuation curves. For instance, \cite{Calzetti00} used the spectra of $39$ local starburst galaxies to define an average attenuation curve, obtained by assuming a simple dust screen geometry for all galaxies in the sample. 
This attenuation law is flatter (``greyer") than the MW, LMC or SMC extinction curves, and it completely lacks the $2175\,\angstrom$ feature. In the decades since, numerous works have applied the method introduced by \cite{Calzetti00} to study galaxy samples out to intermediate redshifts $z\sim 1-3$ \citep{Reddy15,Zeimann15,Scoville15,Salmon16,Tress18,Battisti20,Shivaei22}. Studies analyzing large samples of $z\sim0$ sources (e.g. $230,000$ galaxies in \citealt{Salim18}) find that attenuation curves exhibit a very wide range of slopes, with a strong dependence on the galaxies' optical opacity (more opaque galaxies have flatter curves), and (indirectly) on stellar mass \citep[e.g.][]{Garn2010}. A wide range of UV bump amplitudes is measured, from zero to MW-like, with an average strength $\sim 1/3$ that of the MW bump. Low-z analogs of high-redshift galaxies have an average curve that is somewhat steeper than the SMC curve, with a modest UV bump \citep{DustE22}. At slightly higher redshifts, analyses of $218$ star forming galaxies at $z=1.4-2.6$ from the MOSDEF survey \citep{Reddy15} revealed a positive correlation between galaxy metallicity and attenuation curve slope and UV bump strength (i.e., more massive and metal-rich galaxies show shallower attenuation curves with more pronounced UV bumps, up to $\sim 3 \times$ stronger than the MW bump; \citealt{Shivaei20,Shivaei22}). 

Recent observational studies conducted with the James Webb Space Telescope (JWST, \citealt{JWST_06,JWST_23}) have allowed us to extend these attenuation curve studies out to $z\sim 10$, albeit in inherently biased samples of UV-bright sources \cite[e.g.;][]{Markov23,Markov24,Fisher25}. 
\cite{Witstok23} found evidence of the $2175\,\angstrom$ feature in an individual galaxy at $z=6.7$, indicating rapid dust production and survival of carbonaceous dust at early epochs (at least in some sources, see also \citealt{Lin25}). A complete understanding of the assembly of dust and its effects at these very high redshifts is still lacking, partly due to the limited and somewhat biased (towards the brightest and least attenuated systems at high-$z$) nature of the available observational samples, and due to the various degenerate physical processes that can affect the shape of a galaxy global attenuation curve. These effects include the star-dust geometry, intrinsic global galaxy properties such as metallicity, specific SFR, and stellar mass, as well as dust grain composition and grain size distribution.

\indent Hydrodynamical simulations and semi-analytic models (SAMs) provide self-consistent frameworks to model the co-evolution of dust and galaxies out to early epochs. These models either explicitly track dust processes (for SAMs see e.g., \citealt{Popping17,Vijayan19,Triani20,Dayal22,Mauerhofer23}, and for hydrodynamical simulations, see e.g., \citealt{Choban2022,2024ApJ...974..136S,diCesare23}) or rely on dust-to-metal and dust-to-gas scaling relations (for SAMs, see e.g., \citealt{deLucia2007,Somerville2012}, and for simulations, see e.g., \citealt{Torrey15,Trayford15,Behrens18,Pallottini22,DiMascia24}). Additionally, radiative transfer (RT) calculations, whether applied on-the-fly or in post-processing (typically employing codes such as {\sc{skirt}} \citep{CampsBaes2020} or {\sc{powderday}} \citep{Narayanan21}), enable simulations to quantify (and isolate) the impact of dust on galaxy emission at both global and resolved spatial scales \citep[e.g.;][]{SeonDraine16,Narayanan18,Cochrane19,2023MNRAS.518.5522C,Cochrane23,Cochrane24,Trayford20,Parsotan2021,DiMascia21,DiMascia24,Vijayan24}.

Several studies have explored how radiative transfer effects and dust-star geometry shape the attenuation curve of simulated galaxies. \cite{SeonDraine16} investigated dust attenuation in galactic environments, relying on models of radiative transfer in a spherical, clumpy ISM. They showed that the attenuation curves are primarily determined by the wavelength dependence of absorption, rather than by the underlying extinction curve.
\citet{Narayanan18} and \citet{Trayford20} examined this in hydrodynamical simulations (MUFASA zoom-ins and the EAGLE cosmological box, respectively), finding that attenuation curves steepen when old stellar populations dominate and flatten when young stars remain unobscured. They also found that the $2175\,\angstrom$ bump strength correlates with the fraction of unobscured O- and B-type stars. 
Building on this, \citet{Lin21} used a simple spherical geometry approximation to disentangle the effects of dust-to-stellar geometry and the intrinsic extinction curve. Their results confirmed that scattering and varying optical depths between young and old stellar populations could significantly alter attenuation curve slopes. They also demonstrated that even vastly different extinction curves could produce similar attenuation features due to radiative transfer effects.
Expanding upon these works with more realistic dust-to-gas morphologies, \citet{DiMascia24} investigated how attenuation curves evolve throughout the lifetime of an isolated giant molecular cloud (GMC). They showed that, even at a fixed dust composition and grain size distribution, feedback and turbulence can introduce significant variation, sometimes leading to the disappearance of the $2175\,\angstrom$ bump, even for Milky Way-like dust compositions.
Further complexity arises when considering different dust grain size distributions \citep{Aoyama:2020}. For instance, \citet{DiMascia21} found that in cosmological zoom-in hydrodynamical simulations of $z\sim 6$ quasar host galaxies, removing small grains ($<0.1\,\micron$) was necessary to reproduce observed spectral energy distributions (SEDs). 
To bridge the gap between simulations and observations, \citet{Hahn2022} introduced the Empirical Dust Attenuation (EDA) framework, which assigns attenuation curves to simulated galaxies based on empirical constraints, varying with properties such as stellar mass and star formation rate (SFR). 

\begin{figure*}
    \centering
    \includegraphics[width=1.05\textwidth]{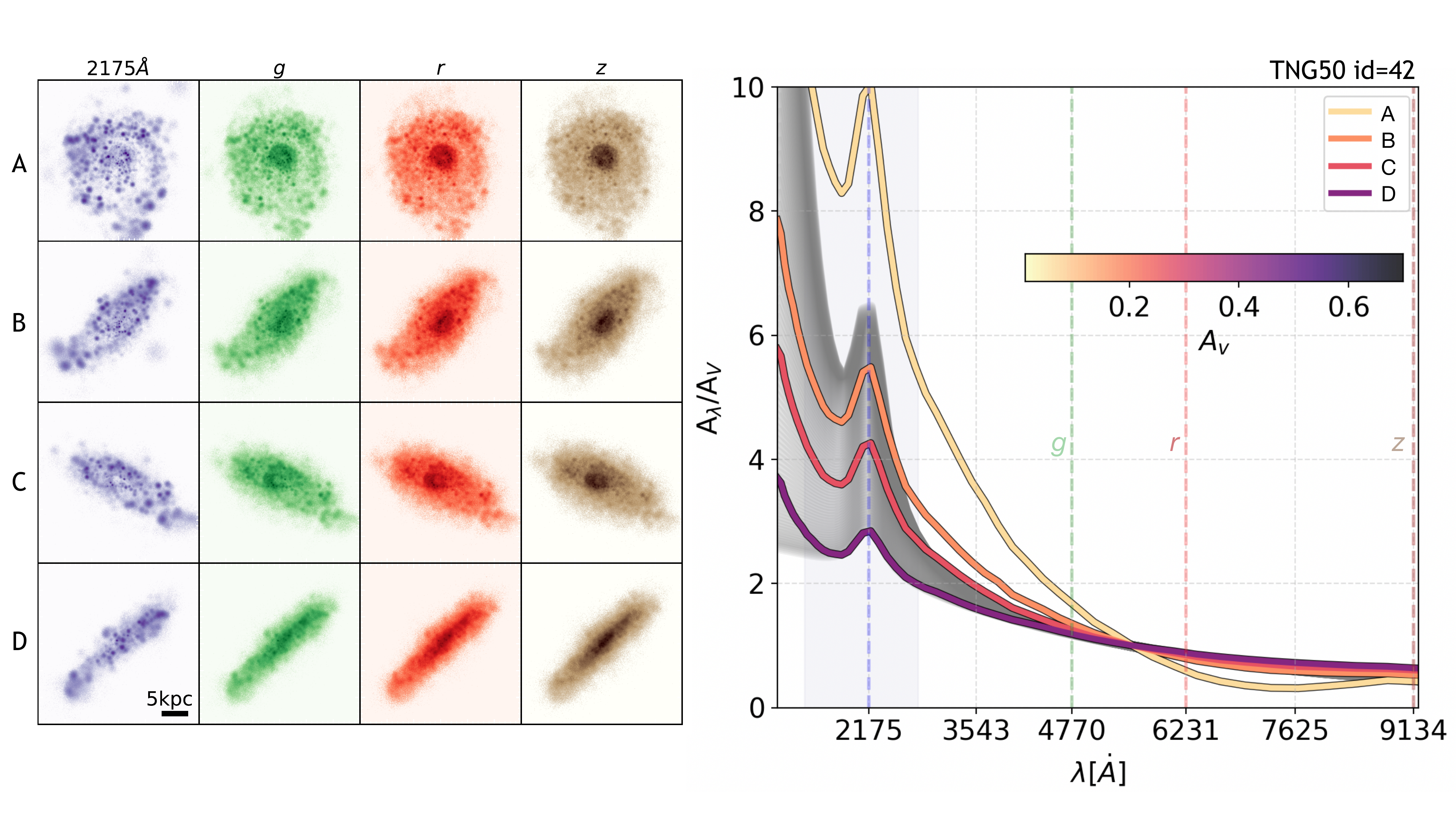}
    \vspace{-0.8cm}
    \caption{\textbf{Left Panel:} Images of a TNG50 galaxy (id 42) observed along different lines of sight,
    with increasing values of visual attenuation ($A_V$) from top to bottom. Each row shows the predicted emission at a different wavelength: $2175\,\angstrom$, and the SDSS $g$, $r$, and $z$ bands.
    \textbf{Right panel:} Corresponding attenuation curves for each line of sight, color-coded according to $A_V$ (see colorbar). The grey shaded area represents the $16^{\rm{th}}–84^{\rm{th}}$ percentile variation of attenuation curves across the entire TNG combined sample, including all sources and lines of sight. The wavelengths corresponding to the filters shown in the left panel are highlighted by vertical dashed lines. }
    \label{example_images}
\end{figure*}

In this paper, we present a physically motivated model for attenuation curves derived from RT post-processing of IllustrisTNG50 and IllustrisTNG100 simulations. We analyze nearly $6400$ galaxies, exploring correlations between global galaxy properties and attenuation curve shapes, which we parametrize using a flexible, empirically motivated four-parameter model. 
We derive probability distribution functions (PDFs) of the best-fitting parameters describing our library of synthetic attenuation curves, along with a set of simulation-based scaling relations that capture the key physical dependencies of the curve shapes. These PDFs and scaling relations can serve as physically motivated priors in SED fitting procedures or as components of large-scale cosmological inference efforts, such as those conducted within the Learning the Universe (LtU)\footnote{\url{http://learning-the-universe.org}} collaboration. The LtU project employs a Bayesian forward modeling approach to reconstruct the initial conditions and governing laws of the Universe. This requires generating synthetic observables and astrophysical parameters, such as the dust attenuation effects studied here. By integrating advanced galaxy formation simulations, sub-grid models, machine learning acceleration, and simulation-based inference techniques, the LtU collaboration aims to efficiently infer the posterior distributions of cosmological and astrophysical parameters from observational data, overcoming the computational challenges posed by high-dimensional parameter spaces.

The remainder of the paper is structured as follows. In Sec.~\ref{Sect_Methods}, we discuss the simulations and radiative transfer post-processing setup employed in this study, as well as the observational dataset used for comparison. In Sec.~\ref{fitting_attenuation_curves}, we present the results from fitting the synthetic attenuation curves of TNG galaxies and compare them with observations of local galaxies. In Sec.~\ref{Sect_Dust_GalaxyProp_Correl}, we investigate correlations between global galaxy properties and dust attenuation curve parameters in the simulated galaxies and compare our findings with observational results. In Sec.~\ref{Sect_Analytical_Model}, we derive simulation-based fits for scaling relations aimed at predicting a galaxy's attenuation curve based on a single intrinsic or observable property. Finally, in Sec.~\ref{Sect_discussion}, we compare our results with literature findings from both theory and observations and discuss the caveats of our analysis. We conclude with a summary in Sec.~\ref{summary}.

\begin{figure*}[t]
    \centering
    \includegraphics[width=0.495\linewidth]{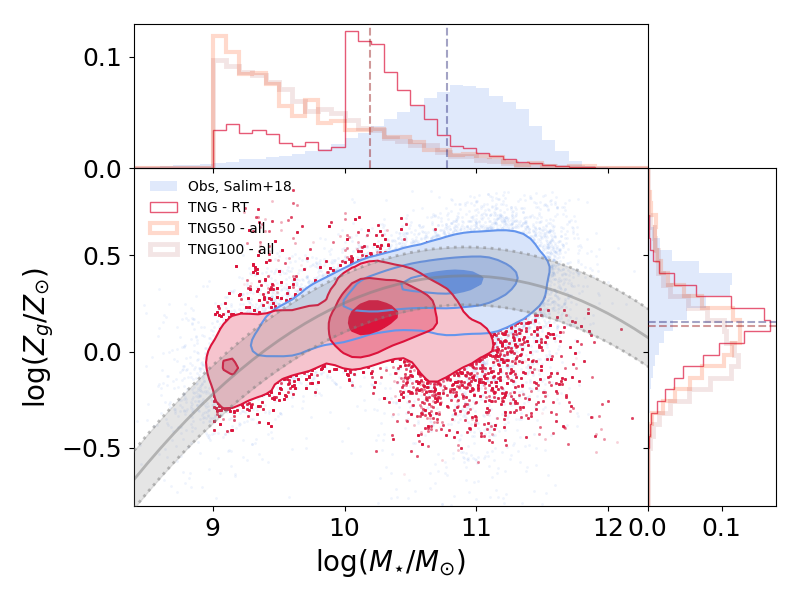}
    \includegraphics[width=0.495\linewidth]{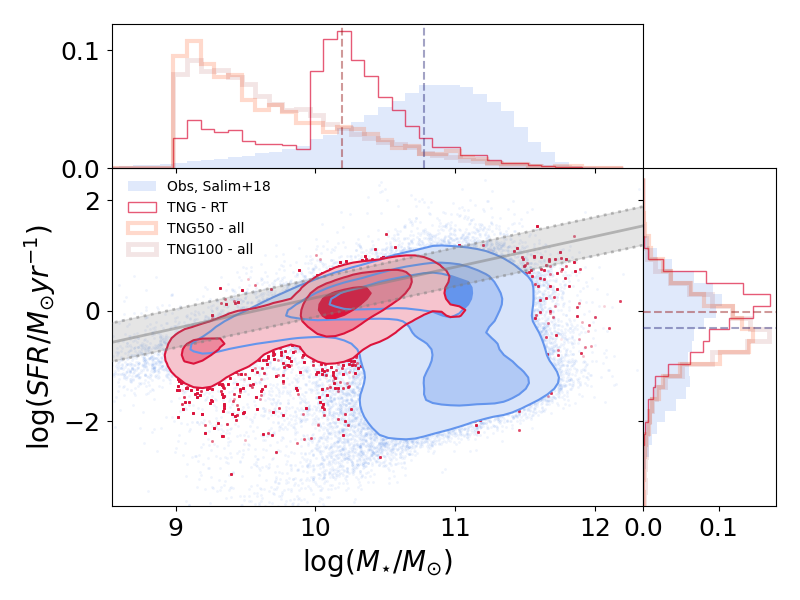}
    \caption{Comparison of the stellar mass-metallicity distribution (MZ; \textbf{left panel}) and stellar mass-SFR relation (MS; \textbf{right panel}) between the observed GSWLC \citep{Salim18} and simulated TNG galaxy samples. The contours correspond to the $16^{\rm{th}}$, $50^{\rm{th}}$, and $84^{\rm{th}}$ percentiles. Vertical dashed lines indicate the median values for metallicity, stellar mass, and SFR for the simulated (red) and observed (blue) samples. The empty histograms show the distributions normalized to the number of galaxies in the TNG50 (orange) and TNG100 (dark red) simulations. The solid gray line represents the local MZ relation and star-forming MS from \cite{Lopez13}, with the shaded region indicating the $1\,\sigma$ scatter.}
    \label{Fig_Mstar-Z_Obs_Comp}
\end{figure*}

\section{Methods \& Data}\label{Sect_Methods}
\noindent In this section, we describe the simulations, radiative transfer post-processing, and fitting techniques used to generate and analyze synthetic attenuation curves, as well as the observational dataset employed for comparison throughout this study.

\subsection{TNG simulations and radiative transfer post processing}
To produce our synthetic attenuation curves, we make use of the IllustrisTNG\footnote{\url{https://www.tng-project.org}} simulations. The IllustrisTNG simulations have been shown to reproduce many global intrinsic properties of local and low-redshift galaxies, including stellar mass functions \citep{Pillepich18}, the star formation main sequence \citep{2019MNRAS.485.4817D}, and the mass-metallicity relation \citep{Torrey19}. Furthermore, they statistically reproduce galaxies' emission properties, as probed by luminosity functions and color distributions \citep[e.g.][]{Torrey15,Pillepich18,Nelson19}. While the TNG simulations do not resolve star-forming regions in galaxies, simulations that achieve this typically do so at the expense of sample size and parameter space coverage, reinforcing our choice of TNG for this exploratory study.

We select $1858$ sources from the TNG50 \citep{Pillepich19} suite and $4535$ sources drawn from the TNG100 \citep{Pillepich18} suite, all at $z=0.07$ (snapshot 93). The range of stellar masses spanned by this galaxy sample is $9 \leq \log_{10} (M_{\star}/\rm{M_{\odot}}) \leq 11.9$, the star formation rates (averaged over 100 Myr) are in the range $-3.69 \leq \log_{10} (\mathrm{SFR/M_{\odot}yr^{-1}}) \leq 1.34$, the gas phase metallicities span $-0.63 \leq \log_{10} (Z/\rm{Z_{\odot}}) \leq 0.83$, and stellar ages span $0.17 <\mathrm{Age/Gyr}< 10$. We show the comparison of our selected sources to the full TNG50 sample and TNG 100 sample in Fig. \ref{Fig_Mstar-Z_Obs_Comp}. The peak around $M_{\star} = 10^{10}\,\rm M_{\odot}$ in the stellar mass distribution of our selected galaxies results from different selection criteria for the TNG50 and TNG100 samples. We include all TNG50 galaxies with non-negligible SFR or metal content, while for TNG100, we impose a mass cut of $M_{\star} > 10^{10}\,\rm M_{\odot}$ to ensure a well-populated massive end, aiming for a sample representative of SDSS-like sources (see the next paragraph for details). Since the 1D histograms in Fig.~\ref{Fig_Mstar-Z_Obs_Comp} are normalized by sample size, both TNG50 and TNG100 contain more low-mass galaxies. However, our sub-sample peaks at $10^{10}\,\rm{M_{\odot}}$ due to our selection criteria.

\indent We post-process the TNG galaxies with the radiative transfer code {\sc{skirt}}\footnote{\url{https://skirt.ugent.be/root/_home.html}} \citep{Camps2015,CampsBaes2020} to model the continuum emission, including the effects of dust attenuation, from far-UV to far-infrared wavelengths. Our procedure broadly follows the methods described by \cite{Schulz20} and \cite{Popping2022_tng}. We briefly summarize the technical details here.\\
\indent The emission from star particles is modeled according to their ages and metallicities, using \citet{Bruzual2003} stellar population synthesis models. The IllustrisTNG simulation suite does not directly follow the dust abundance of gas cells. Hence, we assume a dust-to-gas mass ratio (D) for gas cells that scales linearly with the gas-phase metallicity:
\begin{equation}\label{eq_RR14}
    D = \frac{1}{163}\ \frac{Z}{\mathrm{Z_{\odot}}}\ ,
\end{equation}
based on the empirical results from \cite{RemyRuyer14}. This scaling has been shown to hold out to cosmic noon \citep{Shapley20,Peroux20,Popping22}.

A dust abundance is only assigned to gas cells that are star-forming or have a temperature less than $75,000\,\rm{K}$. We model a mixture of graphite, silicate and PAH grains, with sizes in the range $a=10^{-4}-10\,\mu\rm{m}$ according to \cite{Weingartner01} Milky Way dust model. Following \cite{Popping2022_tng}, we do not model the contribution from young birth clouds. We perform the radiative transfer on an octree dust grid, in which cell sizes are adjusted according to the dust density distribution, with the condition that no dust cell may contain more than $0.0001\%$ of the total dust mass of the galaxy. We calculate rest-frame far-UV to far-IR emission (at 200 wavelengths) along $51$ lines of sight, evenly distributed in solid angle. Comparing the modeled observed emission along a given line of sight to the intrinsic (dust-free) emission, we calculate integrated dust attenuation curves along each line of sight. A schematic view of one of the TNG50 galaxies is shown in Fig.~\ref{example_images}. On the left, we present emission maps of the galaxy in the GALEX near-UV band (centered at $2175\,\angstrom$), followed by the SDSS $g$, $r$, and $z$ filters. On the right, we show the attenuation curves corresponding to four selected lines of sight, chosen to maximize the variation in visual attenuation ($A_V$) along them.

\subsection{GWSLC Observational Dataset}
Throughout the paper, we compare our results with data from the deep GALEX-SDSS-WISE Legacy Catalog (GSWLC, \citealt{Salim18}). The GSWLC covers $7\%$  of the SDSS catalog, totaling $48,401$ sources with physical properties comparable to our combined TNG sample. These observed galaxies span a stellar mass range of $7 \leq \log_{10} (M_{\star}/\rm{M_{\odot}}) \leq 12$, star formation rates (averaged over $100\,\rm{Myr}$) of $-5 \leq \log_{10} (\mathrm{SFR/M_{\odot}yr^{-1}}) \leq 2.6$, gas phase metallicities of $-1.6 \leq \log_{10} (Z/\rm{Z_{\odot}}) \leq 0.8$, and stellar ages of $2 <\mathrm{Age/Gyr}< 9$. All the aforementioned quantities are derived from SED fitting using well-established procedures \citep{Salim16}. Star formation rates (SFRs) are computed on a $100\,\rm{Myr}$ timescale due to the lack of emission line data required for shorter timescale estimates. Galaxy sizes are determined using multiple methods, including isophotal measurements \citep{Salim23} and traditional Sérsic profile fitting applied to NUV, FUV, and optical images \citep{Meert15}. Finally, gas-phase metallicities are estimated using the N2O2 method, which relies on the ratio of [NII] 6584 to [OII] 3727 \citep{Salim18}.

A direct comparison of the metallicity, SFR, and stellar mass distributions between the simulated and observed galaxies is shown in Fig.~\ref{Fig_Mstar-Z_Obs_Comp}. Our sub-sample of RT post-processed TNG galaxies follows the mass-metallicity (MZ) and main sequence (MS) relations derived by \cite{Lopez13} for the SDSS \citep{SDSS_Adelman,SDSS_Abazajian} and GAMA \citep{GAMA} samples. By construction, the GSWLC sample also follows these relationships, as it is a sub-sample of the SDSS catalog. However, at the high-mass end ($M_{\star} > 10^{10}\,\rm{M_{\odot}}$), the GSWLC dataset contains a population of quiescent galaxies with low star formation rates ($\mathrm{SFR} < 10^{-1}\,\rm{M_{\odot}/yr}$). While a small fraction of such galaxies is also present in our simulated sample, they are rare, comprising less than the $16^{\rm{th}}$ percentile of the distribution.

Overall, the two datasets exhibit significant overlap, confirming that the simulated galaxies provide a statistically representative match to the observed galaxy population, particularly for star-forming systems.
We emphasize that the goal of the subsequent comparison is not to achieve a one-to-one match between simulated and observed sources. Rather, we aim to assess whether the synthetic attenuation curves derived from post-processed TNG simulations systematically differ from the observed attenuation curves, and to verify whether the relations underpinning our simulation-based scaling relations hold true for the observational sample.
The GSWLC-inferred attenuation curves are particularly well-suited for this comparison, as they represent the global attenuation properties of a statistically significant sample of typical local galaxies \citep[see][for further details]{Salim18}. As a sub-sample of the extensively studied SDSS catalog, the GSWLC spans a broad parameter space, offering a much more meaningful benchmark than the limited local extinction curves (SMC, MW, LMC), which are derived from individual stars and lines of sight.

\begin{figure*}
    \centering
    \includegraphics[width=0.97\textwidth]{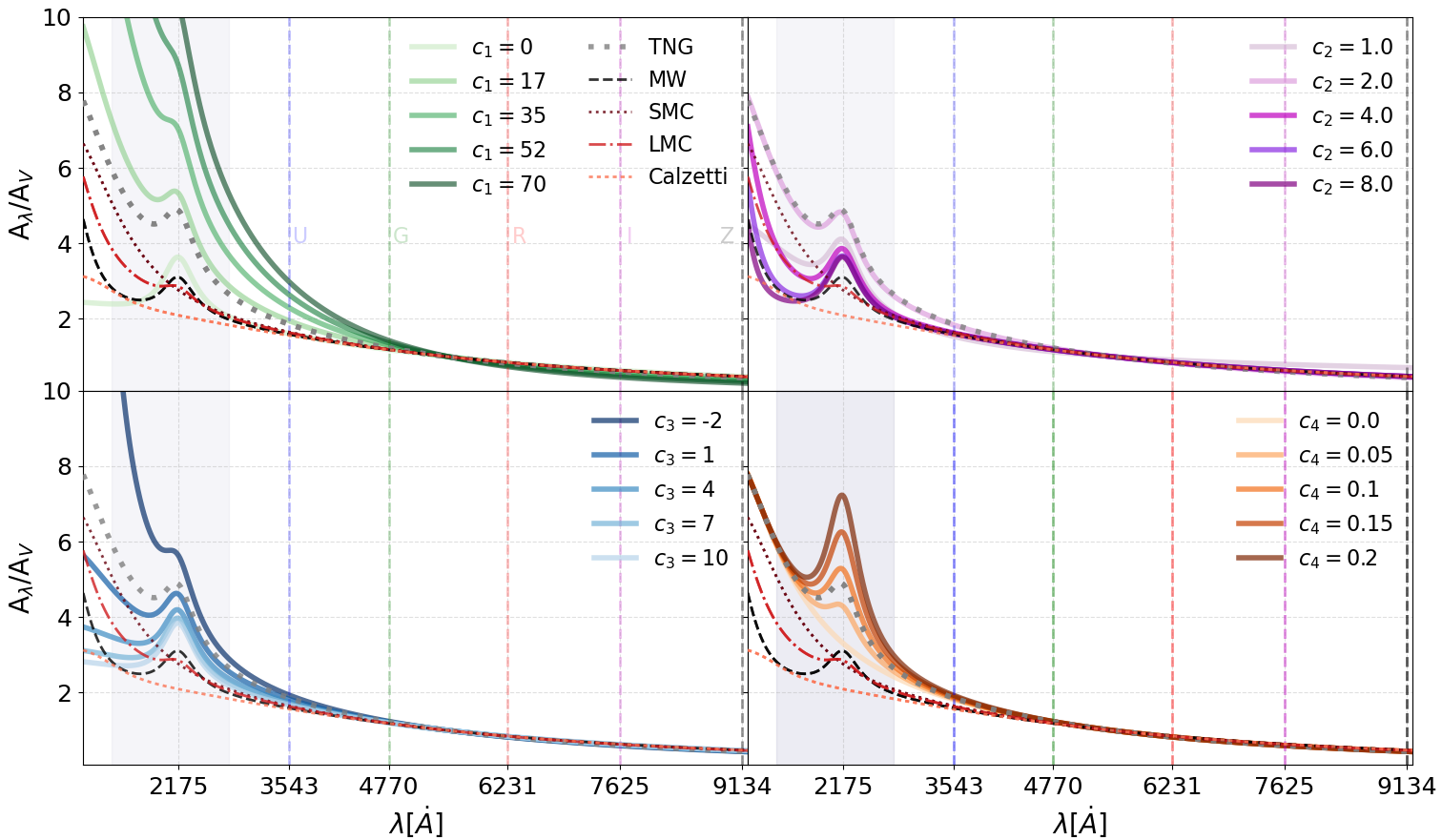}
    \caption{Summary plot illustrating the effect of independently varying each of the attenuation curve parameters $(c_1,c_2,c_3,c_4)$ from eq.~\ref{eq_Li08}, starting from the median attenuation curve of the combined TNG galaxy sample, shown as the thick grey dotted line. The median parameters are $[c_1,c_2,c_3,c_4]=[12.8, 6.52, -0.11, 0.08]$ (see Tab.~\ref{Table_best_fit_dust_par_TNG_Salim}). The vertical colored lines indicate the central wavelengths of the SDSS filters ($u$, $g$, $r$, $i$, $z$), while the gray shaded region highlights the wavelength coverage of the GALEX bands.
}
    \label{Fig_Param_Att_Curve}
\end{figure*}

\subsection{Dust attenuation curve parametrization} 
For each TNG galaxy, we fit dust attenuation curves along the $51$ lines of sight using the flexible parametrization proposed by \cite{Li08}. We do the same for the observed galaxies from the GSWLC sample, re-fitting the attenuation curves derived originally from SED fitting in \cite{Salim18}. With parameter adjustment, our adopted parametrization can describe a range of standard dust curves (Calzetti, MW, SMC and LMC), as well as non-conventional ones possibly resulting from complex dust and star geometries. The analytical expression derived by \cite{Li08}, normalized to the attenuation in the $V-$band ($0.55\,\mu\rm{m}$) $ A_{\rm V}$, is:
\begin{equation}\label{eq_Li08}
\begin{split}   
A_{\lambda}/ A_{\rm V} &= \frac{c_1}{(\lambda/0.08)^{c_2}+(0.08/\lambda)^{c_2} +c_3} \\ 
& +  \frac{233[1-c_1/(0.145^{-c_2}+0.145^{c_2}+c_3)-c_4/4.60]}{(\lambda/0.046)^2+(0.046/\lambda)^2+90} \\ 
& + \frac{c_4}{(\lambda/0.2175)^2+(0.2175/\lambda)^2-1.95},
\end{split}
\end{equation}
where $c_1$, $c_2$, $c_3$ and $c_4$ are dimensionless parameters and $\lambda$ is the wavelength in $\mu \rm{m}$ units. The three terms in eq.~\ref{eq_Li08} describe i) the far ultraviolet (FUV) attenuation rise, ii) the attenuation in the optical and near infrared (NIR) range, and iii) the $2175\,\angstrom$ bump strength, respectively. 
The $c_1-c_4$ parameters reproducing the Calzetti, SMC, LMC, MW templates are reported in Tab.~\ref{Table_Known_Att_Curves}. The reported values correspond to the best-fit values of the fiducial empirical extinction/attenuation curves, but it is worth mentioning, that variations due to different sightlines are observed in the MW, SMC and LMC \citep{Salim20}. Next, we discuss the impact of each parameter on the attenuation curve.

\begin{table}[h]
\centering
 \caption{Model fit parameters for the Calzetti, SMC, MW, and the LMC templates, drawn from \cite{Li08}.
 \label{Table_Known_Att_Curves}
 }
\begin{tabular}{ccccc}
 \hline \hline
  \noalign{\smallskip}
   & Calzetti & SMC & MW & LMC  \\
   \noalign{\smallskip}
 \hline
 \hline
 \noalign{\smallskip}
$c_1$ & 44.9 & 38.7 & 14.4 & 4.47 \\
 \noalign{\smallskip}
$c_2$ & 7.56  &  3.83 & 6.52 &  2.39\\ %
 \noalign{\smallskip}
 $c_3$ & 61.2 &  6.34 & 2.04 &  -0.988 \\ %
 \noalign{\smallskip}
$c_4$ & 0& 0   & 0.0519 & 0.0221\\ %
 \noalign{\smallskip}
\hline
\end{tabular}
\end{table}

In Fig.~\ref{Fig_Param_Att_Curve}, we show the effect on $A_{\lambda}/A_{\rm V}$ resulting from changing one of the four parameters $(c_1,c_2,c_3,c_4)$ at a time. 
Increasing $c_1$ results in a steeper UV rise at $\lambda < 2175\,\angstrom$.
However, $A_{\lambda}/A_{\rm V}$ is also impacted by $c_3$ at shorter wavelengths ($\lambda<1500\,\angstrom$), introducing a degeneracy. 
Decreasing $c_3$ results in a steeper FUV rise. In fact, the Calzetti attenuation curve, which is the most shallow among the local empirical attenuation curves, is actually characterized by the largest $c_1=44.9$, but also the largest $c_3=61.2$. 
The parameter $c_2$, sets the global slope of the attenuation curve, with values $0<c_2<1$ characterizing exotic $A_{\lambda}/A_{\rm V}$ curves with greater attenuation at longer wavelengths. Such a scenario can only result from scattering in peculiar stellar-to-dust geometries, since the extinction cross section for both silicates and carbonaceous grains is monotonically decreasing (with the only exception being the $2175\,\angstrom$ feature) with wavelength at $\lambda>700\,\angstrom$ \citep{Draine03}.
For values $c_2>1$, the dependence of the slope of the attenuation curve on $c_2$ is non-monotonic. Larger $c_2$ values in the range $1<c_2<2$ result in steeper curves, but for $c_2>2$, larger $c_2$ values result in shallower $A_{\lambda}/A_{\rm V}$. It is worth noting that all empirical extinction/attenuation curves (MW, SMC, LMC, Calzetti) have $c_2>2$, thus $c_2$ in this range is inversely correlated with the steepness of $A_{\lambda}/A_{\rm V}$. Finally, the parameter $c_4$ is straightforward to interpret: $c_4=0$ corresponds to no $2175\,\angstrom$ bump, while larger $c_4$ values indicate a more pronounced $2175\,\angstrom$ feature.

\begin{figure*}
    \centering
    \includegraphics[width=0.999\textwidth]{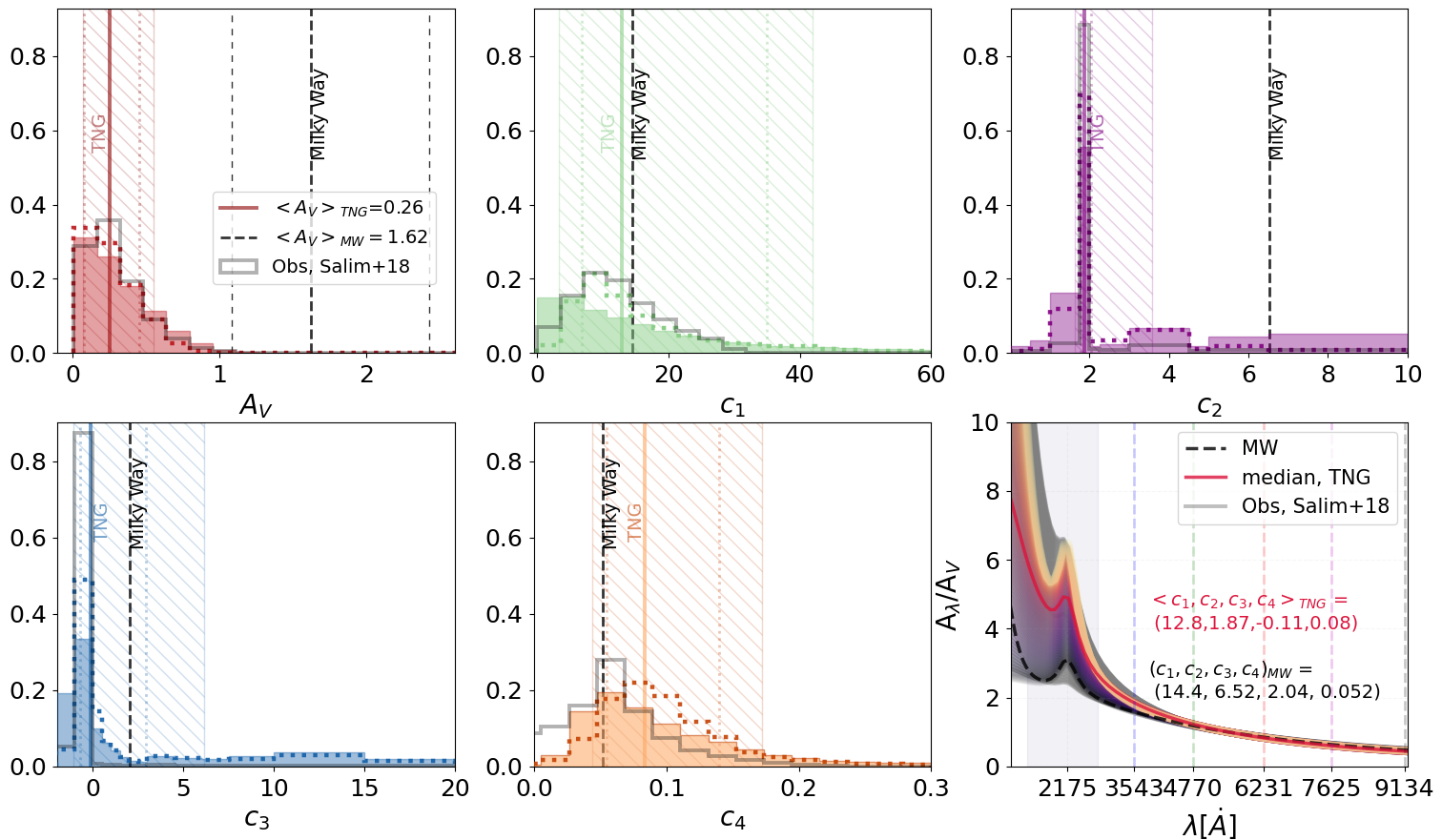}
    \caption{\textbf{Panels 1-5}: distribution of the best-fitting values of $ A_{\rm V}$ and parameters $c_1,c_2,c_3,c_4$ for the TNG galaxies analyzed. The filled (dashed) histograms show the PDF of the (median) values considering (over) all lines for all sources. \textbf{Panel 6}: attenuation curves corresponding to the median, $16^{\rm{th}}$ and $84^{\rm{th}}$ percentiles of the attenuation curve parameters values shown in the previous histograms compared to the MW curve (dashed black line). The grey lines show the $10^{\rm{th}}-84^{\rm{th}}$ percentiles of the attenuation curves from the GSWLC observed sample by \cite{Salim18}.}
    \label{Fig_Param-Hist_TNG}
\end{figure*}
\begin{table*}[!t]
    \centering
    \renewcommand{\arraystretch}{1.28}
     \caption{Median values and percentile ranges of the parameters $A_{\rm V}, c_1, c_2, c_3$ and $ c_4$, obtained from {\it{emcee}} fits to the synthetic attenuation curves generated for the TNG50 and TNG100 galaxies. For each parameter, the first row shows the median value, while the second row provides the $16^{\rm{th}}$ and $84^{\rm{th}}$ percentiles in parentheses. These values incorporate all $51$ lines of sight for each of the TNG galaxies. The final column displays the corresponding values for the measured attenuation curves from the \cite{Salim18} GSWLC sample, as shown in the right-hand panel of Fig.~\ref{Fig_Mstar-Z_Obs_Comp}.
     \label{Table_best_fit_dust_par_TNG_Salim}
     }
\begin{tabular}{lcccc}
\hline
\hline
Parameter & TNG50 & TNG100 & TNG (combined) & Obs (Salim+18) \\
\hline
\multirow{2}{*}{$A_{\rm V}$} 
& 0.17 & 0.32 & 0.26 & 0.24 \\
& (0.06, 0.38) & (0.10, 0.61) & (0.08, 0.55) & (0.13, 0.47) \\
\multirow{2}{*}{$c_1$} 
& 12.8 & 12.8 & 12.8 & 11.3 \\
& (3.8, 39.6) & (3.0, 43.2) & (3.3, 41.9) & (6.4, 19.8) \\
\multirow{2}{*}{$c_2$} 
& 1.88 & 1.87 & 1.87 & 1.89 \\
& (1.74, 3.27) & (1.57, 3.66) & (1.64, 3.56) & (1.88, 1.90) \\
\multirow{2}{*}{$c_3$} 
& -0.28 & 0.00 & -0.11 & -0.67 \\
& (-1.09, 3.77) & (-1.10, 7.41) & (-1.21, 6.13) & (-0.82, -0.60) \\
\multirow{2}{*}{$c_4$} 
& 0.10 & 0.08 & 0.08 & 0.055 \\
& (0.05, 0.21) & (0.04, 0.16) & (0.04, 0.17) & (0.020, 0.094) \\
\hline
\hline
\end{tabular}
\end{table*}

\section{Fitting the attenuation curves of TNG galaxies}\label{fitting_attenuation_curves}
We fit the simulated and observationally derived attenuation curves with the function described in eq.~\ref{eq_Li08} using the affine-invariant ensemble sampler for Markov chain Monte Carlo (MCMC) proposed by \cite{Goodman10} and implemented in python as the open source code {\it{emcee}} by \cite{emcee}. We choose flat priors for each parameter and the following wide range of values: $0 < c_1 < 200$, $0.0 < c_2 < 100$, $-200 < c_3 < 200$, and $0.0 < c_4 < 0.8$. The values of the parameters describing the most commonly used empirical extinction/attenuation curves (listed in Tab.~\ref{Table_Known_Att_Curves}) are well within these ranges. Nevertheless, we allow for such wide ranges to capture the exotic attenuation curves that can result from complex dust-to-stellar geometries \citep[see e.g.][]{Popping17,Salim18,Narayanan18,SeonDraine16,DiMascia24}.

\begin{figure*}
    \centering
\vspace{-0.5cm}    \includegraphics[width=0.93\linewidth]{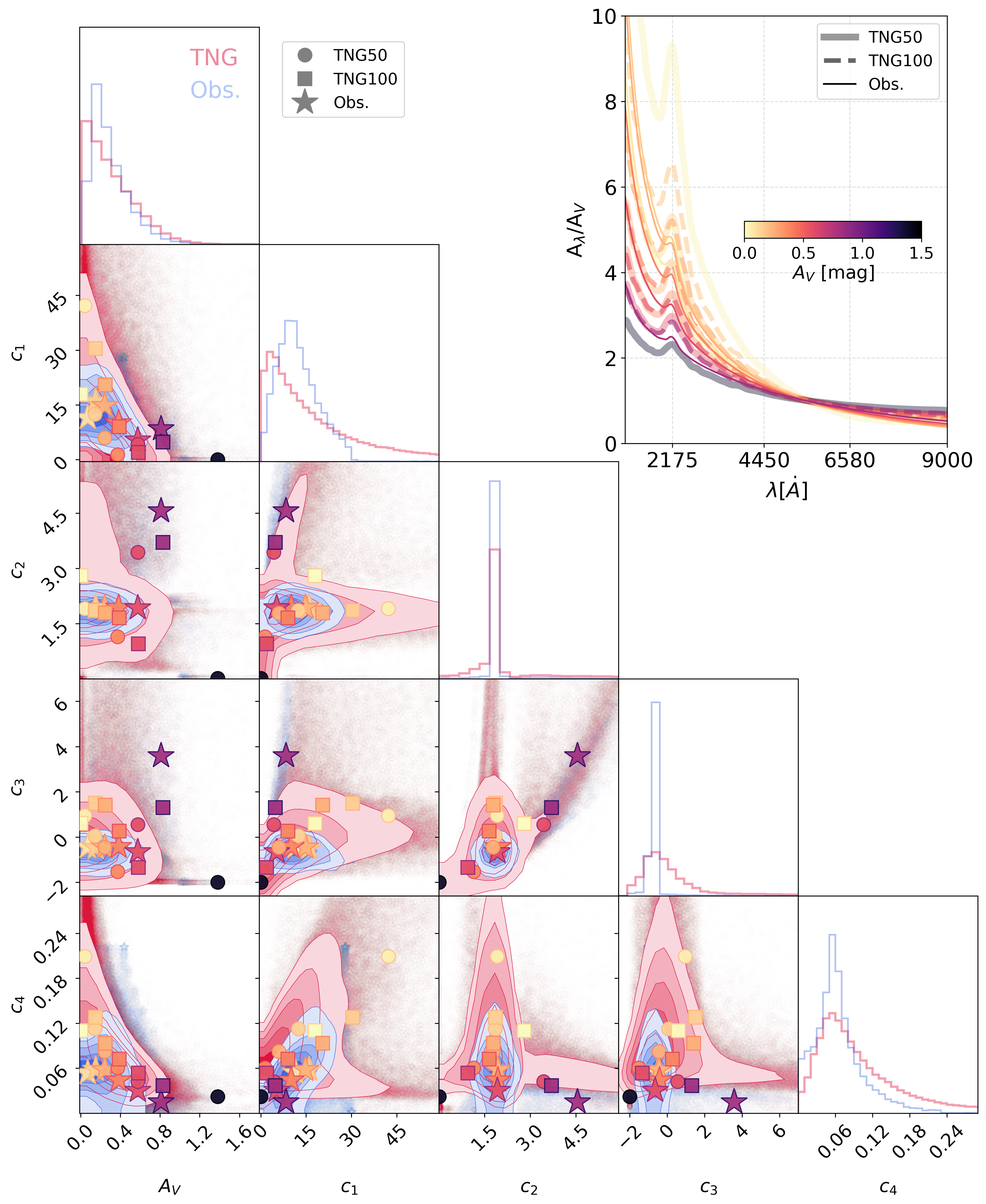}
    \caption{ \textbf{Main Panel:} Corner plot showing the two-dimensional distributions of the best-fit attenuation curve parameters (\(c_1\), \(c_2\), \(c_3\), \(c_4\)) and \(A_V\) for observed galaxies (blue) and simulated galaxies from the TNG simulations (red). The one-dimensional histograms are shown along the diagonal (see also Fig.~\ref{Fig_Param-Hist_TNG} for more detailed 1D distributions). Overlaid are the best-fit parameters derived from fitting the median attenuation curves of TNG50 (circles), TNG100 (hexagons), and the observed sample (stars), binned by attenuation \(A_V\) into six bins: \(A_V < 0.1\), \(0.1 \leq A_V < 0.2\), \(0.2 \leq A_V < 0.3\), \(0.3 \leq A_V < 0.5\), \(0.5 \leq A_V < 0.7\), and \(A_V \geq 0.7\). \textbf{Inset Panel:} Median attenuation curves binned by \(A_V\) for TNG50 (solid, thick lines), TNG100 (dashed, thick lines), and observed galaxies from \citet{Salim18} (solid, thin lines). The colors represent the median \(A_V\) in each bin for each sample (see colorbar), and the symbols (corresponding to the attenuation curve parameters describing each of the curves) in the main panel are consistently colored to match. }
    \label{Fig_Att_Curve_Avbins_Corner}
\end{figure*}

The distributions of the best-fitting parameters $( A_{\rm V},c_1,c_2,c_3,c_4)$ for the 6211 $z\sim0$ simulated galaxies from the TNG suite are summarized in Fig.~\ref{Fig_Param-Hist_TNG} and Fig.~\ref{Fig_Att_Curve_Avbins_Corner}. Note that these are constructed using fits to the 51 different viewing angles for each galaxy (totaling $316,761$ values). The median values and $16^{\rm{th}}$ and $84^{\rm{th}}$ percentiles of the five parameters are provided in Tab.~\ref{Table_best_fit_dust_par_TNG_Salim} for the TNG50, TNG100, and combined samples.
As detailed in Appendix~\ref{Apped_resol}, we confirm that in the overlapping stellar mass range, differences in resolution between TNG50 and TNG100 do not significantly impact the modeled dust attenuation curves, justifying the use of the combined sample. This result is likely due to the pressurized effective equation of state imposed on the ISM that is used in both TNG50 and TNG100, as well as the fact that neither simulation reaches the required resolution to model the small-scale ISM structures, critical for the detailed dust-to-stellar geometry in star-forming regions -- a notable limitation. 
Moreover, dust models often used in SED fitting or forward modeling of simulation data, separate the dust attenuation from the stellar birth clouds and the dust attenuation from the larger gas-to-dust geometry in the galaxy. In this work, we focus on understanding and improving the modeling of the larger-scale dust attenuation in the galaxy.

The median attenuation in the $V-$band is $ A_{\rm V}= 0.26$. 
Notably, the median values of the best-fitting parameters $(c_1,c_2,c_3,c_4)$ for our TNG galaxies deviate significantly from the values of the MW extinction curve, despite our assumption of MW-like optical grain properties and size distribution within the radiative transfer. The fitted parameters are consistent with a steeper curve ($\left<c_3\right>_{\rm TNG}<0$ and $\left<c_{2}\right>_{\rm TNG}\sim 2$) and, on average, a $\sim 1.5$ times stronger $2175\,\angstrom$ feature ($\left<c_4\right>_{\rm TNG}=0.08)$. If all lines of sight are included in the PDFs of the attenuation curves, the MW parameters are still within the  $16^{\rm{th}}-84^{\rm{th}}$ percentile ranges in all cases except  $c_2$ (which is narrowly peaked around $2$). It is well-known that the enforcement of a MW-like chemical composition (silicate and carbonaceous grains) and grain size distribution does not necessarily result in a MW-like attenuation curve. In fact, dust-to-star geometry effects -- which correlate with global galaxy properties -- play a major role in shaping galaxy attenuation curves \citep{SeonDraine16}. 

When considering all lines of sight, the probability distribution functions (PDFs) of the attenuation curve parameters for the simulated galaxies closely resemble those obtained for the observed galaxies in the GSWLC sample, effectively capturing their range of variation. The FUV slope parameter \(c_3\) and overall slope parameter \(c_2\) show an even more pronounced peak in the observations relative to TNG. This is unsurprising given that the measured attenuation curves (which we re-fit here with the 4-parameter model) were obtained from SED fitting relying on a model with only two free parameters, the slope and bump strength (see the parameterization in \citealt{Salim18}). Interestingly, \(c_4\) is marginally lower in the observed galaxies (\(0.055\)) compared to TNG (\(0.08\) for the combined dataset), implying a less pronounced bump. This might be due to the assumption of MW-like dust in all our RT post-processing, and alternative choices will be explored in future work. Nevertheless, in all cases, the differences are within \(1\sigma\), indicating overall consistency between the forward-modeled simulations and observations. 

We find that the variation in the shape of the attenuation curve is mostly associated with the change of $A_V$ along the line of sight. 
This is shown in Fig.~\ref{Fig_Att_Curve_Avbins_Corner}, where we present the median attenuation curves for TNG50, TNG100, and observed galaxies \citep{Salim18} binned by \(A_{\rm V}\).  
We find that the attenuation curves become flatter with increasing \(A_{\rm V}\) across all datasets. The \cite{Salim18} curves are characterized by systematically smaller UV bumps compared to the TNG galaxies across all \(A_{\rm V}\) bins. This might be due to our choice of MW dust composition (which includes PAHs producing the $2175\,\angstrom$ feature in the WD01 dust model).
The percentage of galaxies within each \(A_{\rm V}\) bin across all samples explains why the attenuation curves are mostly steeper than the MW or Calzetti ones. The high \(A_{\rm V}\) bins (\(A_{\rm V} > 0.5\)) always contain less than $10\%$ of the total sources, indicating that low- to intermediate-\(A_{\rm V}\) galaxies dominate the samples. 

In addition to the 1D histograms, Fig.~\ref{Fig_Att_Curve_Avbins_Corner} presents the correlations between different attenuation curve parameters in both simulations and observations. A strong correlation is evident between $A_{\rm V}$ and both $c_1$ and $c_4$, a trend that is further reinforced when considering the best-fit values of the binned attenuation curves. As a result, $c_1$ and $c_4$ are also correlated, albeit less strongly than either is with $A_{\rm V}$. Additionally, we find a trend of increasing $c_3$ with increasing $c_1$. We also observe hints of bimodal behavior in $c_2$ and $c_3$, although this primarily affects the tails of the simulated distribution and is even less pronounced in the observed dataset.
The median values corresponding to the highest $A_{\rm V}$ bins (purple and black points in Fig.~\ref{Fig_Att_Curve_Avbins_Corner}) often fall outside the main contours of the full TNG and GSWLC distribution. This is expected, as our sample is biased towards lower $A_{\rm V}$ sources and lines of sight. Consequently, our results are most applicable to main-sequence, UV-bright galaxies rather than highly dust-obscured sources such as submillimeter galaxies (SMGs). In the next section, we quantitatively investigate the key galaxy properties associated with variations in $A_V$ and thus in the other attenuation curve parameters.

\begin{table*}[ht!]
\centering
\renewcommand{\arraystretch}{0.9}
\begin{tabular}{c|cccccc|ccccc} 
\multicolumn{11}{r}{\textbf{SPEARMAN correlation coefficient (TNG combined)}} \\
\hline
\multicolumn{5}{r}{\textcolor{ceruleanblue}{All los}} & \multicolumn{6}{r}{\textcolor{blue}{Medians over los}} \\
\cline{1-7} \cline{8-12}
&& \textcolor{ceruleanblue}{$c_1$} & \textcolor{ceruleanblue}{$c_2$} & \textcolor{ceruleanblue}{$c_3$} & \textcolor{ceruleanblue}{$c_4$} & \textcolor{ceruleanblue}{$A_{\rm V}$} & \textcolor{blue}{$c_1$} & \textcolor{blue}{$c_2$} & \textcolor{blue}{$c_3$} & \textcolor{blue}{$c_4$} & \textcolor{blue}{$A_{\rm V}$}\\
\multirow{32}{*}{{\rotatebox[origin=c]{90}{\textcolor{seagreen}{\textbf{Galaxy Properties}}}}}
&\textcolor{seagreen}{$\mathrm{SFR_{\rm 10}}$} & -0.134 & 0.030 & 0.029 & -0.166 & 0.436 & -0.239 & -0.016 & 0.017 & -0.133 & 0.540 \\
&\textcolor{seagreen}{$\mathrm{SFR_{\rm 100}}$} & -0.135 & 0.031 & 0.031 & -0.173 & 0.445 & -0.238 & -0.012 & 0.023 & -0.148 & 0.551 \\
&\textcolor{seagreen}{$\mathrm{sSFR_{\rm 10}}$} & -0.138 & 0.035 & 0.029 & -0.171 & 0.438 & -0.248 & -0.008 & 0.019 & -0.143 & 0.547 \\
&\textcolor{seagreen}{$\mathrm{sSFR_{\rm 100}}$} & -0.140 & 0.035 & 0.031 & -0.178 & 0.448 & -0.248 & -0.005 & 0.024 & -0.158 & 0.559 \\
&\textcolor{seagreen}{Age} &  0.149 & -0.065 & 0.002 & 0.158 & -0.184 &  0.303 & -0.066 & 0.017 & 0.232 & -0.343 \\
&\textcolor{seagreen}{$r_{\rm \star,o}$} &  0.050 & -0.020 & -0.060 & 0.163 & -0.137 &  0.122 & -0.072 & -0.116 & 0.307 & -0.210 \\
&\textcolor{seagreen}{$r_{\rm \star,y}$} & -0.083 & 0.238 & -0.122 & 0.156 & -0.381 & -0.073 & 0.259 & -0.293 & 0.342 & -0.560 \\
&\textcolor{seagreen}{$r_{\rm g}$} & -0.079 & 0.086 & 0.050 & -0.154 & 0.013 & -0.017 & 0.140 & 0.062 & -0.106 & -0.132 \\
&\textcolor{seagreen}{$r_{\rm g,SF}$} & -0.071 & 0.211 & -0.066 & 0.075 & -0.380 & -0.017 & 0.259 & -0.142 & 0.204 & -0.559 \\
&\textcolor{seagreen}{$M_{\star}$}&  0.061 & -0.018 & 0.050 & -0.030 & 0.073 &  0.189 & 0.001 & 0.062 & 0.082 & -0.124 \\
&\textcolor{seagreen}{$M_{\rm g}$} & -0.081 & 0.093 & 0.039 & -0.135 & 0.052 & -0.040 & 0.119 & 0.034 & -0.067 & -0.059 \\
&\textcolor{seagreen}{$M_{\rm dust}$} & -0.071 & 0.058 & 0.057 & -0.159 & 0.173 & -0.048 & 0.065 & 0.063 & -0.117 & 0.113 \\
&\textcolor{seagreen}{$M_{\rm dust}(r_{\rm \star,y})$} & -0.052 & 0.002 & -0.015 & -0.008 & 0.212 & -0.066 & -0.102 & -0.125 & 0.158 & 0.156 \\
&\textcolor{seagreen}{$M_{\rm dust}(r_{\rm \star,o})$} &  0.071 & -0.033 & 0.062 & -0.044 & 0.101 &  0.203 & -0.017 & 0.086 & 0.056 & -0.093 \\
&\textcolor{seagreen}{$M_{\rm dust}(r_{\rm g,SF})$} &  0.005 & 0.039 & 0.039 & -0.063 & 0.056 &  0.112 & 0.070 & 0.041 & 0.042 & -0.144 \\
&\textcolor{seagreen}{$Z_{\rm g}$} & 0.061 & -0.146 & 0.113 & -0.131 & 0.492 &  0.027 & -0.199 & 0.208 & -0.231 & 0.639 \\
\cline{2-12}
& \textcolor{seagreen}{$r_{\rm \star,o}/r_{\rm g}$} & 0.108 & -0.116 & -0.078 & 0.228 & -0.056 & 0.093 & -0.215 & -0.115 & 0.263 & 0.045 \\
& \textcolor{seagreen}{$r_{\rm \star,o}/r_{\rm g,SF}$} & 0.125 & -0.298 & 0.035 & 0.033 & 0.379 & 0.117 & -0.393 & 0.101 & -0.026 & 0.534 \\
& \textcolor{seagreen}{$r_{\rm \star,o}/r_{\rm \star,y}$} & 0.127 & \textbf{-0.322} & 0.107 & -0.079 & 0.401 & 0.164 & -0.398 & 0.284 & -0.201 & 0.561 \\
& \textcolor{seagreen}{$r_{\rm \star,y}/r_{\rm g}$} & 0.004 & 0.125 & -0.122 & 0.227 & -0.334 & -0.038 & 0.157 & -0.269 & 0.342 & -0.438 \\
& \textcolor{seagreen}{$r_{\rm \star,y}/r_{\rm g,SF}$} & -0.009 & 0.159 & -0.098 & 0.189 & -0.287 & -0.03 & 0.212 & -0.239 & 0.301 & -0.392 \\
& \textcolor{seagreen}{$M_{\rm g}/M_{\star}$} & -0.173 & 0.164 & -0.019 & -0.129 & -0.084 & -0.28 & 0.199 & -0.084 & -0.155 & -0.044 \\
& \textcolor{seagreen}{$M_{\rm dust}/M_{\star}$} & -0.18 & 0.14 & 0.029 & -0.211 & 0.113 & -0.304 & 0.166 & -0.001 & -0.275 & 0.21 \\
& \textcolor{seagreen}{$M_{\rm dust}(r_{\rm \star,y})/M_{\rm dust}(r_{\rm g,SF})$} & -0.076 & 0.146 & -0.113 & 0.16 & -0.185 & -0.172 & 0.088 & -0.3 & 0.288 & -0.165 \\
& \textcolor{seagreen}{$M_{\rm dust}(r_{\rm \star,o})/M_{\rm dust}(r_{\rm g,SF})$} & 0.141 & -0.308 & 0.051 & 0.02 & 0.322 & 0.167 & \textbf{-0.402} & 0.135 & -0.042 & 0.447 \\
& \textcolor{seagreen}{$M_{\rm dust}(r_{\rm \star,y})/M_{\rm dust}(r_{\rm \star,o})$} & -0.15 & 0.248 & -0.118 & 0.093 & -0.211 & -0.26 & 0.201 & -0.322 & 0.231 & -0.208 \\
& \textcolor{seagreen}{$\Sigma_{\rm dust} (r_{\rm g})$} & 0.052 & -0.075 & 0.02 & 0.034 & 0.227 & -0.003 & -0.153 & 0.041 & 0.0 & 0.38 \\
& \textcolor{seagreen}{$\Sigma_{\rm dust} (r_{\rm \star,y})$} & 0.11 & -0.275 & 0.155 & -0.191 & 0.552 & 0.189 & -0.368 & 0.36 & -0.314 & 0.707 \\
& \textcolor{seagreen}{$\Sigma_{\rm dust} (r_{\rm \star,o})$} & 0.037 & -0.026 & 0.116 & -0.184 & 0.246 & 0.111 & 0.032 & 0.207 & -0.222 & 0.135 \\
& \textcolor{seagreen}{$\Sigma_{\rm dust} (r_{\rm g,SF})$} & 0.124 & -0.244 & 0.113 & -0.112 & 0.505 & 0.188 & -0.327 & 0.242 & -0.188 & 0.616 \\
& \textcolor{seagreen}{$\Sigma_{\rm dust} (r_{\rm \star,y})/\Sigma_{\rm dust} (r_{\rm \star,o})$} & 0.119 & -0.319 & 0.104 & -0.079 & 0.391 & 0.149 & -0.396 & 0.277 & -0.2 & 0.552 \\
& \textcolor{seagreen}{$\Sigma_{\rm dust} (r_{\rm \star,y})/\Sigma_{\rm dust} (r_{\rm g,SF})$} & -0.037 & -0.107 & 0.075 & -0.189 & 0.225 & -0.049 & -0.144 & 0.179 & -0.289 & 0.312 \\
& \textcolor{seagreen}{$\Sigma_{\rm dust} (r_{\rm g})/\Sigma_{\rm dust} (r_{\rm \star,y})$} & -0.063 & 0.204 & -0.114 & 0.164 & -0.361 & -0.157 & 0.287 & -0.248 & 0.222 & -0.422 \\
& \textcolor{seagreen}{$\Sigma_{\rm dust} (r_{\rm \star,o})/\Sigma_{\rm dust} (r_{\rm g,SF})$} & -0.123 & 0.296 & -0.035 & -0.033 & -0.382 & -0.113 & 0.392 & -0.099 & 0.025 & -0.538 \\
& \textcolor{seagreen}{$\Sigma_{\rm dust} (r_{\rm \star,o})/\Sigma_{\rm dust} (r_{\rm g})$} & -0.003 & 0.029 & 0.078 & -0.154 & 0.045 & 0.084 & 0.104 & 0.124 & -0.15 & -0.114 \\
& \textcolor{seagreen}{$\Sigma_{\rm SFR,y}$} & 0.020 & -0.199 & {\cellcolor{asparagus}\textcolor{bostonuniversityred}{\textbf{0.165}}} & -0.285 & {\cellcolor{asparagus}\textbf{0.615}} & 0.005 & -0.201 & {\cellcolor{asparagus}\textbf{0.379}} & \textbf{-0.471} & {\cellcolor{asparagus}\textbf{0.801}} \\
& \textcolor{seagreen}{$\Sigma_{\rm SFR,o}$} &  {\cellcolor{asparagus}\textcolor{bostonuniversityred}{\textbf{-0.188}}} & 0.066 & 0.085 & -0.339 & 0.572 & {\cellcolor{asparagus}\textbf{-0.333}} & 0.08 & 0.115 & -0.423 & 0.707 \\
& \textcolor{seagreen}{$\Sigma_{\rm SFR}$} & -0.185 & 0.061 & 0.091 & \textbf{-0.346} & 0.583 & -0.327 & 0.074 & 0.128 & -0.435 & 0.719 \\
& \textcolor{seagreen}{$\Sigma_{\rm g}$} & 0.078 & -0.085 & -0.049 & 0.153 & -0.011 & 0.016 & -0.14 & -0.062 & 0.106 & 0.134 \\
& \textcolor{seagreen}{$\Sigma_{\rm g,SF} $} & 0.07 & -0.212 & 0.069 & -0.08 & 0.388 & 0.016 & -0.26 & 0.146 & -0.209 & 0.566 \\
& \textcolor{seagreen}{$\kappa_{s,y}$} & -0.066 & -0.044 & 0.117 & -0.27 & 0.39 & -0.049 & -0.043 & 0.242 & -0.349 & 0.472 \\
& \textcolor{seagreen}{$\kappa_{s,o}$} & -0.131 & 0.063 & 0.051 & -0.21 & 0.264 & -0.168 & 0.065 & 0.03 & -0.181 & 0.249 \\
& \textcolor{seagreen}{$\kappa_{s}$} & -0.134 & 0.079 & 0.063 & -0.228 & 0.251 & -0.166 & 0.102 & 0.063 & -0.215 & 0.231 \\

\hline
\multirow{5}{*}{\rotatebox[origin=c]{90}{\textcolor{ceruleanblue}{\textbf{Dust }}}}
&\textcolor{ceruleanblue}{$c_1$} &  & 0.310 & \textbf{0.652} & 0.279 & -0.269 &  & -0.115 & 0.415 & 0.250 & -0.205 \\
&\textcolor{ceruleanblue}{$c_2$} & 0.310 &  & 0.61 & -0.16 & -0.129 & -0.115 &  & 0.237 & -0.248 & -0.208 \\
&\textcolor{ceruleanblue}{$c_3$} & {\cellcolor{babyblue}{\textbf{0.652}}} & \textbf{0.61} &  & -0.313 & 0.121 & {\cellcolor{babyblue}{\textbf{0.415}}} & 0.237 &  &\textbf{ -0.482} & 0.251 \\
&\textcolor{ceruleanblue}{$c_4$} & 0.279 & -0.16 & -0.313 &  & {\cellcolor{babyblue}{\textbf{-0.64}}} & 0.250 & \textbf{-0.248} &\textbf{ -0.482} &  & {\cellcolor{babyblue}{\textbf{-0.471}}} \\
&\textcolor{ceruleanblue}{$A_{\rm V}$} & -0.269 & -0.129 & 0.121 & \textbf{-0.64} &  & -0.205 & -0.208 & 0.251 & -0.471 &   \\

\hline
\end{tabular}
\caption{Spearman correlation coefficient among attenuation curve parameters $(c_1,c_2,c_3,c_4, A_{\rm V})$ and between these parameters and key galaxy properties (see definitions in Tab.~\ref{tab:galaxy_properties_TNG}). We highlight the largest correlation coefficients in bold for each column and color the cells if the highest correlation is conserved when accounting for all lines of sight. When the correlation coefficient is $<0.2$ we color it in red.}
\label{Table_SpearmanCoeff_TNG_all}
\end{table*}

\section{Correlations between fitted attenuation curve parameters and galaxy properties}\label{Sect_Dust_GalaxyProp_Correl}
 
To guide the subsequent statistical analysis, we begin by comparing the intrinsic physical properties and forward-modeled attenuation curves of the TNG galaxies. We systematically examine a broad set of galaxy properties, seeking correlations with the attenuation curve parameters $c_1$, $c_2$, $c_3$, $c_4$, and $A_{\rm V}$. Adding to the properties available in the TNG dataset, we investigate the quantities defined in Tab.~\ref{tab:galaxy_properties_TNG}. We additionally look for correlations among the attenuation curve parameters themselves. We employ the Spearman correlation coefficient to quantify these relationships. Unlike the Pearson correlation coefficient, which measures the strength of a linear relation between two variables, the Spearman coefficient assesses the ranked-order correlation, capturing the strength and direction of a monotonic relation even if it is not linear. All correlations were analyzed using the combined TNG sample. We also computed the correlations separately for the TNG50 and TNG100 samples, but the results did not show significant differences.

We summarize the results in Tab.~\ref{Table_SpearmanCoeff_TNG_all}. We provide the correlation coefficients between galaxy and attenuation curve parameters both accounting for the full line of sight (los) distribution for each TNG source (left side of the table), and considering only the median value (over the full los distribution) for each attenuation curve parameter for each source. The latter case is studied to quantify the effect of peculiar los, which can wash out the correlation between attenuation curve parameters and galaxy properties. 

\begin{figure*}[hbt!]
    \centering
    \includegraphics[width=0.62\textwidth]{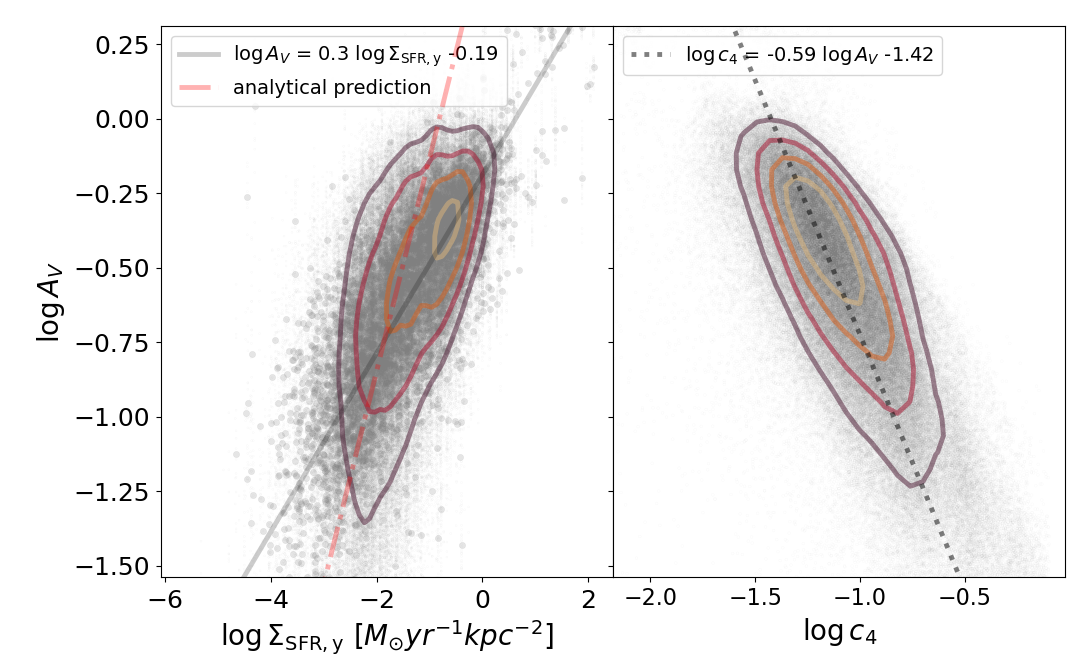}
    \includegraphics[width=0.62\textwidth]{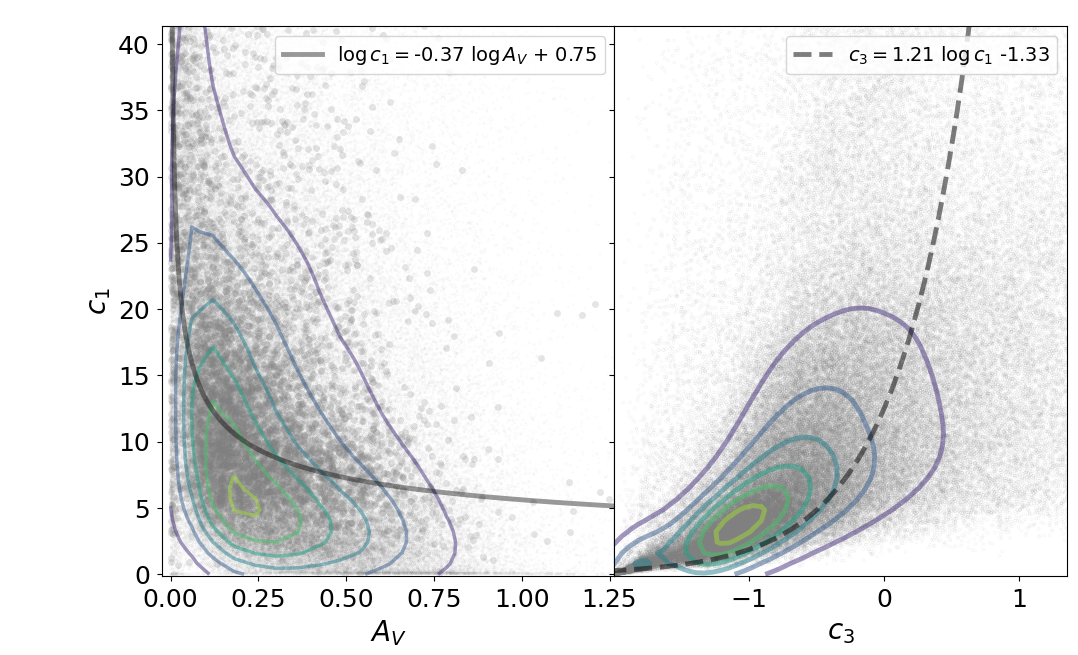}
    \caption{\textbf{Top panel}: Distribution of $A_{\rm V}$ vs. $\Sigma_{\rm SFR, y}$ (left) and $A_{\rm V}$ vs. $c_4$ (right) for TNG galaxies. Larger points in the left-hand panels represent the median value (over all lines of sight) for each source, while smaller symbols show values for individual lines of sight. The black lines represent the best-fit relations derived for TNG galaxies (solid eq.~\ref{eq_Av_SigmaSFRy}, dotted eq.~\ref{c4Av_eq}) whereas the red dotted-dashed line shows the analytical prediction derived in eq.~\ref{Av-sigmaSFR-anal}. \textbf{Lower panel}: The same as the top panel, but for $c_1$ vs. $A_V$, and $c_1$ vs. $c_3$. As for the top panel, the black lines show best-fit relations for TNG galaxies (solid eq.~\ref{c1Av_eq}, dashed eq.~\ref{c1c3_eq}).}
    \label{Fig_Correlations_fits}
\end{figure*}

We find that the strongest correlation between any attenuation curve parameter and any galaxy property is between $ A_{\rm V}$ and the SFR surface density of young star-forming regions (stellar age $\leq 10\,\rm{Myr}$), $\Sigma_{\rm SFR, y}$. This correlation is strong ($\sim 0.6$), even when variations in the attenuation curve parameters due to different lines of sight are taken into account. Although, the correlation becomes tighter when only the median value over the los distribution is considered for each source. For $ A_{\rm V}$, the strongest relation with any other attenuation curve parameter is with $c_4$, which appears to be anti-correlated ($\sim -0.64$). We show the distributions of $ A_{\rm V},c_4,$ and $\Sigma_{\rm SFR,y}$ values found in TNG galaxies in the top panel of Fig.~\ref{Fig_Correlations_fits}. 

The slope parameters $c_1$, $c_2$, and $c_3$ exhibit relatively weak correlations with the physical properties of the galaxy ($\lesssim 0.3$ when the variation with line of sight is taken into account). These mild correlations are consistently observed with $\Sigma_{\rm SFR}$, which, as established, is strongly correlated with $A_{\rm V}$. As seen from binning the attenuation curves in $A_{\rm V}$ bins, the steepness of the curve inversely correlates with $A_{\rm V}$ (i.e. shallower curves for higher $A_V$). Thus, the mild correlation with $\Sigma_{\rm SFR}$ is likely a consequence of its strong correlation with $A_{\rm V}$. 
Furthermore, any correlation identified with $c_2$ is not meaningful due to the highly skewed distribution of $c_2$ values in the TNG galaxies. This is clearly shown in Fig.~\ref{Fig_Param_Att_Curve}, where we show that the change in $c_2$ in the range $2-4$ has a minor impact on the attenuation curve (see the difference between the top two curves in the changing $c_2$ panels). The lack of strong correlations between parameters corresponding to FUV-to-NIR slopes and galaxy properties is not surprising; in fact, the global attenuation curve slope of a galaxy is known to be influenced by the dust-to-stellar geometry in different star-forming regions within the galaxy, which has a nontrivial combined effect on the global attenuation curve. 

Interestingly, when examining the combined TNG sample, we did not find clear trends between any attenuation curve parameter and stellar mass. Specifically, in the complete sample we cannot identify a clear trend in the shape of the attenuation curve and/or $A_V$ with $M_{\star}$, contrary to previous studies (see e.g., \citealt{Salim20}). Indeed, within our sample of TNG galaxies, $M_{\star}$ is poorly correlated with $\Sigma_{\rm SFR,y}$ (or any SFR surface density),
which is the key parameter that appears to affect dust attenuation. 

\begin{figure*}
\centering
    \includegraphics[width=0.45\linewidth]{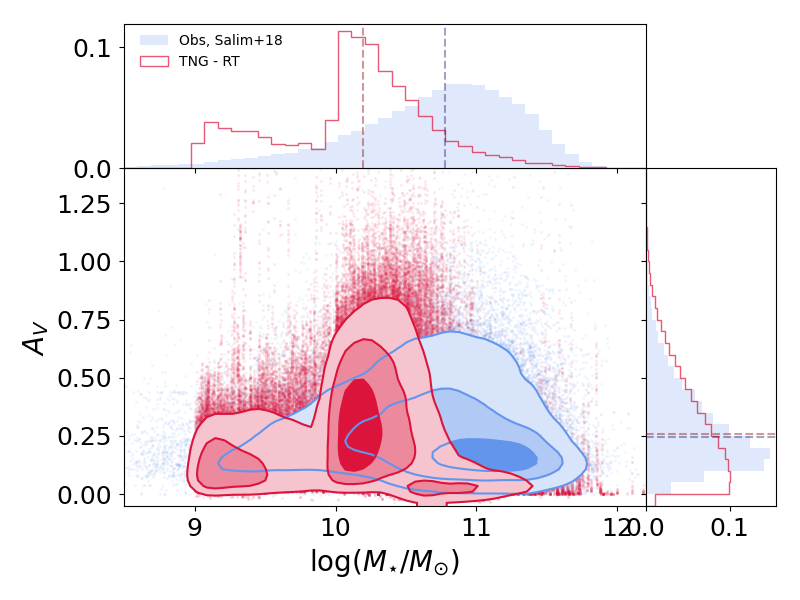}
    \includegraphics[width=0.45\linewidth]{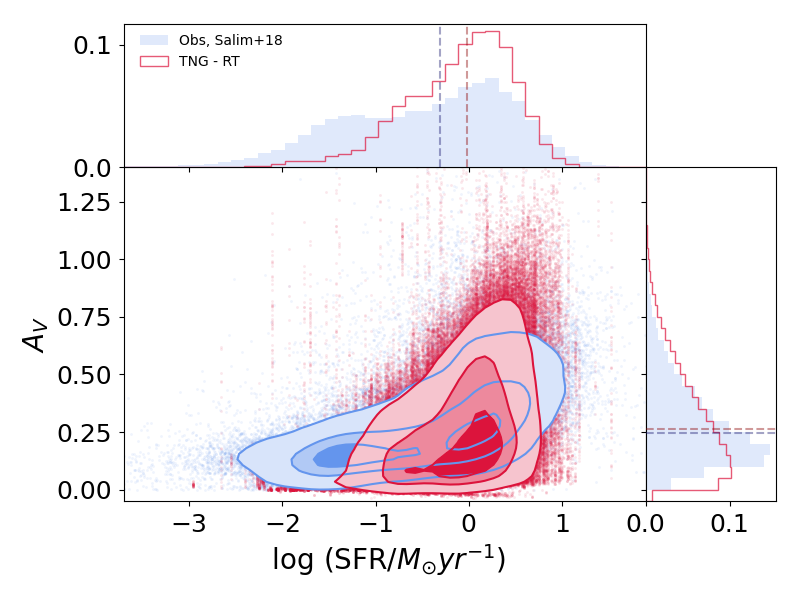}\\
    \includegraphics[width=0.45\linewidth]{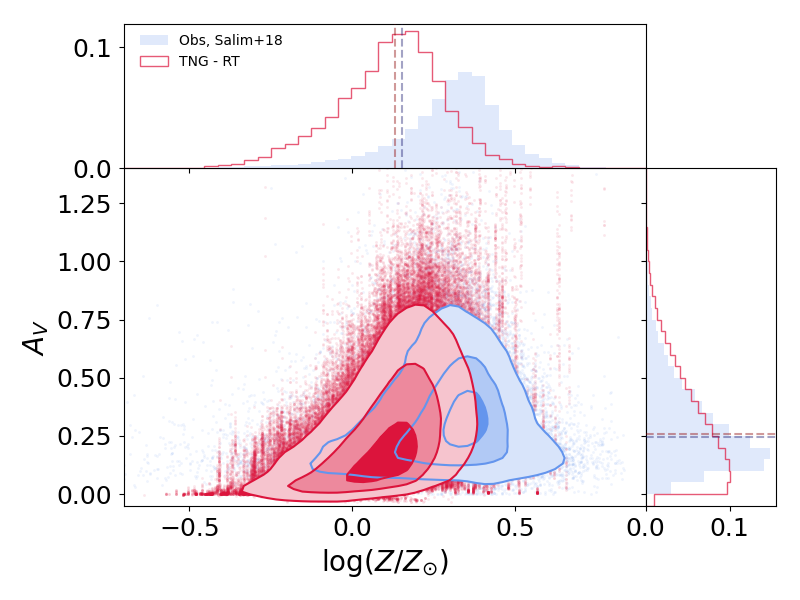}
    \includegraphics[width=0.45\linewidth]{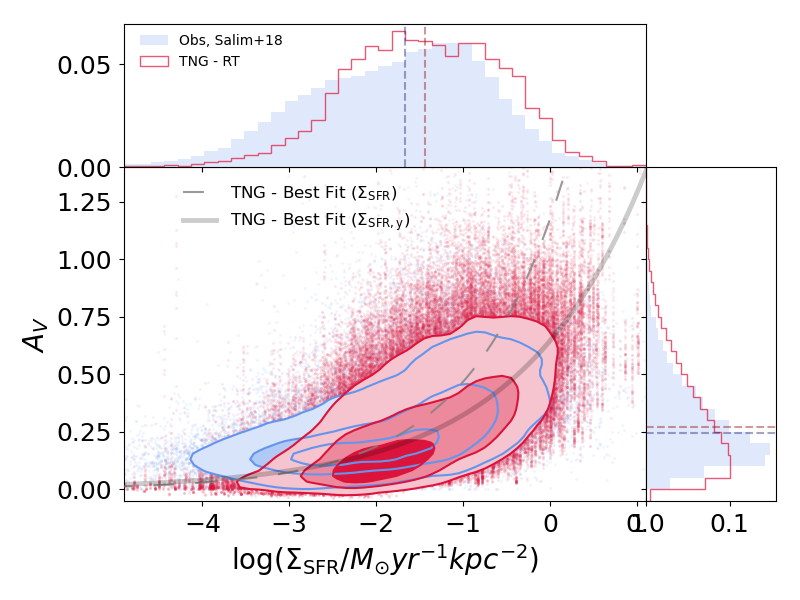}
    \caption{\small{2D PDF of $A_{\rm V}$  as a function of stellar mass \textbf{(upper panel)}, SFR \textbf{(second panel)}, gas metallicity $Z_{\rm g}$ \textbf{(third panel)}, and SFR surface density $\Sigma_{\rm SFR}$ \textbf{(lower panel)} values in the simulated TNG galaxies (in red) and in the previously discussed GSWLC sample (in blue). The dashed lines show the median values in the observed (blue) and simulated (red) samples. The 1D PDFs of $A_{\rm V}$, $M_{\star}$, SFR, $Z_{\rm g}$, and $\Sigma_{\rm SFR}$ are also shown in the horizontal and vertical histograms. 
 }}
    \label{Fig_Obs_Comp_Av}
\end{figure*}

We now turn to looking at correlations between the fitted attenuation curve parameters. We find significant correlations among the different slope parameters $c_1$, $c_2$, and $c_3$, with the strongest correlation between $c_1$ and $c_3$ ($0.65$ when all lines of sight are considered). This is somewhat expected, as both parameters describe UV regime slopes, albeit $c_3$ only characterizes the FUV. Interestingly, the internal correlations among attenuation curve parameters become shallower when considering only the median values across the los distribution for each source, rather than the full distribution of viewing angles. Conversely, this approach tightens the correlations between galaxy properties and attenuation curve parameters.

For the subsequent analytical treatment of attenuation curves, we will rely on the best-fitting relations found when the full distribution of viewing angles is included, even if these correlations are shallower. This approach better matches observational data, where each galaxy has a random orientation and thus a random line of sight (notwithstanding selection effects that may drive preferential viewing angles of certain samples).

\subsection{Comparing attenuation curve-galaxy property correlations in observations and simulations}\label{Sect_Obs_Comparison_Correlations} 

We already compared the distributions of the fitted dust attenuation law parameters \(A_{\rm V}\), \(c_1\), \(c_2\), \(c_3\), and \(c_4\) between the simulated TNG galaxies and the observed sample in Sec.~\ref{fitting_attenuation_curves} and Tab.~\ref{Table_best_fit_dust_par_TNG_Salim}. We find that the attenuation curve parameter distributions for simulated galaxies closely resemble the observed ones, effectively capturing their range of variation. Observations show a sharper peak for \(c_3\) and \(c_2\) and a slightly weaker bump strength (\(c_4 = 0.055\) vs. \(0.08\) in TNG), consistent with the simpler SED fitting model used to obtain the observationally derived attenuation curves. Despite these nuances, the differences remain within \(1\sigma\), indicating overall agreement between simulations and observations.

Following the same steps as in our synthetic attenuation curve analysis (see Sec.~\ref{Sect_Dust_GalaxyProp_Correl}), we look for any correlation between measured global galaxy properties and attenuation curve parameters in the GSWLC sample. The Spearman correlation coefficients for the observed galaxy sample reveal that the visual attenuation, \(A_V\), has the strongest relationships with galaxy properties, showing significant correlations with \(\mathrm{SFR_{100}}\) (\(0.62\)), and the SFR surface densities \(\log \Sigma_{\mathrm{opt}}\) (\(0.62\)), \(\log \Sigma_{\mathrm{FUV}}\) (\(0.61\)), and \(\log \Sigma_{\mathrm{NUV}}\) (\(0.69\)) \citep{Salim23}. The surface area used to normalize the SFR is based on isophotal sizes in the optical FUV and near UV, respectively. We also find a (slightly less, $0.44$) strong correlation between $A_V$ and the $\Sigma_{\rm SFR}$ derived using the effective radius (half-light radius derived from S\'{e}rsic profile fits in the $R$-band by \citealt{Meert15}) rather than the isophotal size.
Among the attenuation curve parameters, \(c_1\) exhibits a moderate anticorrelation with \(A_V\) (\(-0.33\)) and \(\log \Sigma_{\mathrm{opt}}\) (\(-0.12\)). The bump strength parameter \(c_4\) is also correlated with \(c_1\) (\(0.60\)) and \(A_V\) (\(-0.25\)). Conversely, \(c_2\) and \(c_3\), due to their very peaked distributions, show no significant dependencies on galaxy properties. 

Comparing the correlations identified between galaxy properties and attenuation curve parameters reveals encouraging similarities between the observed and simulated samples. In both cases, the strongest correlation is between $A_{\rm V}$ and $\Sigma_{\rm SFR}$. This is evident from Fig.~\ref{Fig_Obs_Comp_Av}, where we show the remarkable consistency in the distributions of $A_{\rm V}$ versus $\Sigma_{\rm SFR}$ between TNG simulated and GSWLC galaxies\footnote{In the plot, we present $\Sigma_{\rm SFR,y}$ for the simulations, as it is the most strongly correlated quantity with $A_{\rm V}$. For the observations, we use $\Sigma_{\rm SFR}$, since $\mathrm{SFR}_{10}$ (the star formation rate averaged over the past $10\,\rm{Myr}$) is not available.}.
In the same Figure, we also see that the $A_{\rm V}$ versus SFR$_{100}$ distributions exhibit very good agreement. Interestingly, neither the simulations nor the observations show a clear monotonic trend between $A_{\rm V}$ and $M_{\star}$; galaxies with the largest stellar masses in both samples tend to have lower $A_{\rm V}$ values. This is consistent with the trend in metallicity shown in Fig.~\ref{Fig_Mstar-Z_Obs_Comp}.

Regarding the other attenuation curve parameters, as seen for the simulated galaxies, we find that the correlations between galaxy properties and slope or bump parameters ($c_1$, $c_2$, $c_3$, $c_4$) are weaker. Regarding internal correlations, we find an anti-correlation between $A_{\rm V}$ and $c_4$ (observed: $-0.25$; simulated: $-0.64$) and $A_{\rm V}$ and $c_1$ (observed: $-0.33$; simulated: $-0.27$), consistent with the TNG predictions. However, a direct comparison for $c_2$ and $c_3$ is limited due to the two-parameter fitting function used in \citet{Salim18} to derive the observed attenuation curves. A full reanalysis of the SED fits for the observational sample is beyond the scope of this work.

\section{Scaling relations for dust attenuation curves}\label{Sect_Analytical_Model}
With the previously identified correlations between attenuation curve parameters and galaxy properties, as well as among the parameters, we can now build a first order analytical model to predict a galaxy's dust attenuation law, given a small number of intrinsic or observable galaxy properties.

Assuming a simplistic spherical or planar dust screen geometry with cold gas radius $r_{\rm g}$, and using the expression for the dust-to-gas ratio $D$ in eq.~\ref{eq_RR14}, $A_{\lambda}$ can be written as:
\begin{equation}\label{Av_expr}
 \begin{aligned}
   A_{\lambda} & =  2.26\ k_{\lambda}\ D\ \left(\frac{\Sigma_{\rm g}}{10^{10}\ \mathrm{M_{\odot} kpc^{-2}}}\right) \\
   & = \frac{4.4\times 10^{-3}}{f_{\mu}}\ k_{\lambda}\ \frac{Z}{\mathrm{Z_{\odot}}}\ \frac{M_{\rm g}}{10^{10}\ \mathrm{M_{\odot}}} \left(\frac{r_{\rm g}}{\mathrm{kpc}}\right)^{-2}\ ,
\end{aligned}
\end{equation}
where $k_{\lambda}$ is the dust extinction cross section per unit dust mass in cgs units ($k_{\rm V} = 3.4822 \times 10^{4}\,\mathrm{cm^2/g}$, for the adopted MW dust model by \citealt{Draine03}\footnote{The reported $k_{\rm V}$ value is obtained by multiplying the extinction cross section per H nucleon, $C_{\rm ext}/H =4.848 \times 10^{-22}$, by the dust mass per H nucleon, $1.398\times 10^{-26}$, as tabulated for the carbonaceous-silicate MW dust model in \cite{Draine03}.}) and $f_{\mu}=(4/3;\, \mathrm{\cos\theta})$ is the geometric factor for a spherical geometry with mixed dust and stars, and a slab geometry with orientation angle $\theta$, respectively \citep{Ferrara22REB}. 
If we substitute the average values of metallicity, gas mass and gas radius for the TNG galaxy sample $(Z/\mathrm{Z_{\odot}},\, M_{\rm g}/\mathrm{M_{\odot}},\,r_{\rm g}/\mathrm{kpc})=(1.27,\,0^{10.6},\,59)$, we infer $ A_{\rm V}=(0.17,0.27)$ for a sphere and randomly oriented slab ($f_{\mu}=0.841$, \citealt{Ferrara22REB}), respectively. These values are remarkably close to the median value inferred from the detailed RT post-processing ($A_{\rm V, TNG} = 0.26$). 

We can understand the correlation of $A_{\rm V}$ with the SFR surface density by combining the definition in eq.~\ref{Av_expr} with the Kennicutt-Schmidt (KS) relation \citep{Kennicutt98}:
\begin{equation}\label{KS}
\frac{\Sigma_{\rm SFR}}{\mathrm{M_{\odot} yr^{-1} kpc^{-2}}} = 10^{-12}\ \kappa_s\ \left(\frac{\Sigma_{\rm g}}{{\mathrm{M_{\odot} kpc^{-2}}}} \right)^{1.4}\ , 
\end{equation}
where $\kappa_s$ is the \quotes{burstiness parameter}. 
Substituting the KS-relation in eq.~\ref{Av_expr}, we obtain the following expression for $A_{\rm V}$:
\begin{equation}\label{Av-sigmaSFR-anal}
\begin{aligned}
    \log A_{\rm V} & = 1.26 + 0.71\ \log \left(\frac{\Sigma_{\rm SFR}}{\mathrm{M_{\odot} yr^{-1} kpc^{-2}}} \right) +\\
    & +\log \left(\frac{Z}{\mathrm{Z_{\odot}}} \right) - 0.71 \log \kappa_s\ .
\end{aligned}
\end{equation}
Considering the average metallicity for TNG galaxies and the average $\log \kappa_s = 1.1$, we derive the relation shown in the left panel of Fig.~\ref{Fig_Correlations_fits}.
We can see that the analytical expression matches the simulations sufficiently well, albeit slightly too steep. This slight mismatch is related to the chosen SFR surface density, $\Sigma_{\rm SFR,y}$, which only traces the young stars. Observationally, such a quantity could be constrained using H$\alpha$ derived SFR measurements \citep{Kennicutt98}. We use this property rather than the global $\Sigma_{\rm SFR}$ (which is the quantity used in the empirical KS relation) for our fiducial model, as $\Sigma_{\rm SFR,y}$ presents a tighter correlation with $A_{\rm V}$ (we stress that, nevertheless, the correlation is also strong with the global $\Sigma_{\rm SFR}$; the correlation factor is as high as $\sim 0.58$).
We provide the best-fitting $A_{\rm V}(\Sigma_{\rm SFR,y})$ relation derived purely from the simulation outputs:
\begin{equation}\label{eq_Av_SigmaSFRy}
    \log A_{\rm V}  = -0.19 + 0.30\ \log \left(\frac{\Sigma_{\rm SFR,y}}{\mathrm{M_{\odot} yr^{-1} kpc^{-2}}} \right)\ .
\end{equation}
If instead we use the global SFR surface density, we find $\log A_{\rm V} = 0.4\,\log \Sigma_{\rm SFR} + 0.07$, which is closer to the expression in eq.~\ref{Av-sigmaSFR-anal}.

As previously discussed, the attenuation in the UV-FUV regime is predominantly governed by the dust-to-stellar geometry and the column density of dust surrounding young star-forming regions vs. the diffuse interstellar medium, which — if not fully dust-obscured — dominate the stellar emission at short wavelengths. Consequently, modeling dust attenuation in this regime based solely on global galaxy properties presents significant challenges.

In Sec.~\ref{fitting_attenuation_curves}, we demonstrated that the key parameter affecting the shapes of attenuation curves across the entire sample — including all lines of sight for each source — is $A_V$. Higher (lower) $A_V$ values are consistently associated with shallower (steeper) slopes. This trend is evident in the internal correlations between attenuation curve parameters: the second strongest (anti)correlation involving $A_V$ is with the UV slope $c_1$ ($-0.269$ when all lines of sight are considered). Notably, this anticorrelation is stronger than any correlation involving global galaxy properties, when the full distribution of lines of sight is taken into account. For this reason, we prioritize this anticorrelation in constructing our best-fit scaling relations.

Using the 2D posteriors derived from the synthetic attenuation curves of the TNG galaxies, we infer the following best-fit relation $\log c_1(\log A_V)$:
\begin{equation}\label{c1Av_eq}
    \log c_1 = -0.37\,\log A_V + 0.75\ .
\end{equation}
This is shown in the bottom left panel of Fig.~\ref{Fig_Correlations_fits}

Having determined $A_V$ and $c_1$, we are left with three additional parameters to infer: $c_2$, $c_3$, and $c_4$. We fix $c_2 = 1.88$, the median value for TNG galaxies. This choice is motivated by the minimal variation in $c_2$ (ranging between $2 < c_2 < 4$), even when fully accounting for variations across lines of sight, resulting in a highly sub-dominant impact on the attenuation curves (see Fig.~\ref{Fig_Param_Att_Curve}).

For the remaining two free parameters, $c_3$ and $c_4$, we turn to the internal correlations between attenuation curve parameters, as they appear to be stronger than any with global galaxy properties. We derive the following relation for $c_3(\log c_1)$:
\begin{equation}\label{c1c3_eq}
    c_3 = 1.21\,\log c_1 - 1.33\ ,
\end{equation}
and for $\log c_4(\log A_{\rm V})$:
\begin{equation}\label{c4Av_eq}
    \log c_4 = -0.59\,\log A_{\rm V} -1.42\ .
\end{equation}
These relations are shown as the black dashed (eq.~\ref{c1c3_eq}) and dotted (eq.~\ref{c4Av_eq}) lines in the central and rightmost panels of Fig.~\ref{Fig_Correlations_fits}.\\ 

\subsection{Testing simulation-based scaling relations against observations}

\begin{figure}
    \centering
    \includegraphics[width=0.9\linewidth]{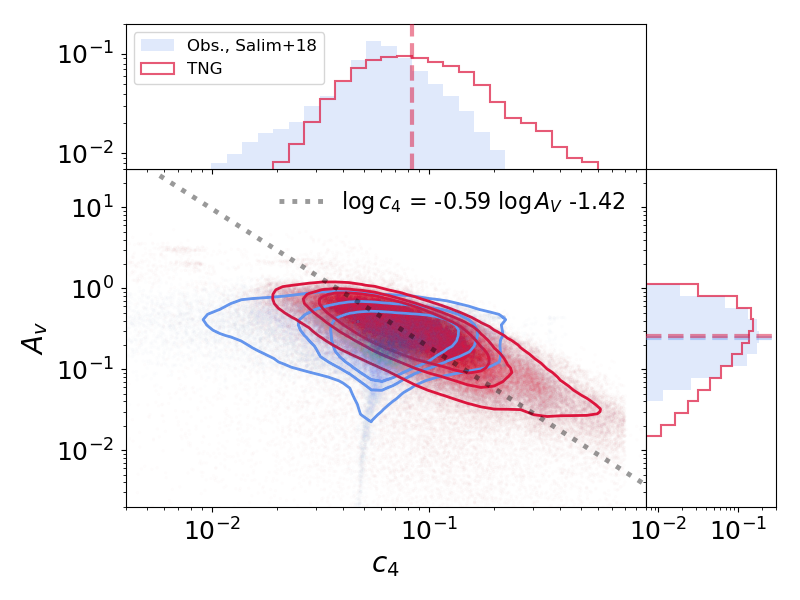}
    \includegraphics[width=0.9\linewidth]{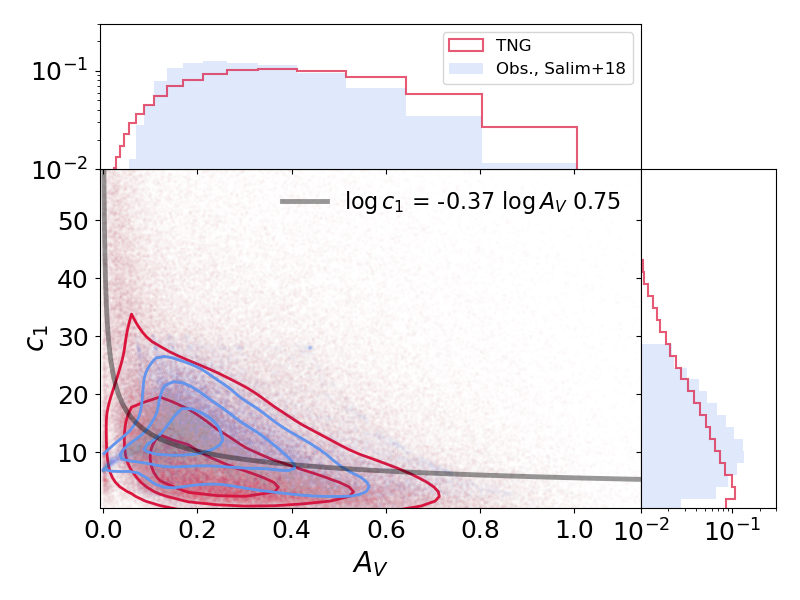}
    \caption{ \textbf{Top Panel:} 2D PDF of $A_{\rm V}$ (visual attenuation) and $c_4$ ($2175\,\angstrom$ bump strength) values in the simulated TNG galaxies (in red) and in the previously discussed GSWLC sample (in blue). The black line shows the best fitting relation identified for TNG galaxies expressed in eq. \ref{c4Av_eq}. \textbf{Bottom Panel:} the same as the top panel, but for $c_1$ (UV slope) and $A_{\rm V}$ values.
 }
    \label{Fig_Obs_Comp_Att_Params}
\end{figure}

We investigate whether the scaling relations derived for TNG sources (eq.~\ref{eq_Av_SigmaSFRy} --~\ref{c4Av_eq}) hold for the observed sample. This comparison is summarized in Fig.~\ref{Fig_Obs_Comp_Att_Params} (and the bottom panel of Fig.~\ref{Fig_Obs_Comp_Av}). We stress that for the TNG sources, we show the full line of sight distribution (51 per source). 

In the bottom panel of Fig.~\ref{Fig_Obs_Comp_Av}, we show that the $\log \Sigma_{\rm SFR,y} (\log A_V)$ relation provided in eq.~\ref{eq_Av_SigmaSFRy}, is a good fit to the observed sample as well. To provide a comprehensive comparison, we also include the analytical predictions based on the more commonly accessible $\Sigma_{\rm SFR}$; the latter also successfully reproduces the observational data. On the other hand, when comparing the simulation-derived $c_4-A_{\rm V}$ relation, observed galaxies are offset towards lower $c_4$ values compared to the simulated sample (see the top panel of Fig.~\ref{Fig_Obs_Comp_Att_Params}). The observations are consistent with a flatter $c_4-A_{\rm V}$ relation, although the anticorrelation between the two parameters remains significant. 
The shallower bump identified in most of the observed sample may indicate differences in the dust composition or grain size distribution, as the bump is typically associated with PAHs, i.e. small carbonaceous grains. 

While we cannot directly study the full distribution of $c_3$ in observations versus simulations due to the limitations in the observational fitting methodology, the median $c_3$ values in the two samples are consistent within $1\sigma$.
Lastly, we find good consistency between the simulation-derived $\log c_1 (\log A_V)$ relation and the observed data (see the bottom panel of Fig. \ref{Fig_Obs_Comp_Att_Params}), as expected given the similar slopes of the attenuation curves binned by $A_V$ shown in Fig.\ref{Fig_Att_Curve_Avbins_Corner}.

In summary, the comparison between the TNG synthetic attenuation curves and the GSWLC observations \citep{Salim18} reveals encouraging consistency in the relationships between galaxy properties and dust attenuation curve parameters, as well as among the dust parameters themselves. We discuss caveats on our analysis in Sec.~\ref{Sect_discussion}. Based on the PDFs of synthetic versus observed attenuation curve parameters, we conclude that the TNG simulations align well with observations, making them a robust foundation for deriving physically motivated priors for attenuation curve parameters.

\begin{figure*}[t]
    \centering
            \centering
            \includegraphics[width=1.0\textwidth]{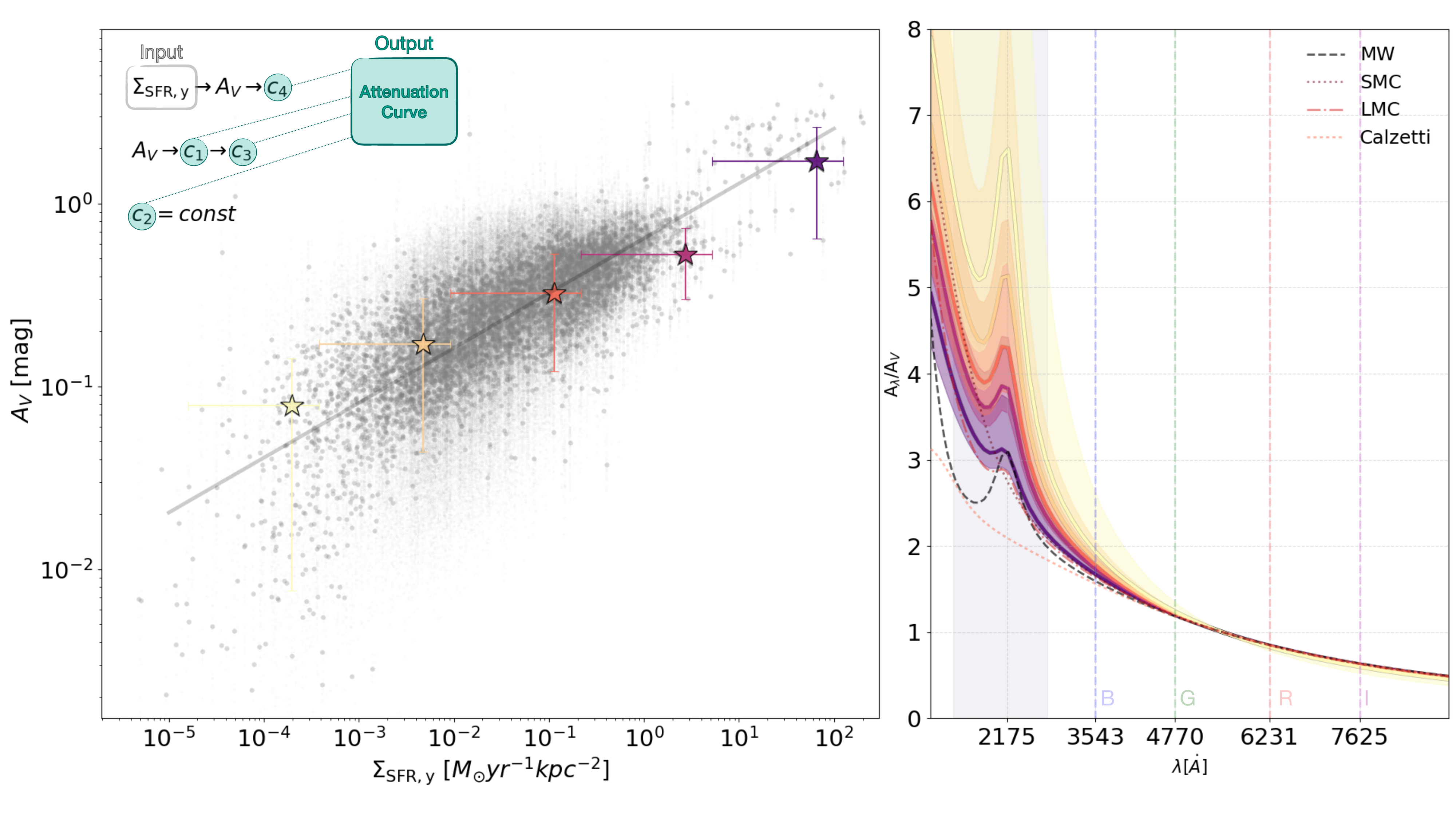}
            \vspace{-0.4cm}
            \captionof{figure}{
                The main panel illustrates the correlation between the visual attenuation (\(A_{\rm V}\)) and star formation rate surface density (\(\Sigma_{\rm SFR, y}\)) for TNG galaxies, with the data points color-coded by stellar mass. The solid line represents the best-fit relation provided in eq.~\ref{eq_Av_SigmaSFRy}. 
                \textbf{Right panel}: this inset presents the predicted attenuation curves (\(A_{\lambda}/A_{\rm V}\)) for the five bins identified by the colored stars in the main panel. The usual local extinction curve templates (MW, SMC, LMC, Calzetti) are shown for comparison. The shaded area represents the $16^{\rm{th}}-84^{\rm{th}}$ percentile variation of the predicted attenuation curves within each bin. The thin blue line shows the attenuation curve for the corresponding \(\Sigma_{\rm dust, y}\) and \(\Sigma_{\rm SFR, y}\) bin, calculated using the UV slope of the MW-like galaxy instead of the median value in the bin (see the \(c_3\) vs \(c_1\) inset).
                \textbf{Top left corner}: the logical scheme of the identified scaling relations is depicted; see the main text for a full description. 
            }
            \label{pretty_plot}
\end{figure*}

\begin{figure*}
    \centering
            \centering
            \vspace{-0.2cm}
            \includegraphics[width=0.9\textwidth]{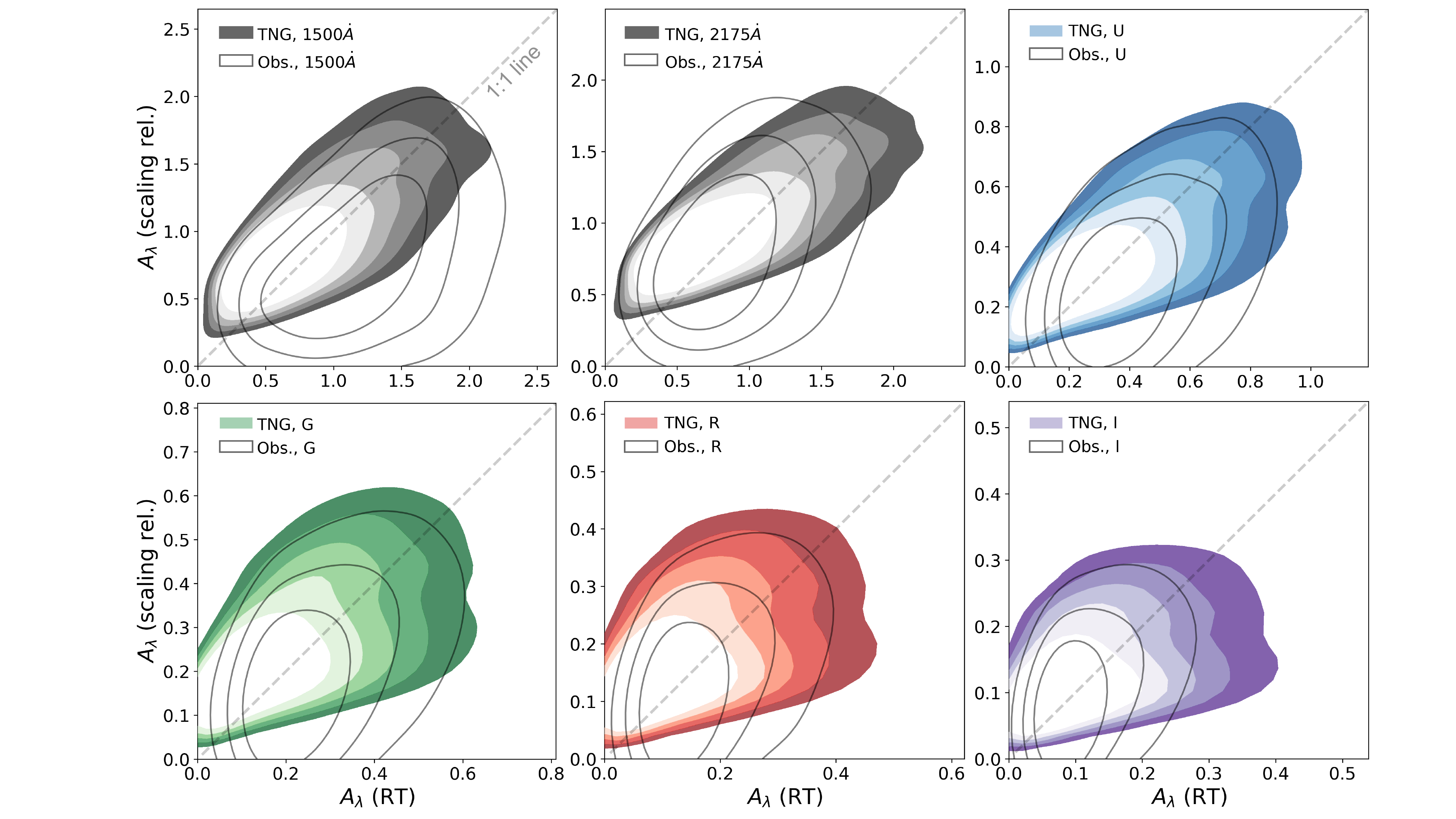}
            \caption{The attenuation predicted by our scaling relations (Eqs.~\ref{eq_Av_SigmaSFRy}-\ref{c4Av_eq}) is shown on the y-axis, while the x-axis represents the true attenuation values derived from radiative transfer simulations. The colored contours illustrate five percentiles, from the $30^{\rm{th}}$ to the $70^{\rm{th}}$, for the combined TNG simulations. Black transparent contours represent the $30^{\rm{th}}$, $50^{\rm{th}}$, and $70^{\rm{th}}$ percentiles for the GSWLC observational catalog \citep{Salim18}. 
            Each panel corresponds to a different photometric band, labeled in the top-left corner. The dashed line indicates the 1:1 bisector, representing perfect agreement between the predicted and true attenuation values.}     
            \label{validation_plot}
\end{figure*}

\subsection{Validation of scaling relation-based attenuation curves}
The $A_{\lambda}/A_{\rm V}$ curves obtained from the single parameter simulation-based scaling relations presented in Sec.~\ref{Sect_Analytical_Model} (outlined schematically in the upper left corner of Fig.~\ref{pretty_plot}) are displayed in the right-hand panel of the same figure.
As anticipated based on the preceding analysis, these attenuation curves derived from the TNG-based scaling relations are generally steeper than the average Milky Way (MW) extinction curve. Galaxies with higher star formation rate surface densities ($\Sigma_{\rm SFR}$) and visual attenuation ($A_V$) exhibit shallower, MW-like attenuation curves with less pronounced $2175\,\angstrom$ bumps (i.e. lower $c_4$ values). Remarkably, the predicted attenuation curves closely follow the trends derived for the median attenuation curves (binned by $A_V$) for the observed GSWLC sample \citep{Salim18}, as shown in Fig.~\ref{Fig_Att_Curve_Avbins_Corner}.

To make the comparison more quantitative, in Fig.~\ref{validation_plot} we show the predicted attenuation across a wide wavelength range, from $1500\,\angstrom$ to the $I$-band (centered around $7625\,\angstrom$), compared to the attenuation $A_{\lambda}$ derived from radiative transfer for the TNG galaxies and from SED fitting for the observations. In all cases, we find very good agreement. We note, that for the observations, we applied the $A_V$-$\Sigma_{\rm SFR}$ relation derived in Eq.~\ref{eq_Av_SigmaSFRy} for the simulations and applied it to the global SFR surface density $\Sigma_{\rm SFR}$.

Within the simulations, we find the best agreement at shorter wavelengths, although these are the most affected by inclination effects. This is because we dedicate three parameters to describe the UV regime while we fix the parameter $c_2$, which mainly influences the longer wavelengths. As previously explained, within the $16^{\rm{th}}-84^{\rm{th}}$ percentile range of the PDF of TNG galaxies, the variation in this parameter is sufficiently small that is should play a subdominant role in shaping the attenuation curves. However, a more complex model can be explored in future work to assess potential refinements in the treatment of longer wavelengths. In any case, the absolute attenuation at longer wavelengths is smaller compared to the UV regime. At $1500\,\angstrom$, the attenuation varies between $0.3 \lesssim A_{\lambda} \lesssim 2$, whereas in the $I$-band, it falls within the range $0 \lesssim A_{\lambda} \lesssim 0.3$. Thus, while the agreement at longer wavelengths is slightly worse, this is not a major concern, given the lower absolute values of attenuation in this regime.

\section{Discussion}\label{Sect_discussion}
In this study, we generated and analyzed a large library of synthetic attenuation curves for local galaxies, via performing radiative transfer on the TNG50 and TNG100 simulations. Additionally, we made detailed and consistent comparisons with attenuation curves derived from the observational GSWLC sample. We identify a strong correlation between $A_{\rm V}$ and the SFR surface density in both simulations and observations. Crucially, we find that the attenuation curve shape is primarily correlated with $A_{\rm V}$, with higher values corresponding to flatter curves and weaker $2175\,\angstrom$ bumps.

We provide a physical motivation for the correlation with $\Sigma_{\rm SFR}$. By definition, $A_{\rm V}$ scales with the dust surface density $\Sigma_{\rm dust}$. Assuming a fixed or metallicity-dependent dust-to-gas ratio, $\Sigma_{\rm dust}$ can be linked to  $\Sigma_{\rm g}$. For galaxies following the Kennicutt-Schmidt relation, $\Sigma_{\rm g}$ can be further connected to the SFR surface density $\Sigma_{\rm SFR}$ (see eq.~\ref{Av-sigmaSFR-anal}). Interestingly, we find the strongest correlation with the surface density of young stars, which is consistent with the idea that the interplay between the geometry and properties of young, actively star-forming regions versus the diffuse interstellar medium (ISM) is a key factor in shaping dust attenuation. The additional dependence on metallicity and burstiness may explain the scatter in the $A_{\rm V}$-$\Sigma_{\rm SFR}$ relation.

The physical origin of the correlation between $ A_{\rm V}$ and the overall attenuation curve slope remains a topic of debate. Several potential explanations have been proposed. \cite{Chevallard13} suggests that at low densities, scattering dominates over absorption. Scattering is more forward-directed in the blue and more isotropic in the red, meaning that blue photons are less likely to escape the galaxy before being absorbed, resulting in a steeper attenuation curve. \cite{Narayanan18} and \cite{Inoue05} argue that differential attenuation between young stars (dominating FUV emission) and old stars (dominating optical/NIR emission) causes variations in the slope. Specifically, at high $A_{\rm V}$, the emission is dominated by stars in low-density pockets across all wavelengths, leading to shallower attenuation curves, as UV photons from young stars remain trapped within their dense birth clouds. In contrast, at lower $A_{\rm V}$, UV radiation more easily leaks out from both diffuse regions and partially transparent birth clouds, resulting in steeper attenuation curves.
\cite{DiMascia21} shows, through analytical arguments, that in a system with stars and dust uniformly distributed in a sphere, the attenuation curve flattens as the $V$-band optical depth increases. \cite{Hirashita20} propose that the slope of the attenuation curve is influenced by the most abundant dust grain size, which varies with the age of the galaxy. However, this effect has been suggested to be sub-dominant to the stellar-to-dust geometry distribution (\citealt{SeonDraine16}, Matsumoto et al. in prep.); thus, in this work, we do not investigate variations in the grain size distribution, which is fixed to the WD01 MW model.

\subsection{Comparison to Previous Theoretical Studies}
Our study's findings align with and expand upon previous theoretical models for dust attenuation using SAMs and hydrodynamical simulations.
In SAMs, dust attenuation is typically modeled by assuming that the $V$-band dust optical depth is proportional to the metallicity-weighted column density of the cold gas and then applying local templates for attenuation curves (e.g. MW, SMC, LMC, or Calzetti). Some studies adopt simple analytic models to account for the star-dust geometry, such as a foreground dust screen or a mixed dust-star slab \citep{Devriendt1999,deLucia2007,Somerville2012,Mauerhofer23}, while others couple their models with the results from radiative transfer simulations applied to simplified bulge- and disk-like geometries \citep{Fontanot09,Lacey2011,Lacey2016}.

An extra layer of complexity is often introduced by accounting for the role of birth clouds on the attenuation of young stellar emission \citep{Fontanot09,Cousin19}, following the approach by \cite{CharlotFall2000}, essentially introducing a different effective optical depth (i.e. $A_V$) for stars with ages less than the timescales typical of the lifetimes of GMCs  ($3-10$ Myr; see e.g. \citealt{Sommovigo20}). Historically, most SAMs have implicitly assumed a constant dust-to-metal ratio, but some recent SAMs have included tracking of the formation and destruction of dust, yielding predictions for the dust-to-metal ratio which vary from galaxy to galaxy and with cosmic epoch \citep[e.g.][]{Popping17,Vijayan19,Dayal22,Mauerhofer23}.

Recently, \cite{Zhao2024} developed a semi-analytic model to study the dust distribution in galaxies at redshifts $z \gtrsim 5$. They calibrated their model using the infrared excess-UV slope (IRX-$\beta$) relation and ALMA observations of dust emission. Their findings suggest that to match observed dust emission, galaxies must retain most of their produced dust. However, if this dust is spherically distributed, the modest UV attenuation observed implies that the dust must be more extended than the stellar component. This discrepancy can be reconciled if the dust is distributed anisotropically, with covering fractions ranging from approximately 0.2 to 0.7 in bright galaxies and less than 0.1 in fainter ones. This is consistent with eq.~\ref{Av_expr}, where, under the simplistic assumption of a dust screen geometry, in order to reproduce the simulated galaxies' attenuation, we need to use the gas radius $r_g$ (half-mass radius for cold gas), which is on average $>10$ times more extended than the stellar radius (both $r_{\star,y}$ or $r_{\star,o}$). This requires further investigation, as in observational studies, $A_V$ is often computed using the observed UV or optical size of galaxies (with a slab or spherical screen approximation). This could result in the severe overestimation of the true global $A_V$ of a galaxy. 

Hydrodynamical simulations coupled with radiative transfer calculations provide a more realistic treatment of inclination effects and the complex spatial distribution of dust, gas, and stars.
Several studies have proposed simplified theoretical frameworks to model attenuation curves, providing valuable insights while motivating more physically grounded approaches like the one explored here.
\cite{Jonsson06} utilized hydrodynamic simulations of major galaxy mergers, post-processed with the SUNRISE Monte Carlo RT code, to calculate the effects of dust, assuming a Milky Way (MW) dust model. They found that the attenuation is well described by a slab model with mixed stars and dust, where the effective optical depth scales as a power law with the star formation rate (SFR), baryonic mass, luminosity, and metallicity (similar to what we find in eq.~\ref{Av_expr}). The dependence on SFR is consistent with our findings, although we do not see clear evidence of trends with gas mass or stellar mass. The correlation with metallicity, while subdominant compared to that with dust or SFR surface densities, emerges as the third strongest among the global properties we examined. Future work will focus on assessing how this or other secondary parameters might contribute to the scatter observed around our set of simulation-based scaling relations.

Some previous works have combined cosmological hydrodynamic simulations and radiative transfer tools to generate large statistical samples of synthetic attenuation curves and performed exploratory studies similar to the one presented here. Focusing on cosmological zoom-ins of eight haloes from the MUFASA suite, \citet{Narayanan18} fit the attenuation curves produced from RT post-processing of such halos, identifying a strong correlation between the slopes of normalized attenuation laws and the complexities of star-dust geometry, with variations in the $2175\,\angstrom$ UV bump strength influenced by the fraction of unobscured O and B stars (with the caveat that no treatment of birth clouds is included). They then apply these findings to derive the median curve and expected dispersion at integer redshifts from $z=0-6$. They find that at earlier epochs, attenuation curves become greyer due to the reduced dispersion in star-to-dust geometry, as well as a narrower distribution in median stellar ages with redshift. 

\citet{Trayford20} used the EAGLE cosmological simulation suite to investigate how attenuation curves vary with galaxy properties. They performed RT calculations on approximately $10,000$ galaxies at $z = 0-2$, utilizing the \textsc{skirt} code. They found that attenuation curves become shallower with increasing specific star formation rate and dust surface density, highlighting the crucial role of dust geometry and distribution in shaping attenuation laws. This is consistent with our finding of a positive correlation between $A_V$, $\Sigma_{\rm SFR,y}$, and $c_1$, demonstrating that this result remains robust despite differences in the sub-grid assumptions in the simulations.

Although, we do not find any strong dependence on galaxy age in our study (especially when the full line-of-sight distribution is accounted for; we find correlation coefficients $<0.2$), we do see a correlation between slope parameters and quantities related to the young stellar component size. In addition, when only the median line-of-sight is considered for each galaxy, we identify an anticorrelation between $A_V$ and the age, with younger galaxies having lower $A_V$ and thus, in principle, steeper curves. However, this is not identified as the dominant driver of the scatter in our attenuation curves, which rather vary significantly depending on the $A_V$ of the specific line of sight and on the galaxy's SFR.
 
\cite{Hahn2022} introduced a hybrid method, the Empirical Dust Attenuation (EDA) framework, which shares some conceptual similarities with the approach used here. However, our method relies on simulations post-processed with radiative transfer rather than using observations as a starting point. The EDA employs an observationally motivated parameterization based on \cite{Noll2009} and was applied to three hydrodynamical simulations (SIMBA, TNG, and EAGLE) to forward-model UV and optical color-magnitude relations. Their approach assumes a slab geometry, applies a single inclination angle to all sources, and assigns attenuation curve parameters based on the simulated galaxy's $M_{\star}$ and sSFR. They demonstrated that dust attenuation is essential for reproducing observed color-magnitude relations and that quiescent galaxies exhibit lower amplitude and shallower curves compared to star-forming galaxies. However, they noted the significant degeneracy between dust attenuation and the underlying galaxy formation physics. While our sample mostly includes main-sequence galaxies (see Fig.~\ref{Fig_Mstar-Z_Obs_Comp}), our study arrives at a consistent and complementary conclusion regarding the degeneracy between galaxy parameters and line-of-sight effects, highlighting the complexity of assigning a single attenuation curve to a given observed source. We will dedicate future work to modeling higher-dimensional parameter dependencies and the scatter around the identified scaling relations (eq.~\ref{eq_Av_SigmaSFRy}-\ref{c4Av_eq}).

Complementary studies of attenuation curves from hydrodynamical simulations include works focused on high-redshift ($z>6$) galaxies—where intrinsic galaxy properties and large-scale processes such as cosmic accretion differ significantly from the sources considered here—as well as analyses of isolated ISM patches or GMCs. These works are particularly crucial, as at the high redshift end, probing the dust content of galaxies via direct detection of dust emission requires extremely large commitments of telescope time \citep[see e.g.][]{Inami22}.
\cite{Pallottini22} analyzed the emission properties of high-redshift galaxies using the SERRA cosmological zoom-in simulations combined with \textsc{skirt} post-processing. They showed that highly obscured star-forming regions (essentially GMCs, which remain unresolved even in these high-resolution zoom-ins) dominate the overall attenuation curve shape, with redder UV slopes observed in compact starburst galaxies.
\cite{Mushtaq2023} utilized the FirstLight simulations combined with the POLARIS RT code \citep{2016A&A...593A..87R} to study dust attenuation in galaxies at $z=6$ to $z=8$. They found that high-mass galaxies exhibit flatter attenuation curves and higher UV attenuation, consistent with previous observational results at $z\sim0$ \citep{Salim18}.

Regarding the small sub-galactic scales that remain largely unexplored in most of the works discussed so far, \cite{DiMascia24} investigated attenuation curves within a single molecular cloud of mass $10^5\,\rm{M_\odot}$. They employed radiation-hydrodynamic simulations with self-consistent chemistry treatment, resolving scales down to $0.06\,\rm{pc}$. These simulations were post-processed with the same code used here, \textsc{skirt}. The authors demonstrated that the attenuation curve evolved away from a MW-like extinction curve as a function of line-of-sight orientation and cloud evolutionary stage, despite the assumed MW dust model. This is consistent with previous findings by \cite{SeonDraine16}, who showed that Calzetti-like attenuation curves can arise in individual ISM patches, even in the presence of PAHs, depending on the local stellar-to-dust geometry. 
This consistent finding — of increasing attenuation curve complexity as more physics is incorporated — highlights the importance of moving beyond fixed local templates when generating galaxy catalogs, color diagrams, and luminosity functions in upcoming theoretical and observational statistical studies.

\subsection{Comparison to Previous Observational Studies}

Our analysis highlights several key trends in attenuation curves that we now compare to previous observational studies. One of the most significant findings is the strong dependence of attenuation curve properties on the line of sight, introducing substantial scatter, particularly in galaxies with low $A_V$ (see Fig.~\ref{example_images} and \ref{Fig_Example_EMCEE}). This aligns qualitatively with the results of \citet{Salim18}, who found that, within the Milky Way, steeper extinction curves tend to occur along sightlines that avoid dense giant molecular clouds (GMCs). However, since the TNG simulations do not explicitly model such cold, dense clouds in the ISM, the inclination and line-of-sight effects we observe should not be directly equated to the Milky Way case. Nevertheless, the overall consistency of our results with the GSWLC dataset suggests that our model captures these effects at a statistical level.

Furthermore, we find no robust trends between the attenuation curve shape and stellar mass or age. Instead, the most robust correlations arise with quantities related to $\Sigma_{\rm SFR}$, which encodes information about SF regions' density and morphology. In agreement with our findings, also relying on the GSWLC catalog, \citet{Battisti2016} previously reported that $A_V$ and $\Sigma_{\rm SFR}$ are among the primary drivers of attenuation variations, with higher-density star-forming environments producing flatter attenuation curves.
Similarly, \citet{Hamed2023} observed that in GSWLC galaxies, the steepness of attenuation curves depends on the ratio of stellar size to dust size. We also investigated the correlation between attenuation curve parameters and the relative spatial extent of stellar and gas components in TNG (see Tab.~\ref{tab:galaxy_properties_TNG}). Specifically, we find a positive correlation between $A_V$ and the ratio of the old stellar component size to the gas size, as well as with the ratio of the old to young stellar population sizes. In contrast, we find an anticorrelation with the size of the young stellar component, suggesting that more compact star-forming regions are associated with steeper attenuation curves.

The dependencies of the remaining attenuation curve parameters with these size ratios are less clear. The next strongest correlation is an anticorrelation between $c_2$ (the optical-to-NIR slope) and the $r_{\star,o}/r_{\star,y}$ ratio. However, as established previously, the distribution of $c_2$ is sharply peaked, making these correlations difficult to interpret and their impact on attenuation curve shapes subdominant. Additionally, we note that the resolution of the TNG galaxies is limited, particularly when resolving the morphologies of young and old stellar populations. Follow-up studies employing higher-resolution simulations will be necessary to further investigate these relationships and better constrain the physical drivers of these correlations.

Observations at slightly higher redshifts provide further context. \citet{Barisic2020} found that attenuation curves at $z\sim0.8$ are often significantly steeper than the MW and LMC curves, aligning with our findings that high $\Sigma_{\rm SFR}$ galaxies (often found at high redshift) tend to exhibit steep attenuation curves. Similarly, \citet{Lorenz2023} examined dust attenuation in star-forming galaxies at $1.3\leq z \leq 2.6$ and found that both stellar and nebular attenuation increased with galaxy mass, though their study did not identify strong inclination-dependent trends, unlike what we observe in simulations. This lack of inclination effects may, however, stem from the challenge of assessing such dependencies in unresolved studies of intermediate- and high-redshift galaxies, where integrated light measurements may wash out spatial variations.

\citet{Shivaei2020} investigated the variation of dust attenuation curves with metallicity using data from the MOSDEF survey, focusing on galaxies at $z \sim 1.4$ to $2.6$. They found that higher metallicity galaxies exhibit shallower attenuation curves and a more pronounced UV $2175\,\angstrom$ bump, while lower metallicity galaxies have steeper slopes and lack a significant UV bump. This aligns with our finding that higher metallicity correlates with shallower slopes. However, in our study, we observe that higher metallicity indirectly implies smaller bumps, as $c_4$ anti-correlates with $A_V$. To solidify this result, we need to assess how it changes when non-Milky Way dust compositions are assumed, given that different dust models can inherently produce varying UV bump strengths independent of the line of sight. However, \cite{SeonDraine16} showed that the assumed extinction curve plays a subdominant role compared to geometric effects.. Additionally, it is important to note that the MOSDEF survey targets higher redshift galaxies -- with higher SFR (up to $100\,\rm{M_{\odot}/yr}$ and slightly lower metallicity (down to $0.3\,\mathrm{Z_{\odot}}$, see \citealt{Reddy15}) -  than those analyzed here, which may influence the observed relationships. 

Thanks to recent JWST observations, attenuation curve studies have now been extended to unprecedentedly high redshift, out to $z\sim 10$. \citet{Fisher25} found that the attenuation curves of massive Lyman-break galaxies at $z\approx7$ were, on average, slightly flatter than those seen in local sources, with some galaxies showing evidence for a $2175\,\angstrom$ dust bump. This suggests that our attenuation curve study, which focuses on local main-sequence galaxies that are modestly attenuated in the UV, may also be relevant for high-redshift populations, particularly given the low $A_V$ values routinely measured at early cosmic epochs (but see also \citealt{2020MNRAS.495.4747S} for attenuation curves produced for IllustrisTNG galaxies at $z=2-6$). Supporting this, \citet{Markov23, Markov24} inferred flatter attenuation curves from high-redshift observations, attributing this to the evolution of grain sizes—from larger grains produced by stellar sources at $z\sim12$ to smaller grains reprocessed in the ISM at later epochs ($z\sim2$). For comparison with our study, we stress that, independent of the underlying grain size distribution and composition (which we keep fixed to MW-like dust), we find that the steepness and bump strength of attenuation curves can vary significantly depending on the line of sight and on parameters such as $A_V$ and $\Sigma_{\rm SFR}$. This highlights the importance of testing the robustness of evolutionary trends in large, unbiased (e.g. by a UV brightness selection) samples, and of accounting for the internal evolution of galaxy properties across cosmic time (see Markov et al. in prep.), as the interplay between these properties and the attenuation curve shapes is not straightforward.

\citet{Witstok23} found strong evidence of a $2175\,\angstrom$ bump in absorption at $z\sim6-7$, based on spectroscopic measurements, indicating the presence of small grains in some observed UV-bright galaxies at these epochs (see also \citealt{Lin25}). This result strengthens and complements the findings of \citet{Markov23} and \citet{Fisher25}, who identified similar features in some of their observed galaxies using attenuation curves derived from SED fitting. Together, these studies suggest that dust grain properties and attenuation curve shapes may vary substantially not only with cosmic time but also with galaxy morphology and environment, reinforcing the need for physically motivated models that account for variations in geometry, composition, and grain size distributions across diverse galaxy populations.

\subsection{Caveats in Our Analaysis}

While our analysis provides valuable insights into attenuation curve properties, several caveats must be considered. A primary limitation stems from the resolution of the simulations used. Although the TNG simulations successfully reproduce statistical distributions of global galaxy properties within a range of large-scale environments, they do not resolve the cold, dense giant molecular clouds (GMCs) that are expected to contribute significantly to dust obscuration. As a result, although values $A_V>1$ are found in the central regions of TNG galaxies -- where most of the SF is taking place, the metallicity peaks and $\Sigma_{\rm g} \geq 10^8\ \mathrm{M_{\odot}/kpc^{-2}}$  \citep{Torrey19} -- none of the sightlines pass through optically dark GMCs, nor is the effect of parental cloud obscuration \citep[see e.g.][]{DiMascia24} on young stars accounted for. This may explain the overall low $A_V$ values found in our sample. Consequently, our results are most applicable to main-sequence galaxies that are only modestly attenuated in the UV, rather than to highly dust-obscured systems such as submillimeter galaxies (SMGs) or ultra-luminous infrared galaxies (ULIRGs). However, given that low $A_V$ values are routinely measured in high-redshift, UV-bright sources, our findings may be relevant in that regime as well. That said, we note that most of the galaxies in our sample have relatively high stellar masses ($M_{\star} \sim 10^{10}\,\rm{M_{\odot}}$) and metallicities above $0.3\, \rm{Z_{\odot}}$. As discussed earlier, these properties do not correlate strongly with the shape of the attenuation curve, which is primarily correlated with $\Sigma_{\rm SFR}$. A future study will focus on extending this analysis to high-redshift galaxies.

Another major caveat relates to the physical assumptions in our dust modeling. We adopted a simplified prescription for the gas-to-metal ratio and dust composition, which, as already mentioned, may impact the predicted attenuation properties. Future work employing higher-resolution zoom-in simulations coupled with new subgrid models for dense star-forming clouds will be essential to improve the physical accuracy of our attenuation curve predictions across different galaxy populations. Exploring the effects of different dust models on the scaling relations and the derived synthetic attenuation curves will help solidify our conclusions. A consensus on the relative impact of these effects (dust chemical composition and grain size distribution versus dust-to-stellar geometry and scattering) has yet to be reached, with some studies suggesting that microscopic grain properties play a subdominant role (see Matsumoto et al., in prep., \citealt{SeonDraine16}) and others finding that the assumed extinction curve can strongly impact even statistical properties such as the bright end of the luminosity function \citep{2023MNRAS.519.5987L}.

Finally, uncertainties in the observational quantities derived from SED fitting could introduce systematic biases when comparing them to simulations. These uncertainties include (but are not limited to): dependence on star formation history prescription \citep[e.g.][]{Iyer17,Leja2019,Suess_2022,Pacifici_2023,Narayanan_2024,Cochrane_2025} and stellar population synthesis model \citep{Jones2022}, degeneracies between dust attenuation treatment and inferred galaxy properties \citep{Salim20,DustE22,Lower22,Hamed2023}, and uncertainties in metallicity measurements based on the strong emission line ratios \citep{Maraston01,Poetrodjojo21,Nakajima22,Gibson22,Newman25}. Further issues may arise when deriving dust attenuation laws from SED fitting. \cite{Qin2022} suggest that the observationally inferred relation between $A_V$ and $\delta$ may be influenced by degeneracies in the fitting process, although \cite{Meldorf2024} argue otherwise. Furthermore, assumptions about the star formation history (SFH) can significantly impact the derived attenuation laws. For example, \cite{Belles2023} found that assuming a double power-law SFH resulted in steeper attenuation curves compared to an exponentially declining SFH in studies of local galaxies. Strategies to mitigate such biases, such as fixing stellar metallicity at solar values and minimizing variability in recent SFH bins, have been proposed by \cite{Osborne2024}.

\section{Summary}\label{summary}

In this study, we investigate the impact of dust on the FUV-to-NIR spectral energy distributions of galaxies, focusing on how dust attenuation curves are shaped by global galaxy properties (e.g., stellar mass, SFR, metallicity, and stellar age) and by the spatial distribution of stars and dust. Using the TNG50 and TNG100 simulations, we analyze a sample of $\sim 6400$ galaxies at $z = 0.07$, spanning a broad range in stellar mass, SFR, and metallicity. Synthetic attenuation curves are generated for these galaxies using the RT code \textsc{skirt} \citep{CampsBaes2020}, accounting for orientation effects across 51 lines of sight per source. We fit these curves using a flexible four-parameter model $(A_{\rm V}, c_1, c_2, c_3, c_4)$ that captures both standard (e.g., SMC, MW, and Calzetti) and more diverse attenuation curves. Our key results are presented here.

\begin{itemize}[topsep=2pt, itemsep=3pt, leftmargin=15pt, label=$\square$]
    \item \textbf{Characteristic dust attenuation in simulated galaxies:}  
    The attenuation curves of TNG galaxies exhibit a range of shapes, with most galaxies displaying moderate overall attenuation ($A_{\rm V} \sim 0.26$), a steep UV slope ($c_1 \sim 12.8$), and a stronger $2175\,\angstrom$ bump than the Milky Way ($c_4 \sim 0.08$). These trends align with those inferred for typical observed star-forming galaxies \citep{Salim18}. The diversity in attenuation curves, driven by variations in both galaxy properties and line-of-sight effects, even for a fixed dust composition, underscores the need to move beyond local templates when modeling dust attenuation.
    
    \item \textbf{Dependence on star formation and dust geometry:}  
    Galaxies with higher star formation rate surface densities ($\Sigma_{\rm SFR}$) tend to be more highly dust attenuated, following a tight relation provided in eq.~\ref{eq_Av_SigmaSFRy}. This is consistent with the idea that denser, actively star-forming galaxies have larger dust and gas content. The attenuation curves also become shallower for galaxies with higher $\Sigma_{\rm SFR}$, as reflected in the observed anti-correlation between $A_{\rm V}$ and the UV slope  (see Eq.~\ref{c1Av_eq}).
    
    \item \textbf{Internal consistency in attenuation curve shapes:}  
    The UV slope ($c_1$) correlates with the FUV slope ($c_3$), while the $2175\,\angstrom$ bump strength ($c_4$) is negatively correlated with $A_{\rm V}$. These relationships (eq.~\ref{c1c3_eq} and eq.~\ref{c4Av_eq}, respectively) provide insight into the intrinsic connections between different dust attenuation properties and allow for predictive modeling of attenuation curves.
    
    \item \textbf{Predictive scaling relations for observational applications:}  
    Given the minimal impact of the optical-to-NIR slope ($c_2$) on the shape of attenuation for the simulated galaxies, we fix it to the median value ($1.87$) in our predictive framework (see eq.~\ref{eq_Av_SigmaSFRy}-\ref{c4Av_eq}). This allows us to estimate attenuation curves using only $\Sigma_{\rm SFR}$ and/or $A_{\rm V}$, from which we can infer the UV and FUV slopes ($c_1, c_3$) and bump strength ($c_4$). These relations can be directly incorporated into inference procedures, and hence provide a practical tool for analyzing observational datasets.
    
    \item \textbf{Comparison to observed attenuation curves:}  
    The synthetic attenuation curves derived for the TNG galaxies closely match observational trends from the GALEX-SDSS-WISE Legacy Catalog over the $1500-8000\,\angstrom$ wavelength range. The primary discrepancy lies in the $2175\,\angstrom$ bump strength, which is systematically lower in the observations (albeit still within $1\sigma$ from the median TNG value), potentially reflecting differences in dust composition or grain size distributions.
\end{itemize}

In summary, we provide the most extensive synthetic attenuation curve study performed to date in local galaxies \citep[but see also][]{Trayford20}. Having validated our model with observations, the prior ranges and best-fitting relations derived here can be leveraged in inference studies where dust models are needed to produce galaxy color diagrams or luminosity functions \citep{Lovell24}, or in SED fitting procedures employing parametric attenuation curves \citep[e.g.][]{DustE22,Markov23,Markov24,Fisher25}. Future studies using higher-resolution simulations and dynamic dust models \citep[e.g.][]{Hirashita20,Hu_2019,Hu_2023,Choban2022,Choban2024,Choban2024b} will be crucial to further refine these results and bridge scales from ISM structures to galaxy-wide properties.

\begin{figure*}
    \includegraphics[width=0.3\linewidth]{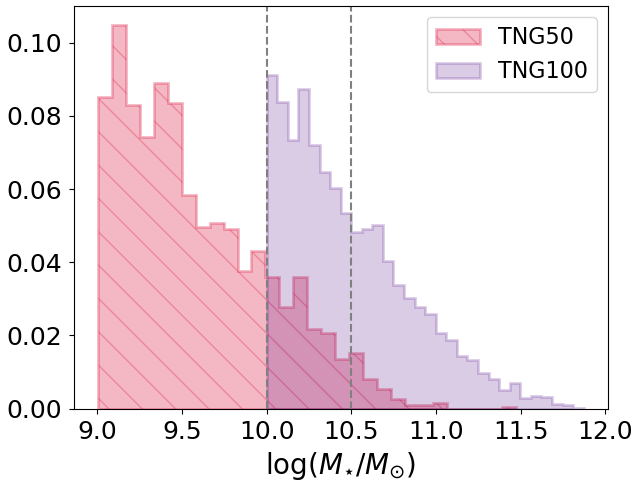}\\
    \includegraphics[width=0.9\linewidth]{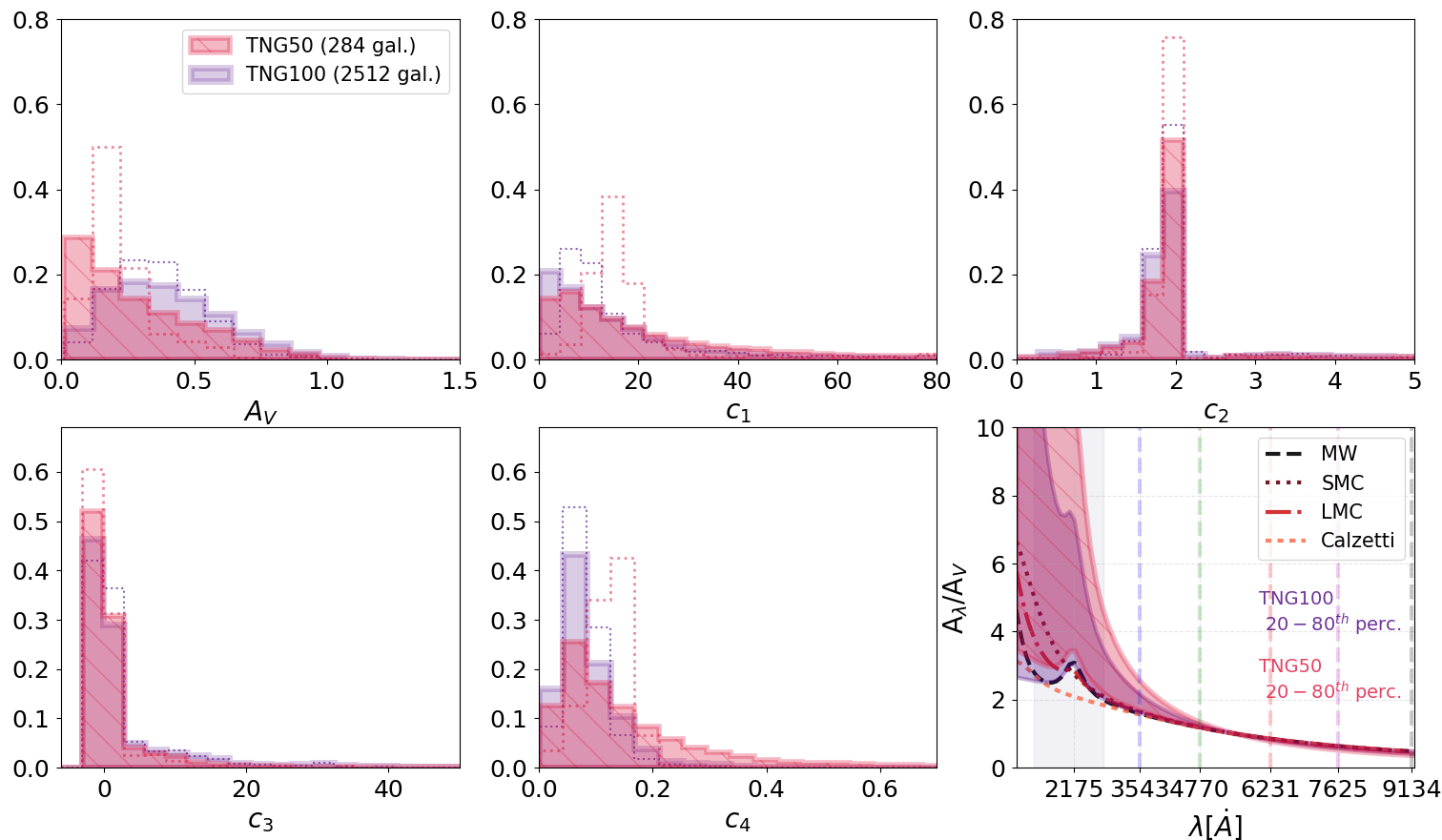}
    \caption{\textbf{Top panel}: PDF of the stellar masses in the considered TNG50 and TNG100 simulated samples. We highlight the overlap region ($10 < \log_{10}(M_{\star}/\rm{M_{\odot}}) < 10.5$) where we compare the attenuation curve properties to investigate resolution effects. \textbf{Bottom panels}: Same as Fig.~\ref{Fig_Param-Hist_TNG} but for TNG50 (red) and TNG100 (purple) galaxies with stellar masses in the interval $10 < \log_{10}(M_{\star}/\rm{M_{\odot}}) < 10.5$. The $20^{\rm{th}}-80^{\rm{th}}$ percentile range spanned by the corresponding attenuation curves is shown in the bottom right panel.}
    \label{consist}
\end{figure*}

\section*{Acknowledgments}
We thank Samir Salim for kindly providing data from his study. We also acknowledge Hiddo Algera, Fabio di Mascia, Hanae Inami, Joel Leja, Gautam Nagaraj, and Erica Nelson for their valuable discussions and insights. This work was supported in part by funding from the Simons Foundation through the Simon’s Collaboration on Learning the Universe. The Flatiron Institute is supported by the Simons Foundation. The radiative transfer simulations presented in this work were run on the Flatiron Institute’s research computing facilities (Popeye). 
RKC is grateful for support from the Leverhulme Trust via the Leverhulme Early Career Fellowship.
We acknowledge usage of the Python programming language \citep{python2,python3}, Astropy \citep{astropy}, Matplotlib \citep{matplotlib}, NumPy \citep{numpy}, and SciPy \citep{scipy}.

\appendix{}
\section{Effect of changing Mass Resolution between TNG50 and TNG100}\label{Apped_resol}
Using galaxies in the overlapping stellar mass range ($10 < \log_{10}(M_{\star}/\rm{M_{\odot}}) < 10.5$), we verify that the change in resolution between the TNG50 and TNG100 simulations does not introduce systematic differences in the modeled dust attenuation curves. As shown in Fig.~\ref{consist}, we compare the attenuation curve parameters and their corresponding $20^{\rm{th}}-80^{\rm{th}}$ percentile ranges in both simulations. We find no major discrepancy between TNG50 and TNG100, aside from TNG50 galaxies being slightly more skewed towards lower $A_V$ values and thus slightly steeper attenuation curves. This comparison demonstrates the robustness of our results across the two different simulation resolutions and justifies the use of the combined TNG50+TNG100 sample in the analysis of the simulations. 

It is important to stress that neither TNG50 nor TNG100 resolves the multiphase ISM nor -- more importantly -- the galaxy's star-forming regions. Indeed, the gas cells minimum softening length is $185\,\rm{pc}$ in TNG100 and $74\,\rm{pc}$ in TNG50 \citep{Pillepich18,Pillepich19}. Typical sizes for local giant molecular clouds are $\simlt 10\,\rm{pc}$ \cite[see e.g.][]{Zucker_2021}. This implies that neither simulation can capture dust-to-stellar geometry effects at the scale of birth clouds, similar to most other galaxy formation simulations, including zoom-ins \citep[see e.g.][which provides attenuation curves from zoom-in simulations at approximately similar resolution]{Narayanan18}. This limitation is significant, as such small-scale structures strongly influence the global attenuation curve of galaxies (see \citealt{DiMascia24} for a study of attenuation curves in isolated SF cloud throughout its lifetime). However, achieving such high spatial resolution typically comes at the expense of the sample size and parameters space probed in terms of global galaxy properties. The study presented here analyzes the largest library of synthetic attenuation curves ever produced, both in terms of inclination angles and a number of local simulated sources. A comparison and validation of our results with higher spatial resolution simulations is deferred to future work.

\begin{table*}
    \centering
    \caption{List of TNG galaxy properties analyzed in this study and their definitions.}
    \begin{tabular}{ll}
        \hline
        \textbf{Property} & \textbf{Definition / Derivation} \\
        \hline
        Star formation rate (SFR) & Averaged over $10\,\rm{Myr}$ ($\mathrm{SFR_{10}}$) and $100\,\rm{Myr}$ ($\mathrm{SFR_{100}}$) \\
        Specific SFR (sSFR) & $\rm{sSFR}_{10}=\mathrm{SFR_{10}}/M_{\star}$; $\rm{sSFR}_{100}=\mathrm{SFR_{100}}/M_{\star}$\\ 
        Stellar age & Mass-weighted mean stellar age \\
        Effective radii & Mass weighted radius of young stars ($r_{\star,y}$; age $<10\,\rm{Myr}$), and old stars ($r_{\star,o}$; age $>10\,\rm{Myr}$)\\
        Gas radii & Half-mass radius for cold gas ($r_{\rm g}$) and star-forming gas ($r_{\rm g,SF}$)\\
        Stellar radius & Half-mass radius for all stars ($r_{\star}$)\\
        Stellar mass ($M_{\star}$) & Total mass of stars within subhalo\\
        Gas mass ($M_{\rm g}$) & Total mass of gas within subhalo\\
        Dust mass ($M_{\rm dust}$) & Total dust mass within subhalo, assuming a fixed dust-to-metal ratio as in eq. \ref{eq_RR14}\\  
        Dust mass within radii & $M_{\rm dust}(r_{\star,y})$; $M_{\rm dust}(r_{\star,o})$; $M_{\rm dust}(r_{\rm g,SF})$ \\
        Gas-phase metallicity ($Z_g$) & Mass-weighted metallicity of the gas component \\
        Dust surface densities at radii & $\Sigma_{\rm dust}(r_{\rm g})=M_{\rm dust}/\pi r_{\rm g}^2$; $\Sigma_{\rm dust}(r_{\star,y})=M_{\rm dust}(r_{\star,y})/\pi r_{\star,y}^2$; \\
        & $\Sigma_{\rm dust}(r_{\star,o})=M_{\rm dust}(r_{\star,o})/\pi r_{\star,o}^2$; \\
        & $\Sigma_{\rm dust} (r_{\rm g,SF})=M_{\rm dust}(r_{\rm g,SF})/\pi r_{\rm g,SF}^2$; \\
        SFR surface densities at radii & $\Sigma_{\rm SFR,y}=\mathrm{SFR}_{10}/\pi r_{\star,y}^2$; $\Sigma_{\rm SFR,o}=\mathrm{SFR}_{100}/\pi r_{\star,o}^2$; $\Sigma_{\rm SFR}=\mathrm{SFR}_{100}/\pi r_{\star}^2$\\
        Gas surface densities at radii & $\Sigma_{\rm g} =M_{\rm g}/\pi r_{\rm g}^2$; $\Sigma_{\rm g,SF}=M_{\rm g}(r_{\rm g,SF})/\pi r_{\rm g,SF}^2$ \\
        Burstiness parameter ($\kappa_s$) & Deviation from the Kennicutt-Schmidt relation: $\kappa_s = 10^{12}\ \Sigma_{\rm SFR}/\Sigma_{\rm g}^{1.4}$;\\ 
        & $\kappa_{s,y} = 10^{12}\ \Sigma_{\rm SFR,y}/\Sigma_{\rm g}^{1.4}$; $\kappa_{s,o} = 10^{12}\ \Sigma_{\rm SFR,o}/\Sigma_{\rm g}^{1.4}$\\
        \hline
    \end{tabular}
    \label{tab:galaxy_properties_TNG}
\end{table*}

\begin{figure*}
    \centering
    \includegraphics[width=0.495\textwidth]{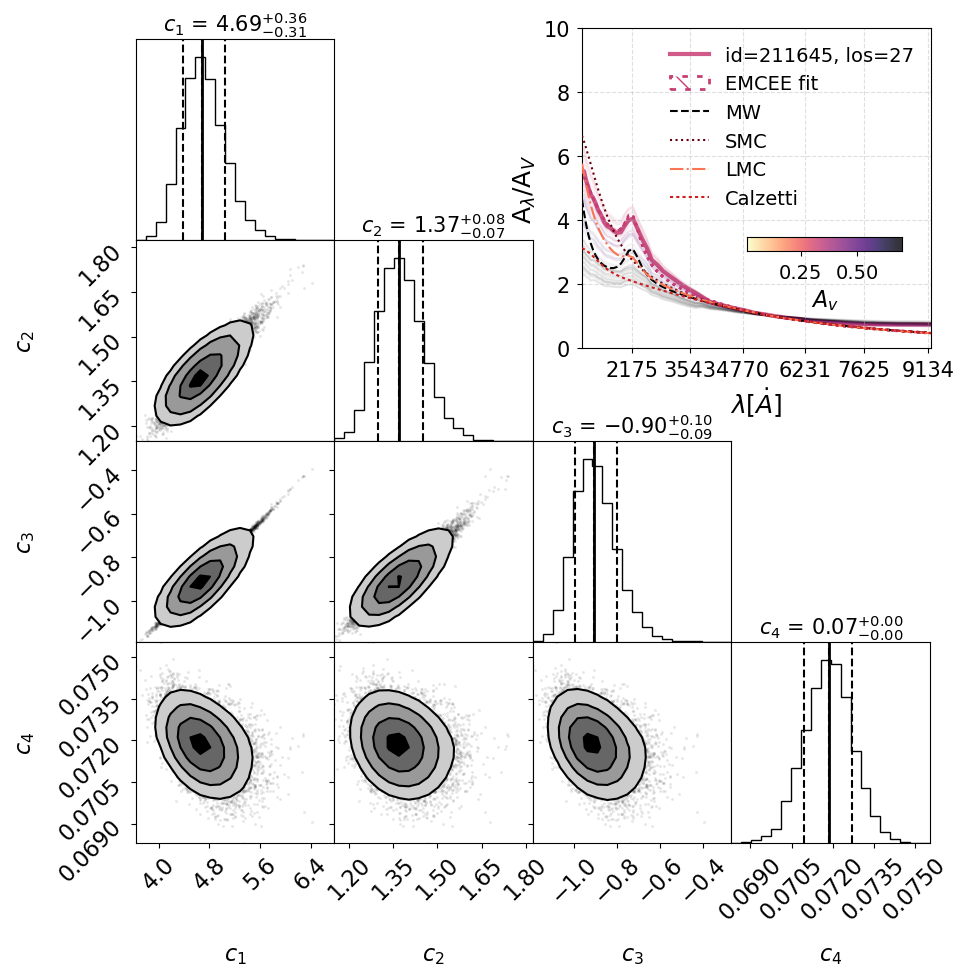}
    \includegraphics[width=0.495\textwidth]{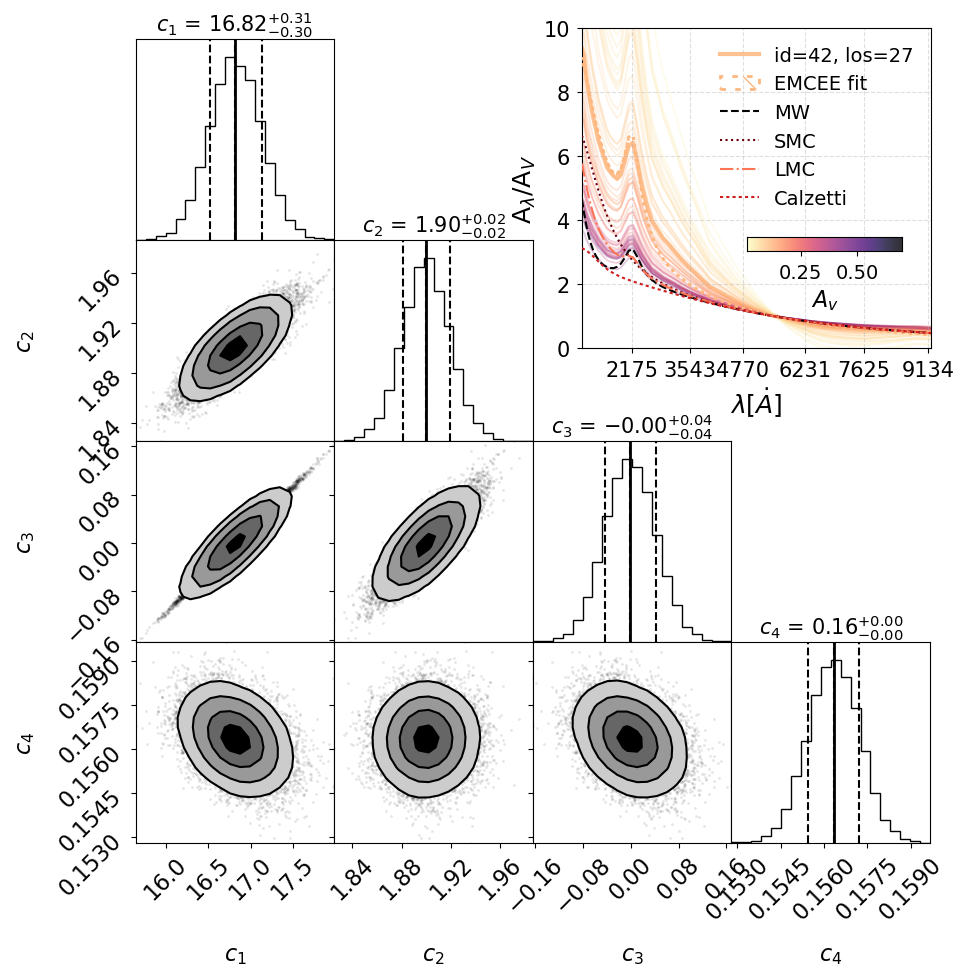}
    \caption{ Posterior plots of the four parameters $(c_1,c_2,c_3,c_4)$ obtained from fitting the attenuation curves (along a randomly selected line of sight, los 27) of two reference TNG50 sources, one with a MW-like attenuation curve (id211645, left) and one with a steeper attenuation curve (id42, right). All parameters are well-constrained. In the 1D posterior plots, the solid and dashed lines show -- respectively -- the median, $16^{\rm{th}}$, and $84^{\rm{th}}$ percentiles of each parameter's posterior distribution. In the inset plots we show the attenuation curves fitted with the {\it emcee} (thick solid line), the corresponding {\it emcee} best-fit (dashed line), and the variation in each source's attenuation curve due to different orientations (solid transparent lines), color coded according to the corresponding $ A_{\rm V}$, see colorbar). For reference, we also show the usual empirical extinction/attenuation curves MW, SMC, LMC and Calzetti.}
    \label{Fig_Example_EMCEE}
\end{figure*}

\section{Effect of changing orientation in TNG50 galaxies}
In Fig.~\ref{Fig_Example_EMCEE} we show two reference TNG50 galaxies, one with a shallow, MW-like attenuation curve (on the left) and the other with a steeper UV rise and stronger bump (on the right); we refer to the latter as the \quotes{steep} source. 
We show the posterior distributions of the parameters $(c_1,c_2,c_3,c_4)$ for the same line of sight for each of the two sources, but also investigate the variation in parameters due to line of sight. 
We find that the MW-like galaxy is characterized by lower $(c_1,c_2,c_4)$ values than the other source and most of the entire TNG50 population (see Tab.~\ref{Table_best_fit_dust_par_TNG_Salim}). On the other hand, for the other reference source, $(c_1,c_2,c_4)$ are very close to the median of the TNG50 population, which indeed tends to show steeper than MW attenuation curves (see the lower right panel in Fig.~\ref{Fig_Param-Hist_TNG}). 

Comparing the attenuation curves derived for different lines of sight, we can see that the UV regime is significantly affected by geometrical effects, with deviations of up to a factor $\gg 2$ in the strength of the bump and the UV slope. This is clearly reflected in the wide variation in the parameters $(c_1,c_3,c_4)$ depending on the line of sight. 
For both sources, the $16^{\rm{th}}-84^{\rm{th}}$ percentile variations — $\Delta c_1 \sim 25$, $\Delta c_3 \sim 8$, and $\Delta c_4 \sim 0.08$ — are much larger than the corresponding best-fit values for the Milky Way itself, for which $(c_{\rm 1,MW},\,c_{\rm 3,MW},\, c_{\rm 4,MW}) = (14.4,\,2.04,\,0.052)$. This indicates that, depending on the line of sight, the inferred attenuation curve can vary from a MW-like to a Calzetti-like curve (for the shallow source), or from a MW-like to a steeper than SMC-like curve (for the steep source; see the inset panels in Fig.~\ref{Fig_Example_EMCEE}).

\bibliographystyle{aasjournal}
\bibliography{bibliography}

\end{document}

%% file: definitions.tex


\def\be{\begin{equation}}
\def\ee{\end{equation}}
\newcommand{\code}[1]{\textsc{#1}}
\newcommand\quotesingle[1]{`{#1}'}
\newcommand\quotes[1]{``{#1}"}
\def\gsim{\lower.5ex\hbox{\gtsima}} 
\def\lsim{\lower.5ex\hbox{\ltsima}} 
\def\gtsima{$\; \buildrel > \over \sim \;$} 
\def\ltsima{$\; \buildrel < \over \sim \;$} \def\gsim{\lower.5ex\hbox{\gtsima}} 
\def\lsim{\lower.5ex\hbox{\ltsima}} 
\def\simgt{\lower.5ex\hbox{\gtsima}} 
\def\simlt{\lower.5ex\hbox{\ltsima}}

\def\msun{{\rm M}_{\odot}}
\def\lsun{{\rm L}_{\odot}}
\def\dsun{{\cal D}_{\odot}}
\def\fsun{\xi_{\odot}}
\def\zsun{{\rm Z}_{\odot}}
\def\msunyr{\msun {\rm yr}^{-1}}
\def\gdens{\msun\,{\rm kpc}^{-2}}
\def\sfrdens{\msun\,{\rm yr}^{-1}\,{\rm kpc}^{-2}}

\def\mum{\mu {\rm m}}
\newcommand{\angstrom}{\mbox{\normalfont\AA}}
\def\cc{\rm cm^{-3}}
\def\uflux{{\rm erg}\,{\rm s}^{-1} {\rm cm}^{-2} }

\def\fdust{\xi_{d}}
\def\fesc{f_{\rm esc}\,}
\def\td{\tau_{sd}}
\def\Sg{$\Sigma_{g}$}
\def\S*{$\Sigma_{\rm SFR}$}
\def\Ssfr{\Sigma_{\rm SFR}}
\def\Sgas{\Sigma_{\rm g}}
\def\Sstar{\Sigma_{\rm *}}
\def\Sesc{\Sigma_{\rm esc}}
\def\Srad{\Sigma_{\rm rad}}

\def\Dsolar{${\cal D}/\dsun$}
\def\Zsolar{$Z/\zsun$}
\def\DDsolar{\left( {{\cal D}\over \dsun} \right)}
\def\ZZsolar{\left( {Z \over \zsun} \right)}
\def\kms{{\rm km\,s}^{-1}\,}
\def\skms{$\sigma_{\rm kms}\,$}

\def\Scii{$\Sigma_{\rm [CII]}$}
\def\Sciimax{$\Sigma_{\rm [CII]}^{\rm max}$}
\def\CII{\hbox{[C~$\scriptstyle\rm II $]~}}
\def\CIII{\hbox{C~$\scriptstyle\rm III $]~}}
\def\OII{\hbox{[O~$\scriptstyle\rm II $]~}}
\def\OIII{\hbox{[O~$\scriptstyle\rm III $]~}}
\def\HH{\hbox{H$_2$}~} 
\def\HI{\hbox{H~$\scriptstyle\rm I\ $}} 
\def\HII{\hbox{H~$\scriptstyle\rm II\ $}} 
\def\CIion{\hbox{C~$\scriptstyle\rm I $~}}
\def\CIIion{\hbox{C~$\scriptstyle\rm II $~}}
\def\CIIIion{\hbox{C~$\scriptstyle\rm III $~}}
\def\CIVion{\hbox{C~$\scriptstyle\rm IV $~}}
\def\nhh{n_{\rm H2}}
\def\nhi{n_{\rm HI}}
\def\nhii{n_{\rm HII}}
\def\fhh{x_{\rm H2}}
\def\fhi{x_{\rm HI}}
\def\fhii{x_{\rm HII}}
\def\fd{f^*_{\rm diss}} 
\def\ks{\kappa_{\rm s}}

\def\cyan{\color{cyan}}
\definecolor{apcolor}{HTML}{b3003b}
\definecolor{afcolor}{HTML}{800080}
\definecolor{lvcolor}{HTML}{DF7401}
\definecolor{mdcolor}{HTML}{01abdf} 
\definecolor{cbcolor}{HTML}{ff0000}
\definecolor{sccolor}{HTML}{cc5500} 
\definecolor{sgcolor}{HTML}{00cc7a}

%% file: main.bbl
\begin{thebibliography}{}
\expandafter\ifx\csname natexlab\endcsname\relax\def\natexlab#1{#1}\fi
\providecommand{\url}[1]{\href{#1}{#1}}
\providecommand{\dodoi}[1]{doi:~\href{http://doi.org/#1}{\nolinkurl{#1}}}
\providecommand{\doeprint}[1]{\href{http://ascl.net/#1}{\nolinkurl{http://ascl.net/#1}}}
\providecommand{\doarXiv}[1]{\href{https://arxiv.org/abs/#1}{\nolinkurl{https://arxiv.org/abs/#1}}}

\bibitem[{{Abazajian} {et~al.}(2009){Abazajian}, {Adelman-McCarthy}, {Ag{\"u}eros}, {Allam}, {Allende Prieto}, {An}, {Anderson}, {Anderson}, {Annis}, {Bahcall}, {Bailer-Jones}, {Barentine}, {Bassett}, {Becker}, {Beers}, {Bell}, {Belokurov}, {Berlind}, {Berman}, {Bernardi}, {Bickerton}, {Bizyaev}, {Blakeslee}, {Blanton}, {Bochanski}, {Boroski}, {Brewington}, {Brinchmann}, {Brinkmann}, {Brunner}, {Budav{\'a}ri}, {Carey}, {Carliles}, {Carr}, {Castander}, {Cinabro}, {Connolly}, {Csabai}, {Cunha}, {Czarapata}, {Davenport}, {de Haas}, {Dilday}, {Doi}, {Eisenstein}, {Evans}, {Evans}, {Fan}, {Friedman}, {Frieman}, {Fukugita}, {G{\"a}nsicke}, {Gates}, {Gillespie}, {Gilmore}, {Gonzalez}, {Gonzalez}, {Grebel}, {Gunn}, {Gy{\"o}ry}, {Hall}, {Harding}, {Harris}, {Harvanek}, {Hawley}, {Hayes}, {Heckman}, {Hendry}, {Hennessy}, {Hindsley}, {Hoblitt}, {Hogan}, {Hogg}, {Holtzman}, {Hyde}, {Ichikawa}, {Ichikawa}, {Im}, {Ivezi{\'c}}, {Jester}, {Jiang}, {Johnson}, {Jorgensen}, {Juri{\'c}}, {Kent}, {Kessler}, {Kleinman}, {Knapp},
  {Konishi}, {Kron}, {Krzesinski}, {Kuropatkin}, {Lampeitl}, {Lebedeva}, {Lee}, {Lee}, {French Leger}, {L{\'e}pine}, {Li}, {Lima}, {Lin}, {Long}, {Loomis}, {Loveday}, {Lupton}, {Magnier}, {Malanushenko}, {Malanushenko}, {Mandelbaum}, {Margon}, {Marriner}, {Mart{\'\i}nez-Delgado}, {Matsubara}, {McGehee}, {McKay}, {Meiksin}, {Morrison}, {Mullally}, {Munn}, {Murphy}, {Nash}, {Nebot}, {Neilsen}, {Newberg}, {Newman}, {Nichol}, {Nicinski}, {Nieto-Santisteban}, {Nitta}, {Okamura}, {Oravetz}, {Ostriker}, {Owen}, {Padmanabhan}, {Pan}, {Park}, {Pauls}, {Peoples}, {Percival}, {Pier}, {Pope}, {Pourbaix}, {Price}, {Purger}, {Quinn}, {Raddick}, {Re Fiorentin}, {Richards}, {Richmond}, {Riess}, {Rix}, {Rockosi}, {Sako}, {Schlegel}, {Schneider}, {Scholz}, {Schreiber}, {Schwope}, {Seljak}, {Sesar}, {Sheldon}, {Shimasaku}, {Sibley}, {Simmons}, {Sivarani}, {Allyn Smith}, {Smith}, {Smol{\v{c}}i{\'c}}, {Snedden}, {Stebbins}, {Steinmetz}, {Stoughton}, {Strauss}, {SubbaRao}, {Suto}, {Szalay}, {Szapudi}, {Szkody}, {Tanaka},
  {Tegmark}, {Teodoro}, {Thakar}, {Tremonti}, {Tucker}, {Uomoto}, {Vanden Berk}, {Vandenberg}, {Vidrih}, {Vogeley}, {Voges}, {Vogt}, {Wadadekar}, {Watters}, {Weinberg}, {West}, {White}, {Wilhite}, {Wonders}, {Yanny}, \& {Yocum}}]{SDSS_Abazajian}
{Abazajian}, K.~N., {Adelman-McCarthy}, J.~K., {Ag{\"u}eros}, M.~A., {et~al.} 2009, \apjs, 182, 543, \dodoi{10.1088/0067-0049/182/2/543}

\bibitem[{{Adelman-McCarthy} {et~al.}(2008){Adelman-McCarthy}, {Ag{\"u}eros}, {Allam}, {Allende Prieto}, {Anderson}, {Anderson}, {Annis}, {Bahcall}, {Bailer-Jones}, {Baldry}, {Barentine}, {Bassett}, {Becker}, {Beers}, {Bell}, {Berlind}, {Bernardi}, {Blanton}, {Bochanski}, {Boroski}, {Brinchmann}, {Brinkmann}, {Brunner}, {Budav{\'a}ri}, {Carliles}, {Carr}, {Castander}, {Cinabro}, {Cool}, {Covey}, {Csabai}, {Cunha}, {Davenport}, {Dilday}, {Doi}, {Eisenstein}, {Evans}, {Fan}, {Finkbeiner}, {Friedman}, {Frieman}, {Fukugita}, {G{\"a}nsicke}, {Gates}, {Gillespie}, {Glazebrook}, {Gray}, {Grebel}, {Gunn}, {Gurbani}, {Hall}, {Harding}, {Harvanek}, {Hawley}, {Hayes}, {Heckman}, {Hendry}, {Hindsley}, {Hirata}, {Hogan}, {Hogg}, {Hyde}, {Ichikawa}, {Ivezi{\'c}}, {Jester}, {Johnson}, {Jorgensen}, {Juri{\'c}}, {Kent}, {Kessler}, {Kleinman}, {Knapp}, {Kron}, {Krzesinski}, {Kuropatkin}, {Lamb}, {Lampeitl}, {Lebedeva}, {Lee}, {French Leger}, {L{\'e}pine}, {Lima}, {Lin}, {Long}, {Loomis}, {Loveday}, {Lupton}, {Malanushenko},
  {Malanushenko}, {Mandelbaum}, {Margon}, {Marriner}, {Mart{\'\i}nez-Delgado}, {Matsubara}, {McGehee}, {McKay}, {Meiksin}, {Morrison}, {Munn}, {Nakajima}, {Neilsen}, {Newberg}, {Nichol}, {Nicinski}, {Nieto-Santisteban}, {Nitta}, {Okamura}, {Owen}, {Oyaizu}, {Padmanabhan}, {Pan}, {Park}, {Peoples}, {Pier}, {Pope}, {Purger}, {Raddick}, {Re Fiorentin}, {Richards}, {Richmond}, {Riess}, {Rix}, {Rockosi}, {Sako}, {Schlegel}, {Schneider}, {Schreiber}, {Schwope}, {Seljak}, {Sesar}, {Sheldon}, {Shimasaku}, {Sivarani}, {Allyn Smith}, {Snedden}, {Steinmetz}, {Strauss}, {SubbaRao}, {Suto}, {Szalay}, {Szapudi}, {Szkody}, {Tegmark}, {Thakar}, {Tremonti}, {Tucker}, {Uomoto}, {Vanden Berk}, {Vandenberg}, {Vidrih}, {Vogeley}, {Voges}, {Vogt}, {Wadadekar}, {Weinberg}, {West}, {White}, {Wilhite}, {Yanny}, {Yocum}, {York}, {Zehavi}, \& {Zucker}}]{SDSS_Adelman}
{Adelman-McCarthy}, J.~K., {Ag{\"u}eros}, M.~A., {Allam}, S.~S., {et~al.} 2008, \apjs, 175, 297, \dodoi{10.1086/524984}

\bibitem[{{Aoyama} {et~al.}(2020){Aoyama}, {Hirashita}, \& {Nagamine}}]{Aoyama:2020}
{Aoyama}, S., {Hirashita}, H., \& {Nagamine}, K. 2020, \mnras, 491, 3844, \dodoi{10.1093/mnras/stz3253}

\bibitem[{{Astropy Collaboration} {et~al.}(2013){Astropy Collaboration}, {Robitaille}, {Tollerud}, {Greenfield}, {Droettboom}, {Bray}, {Aldcroft}, {Davis}, {Ginsburg}, {Price-Whelan}, {Kerzendorf}, {Conley}, {Crighton}, {Barbary}, {Muna}, {Ferguson}, {Grollier}, {Parikh}, {Nair}, {Unther}, {Deil}, {Woillez}, {Conseil}, {Kramer}, {Turner}, {Singer}, {Fox}, {Weaver}, {Zabalza}, {Edwards}, {Azalee Bostroem}, {Burke}, {Casey}, {Crawford}, {Dencheva}, {Ely}, {Jenness}, {Labrie}, {Lim}, {Pierfederici}, {Pontzen}, {Ptak}, {Refsdal}, {Servillat}, \& {Streicher}}]{astropy}
{Astropy Collaboration}, {Robitaille}, T.~P., {Tollerud}, E.~J., {et~al.} 2013, \aap, 558, A33, \dodoi{10.1051/0004-6361/201322068}

\bibitem[{{Bari{\v{s}}i{\'c}} {et~al.}(2020){Bari{\v{s}}i{\'c}}, {Pacifici}, {van der Wel}, {Straatman}, {Bell}, {Bezanson}, {Brammer}, {D'Eugenio}, {Franx}, {van Houdt}, {Maseda}, {Muzzin}, {Sobral}, \& {Wu}}]{Barisic2020}
{Bari{\v{s}}i{\'c}}, I., {Pacifici}, C., {van der Wel}, A., {et~al.} 2020, \apj, 903, 146, \dodoi{10.3847/1538-4357/abba37}

\bibitem[{{Battisti} {et~al.}(2016){Battisti}, {Calzetti}, \& {Chary}}]{Battisti2016}
{Battisti}, A.~J., {Calzetti}, D., \& {Chary}, R.~R. 2016, \apj, 818, 13, \dodoi{10.3847/0004-637X/818/1/13}

\bibitem[{Battisti {et~al.}(2020)Battisti, da~Cunha, Shivaei, Calzetti, \& collaboration)}]{Battisti20}
Battisti, A.~J., da~Cunha, E., Shivaei, I., Calzetti, D., \& collaboration), C. 2020, The Astrophysical Journal, 888, 108, \dodoi{10.3847/1538-4357/ab5fdd}

\bibitem[{{Behrens} {et~al.}(2018){Behrens}, {Pallottini}, {Ferrara}, {Gallerani}, \& {Vallini}}]{Behrens18}
{Behrens}, C., {Pallottini}, A., {Ferrara}, A., {Gallerani}, S., \& {Vallini}, L. 2018, \mnras, 477, 552, \dodoi{10.1093/mnras/sty552}

\bibitem[{{Belles} {et~al.}(2023){Belles}, {Decleir}, {Bowman}, {Hagen}, {Gronwall}, \& {Siegel}}]{Belles2023}
{Belles}, A., {Decleir}, M., {Bowman}, W.~P., {et~al.} 2023, \apj, 953, 54, \dodoi{10.3847/1538-4357/acd332}

\bibitem[{Bouchet {et~al.}(1985)Bouchet, Lequeux, Maurice, Prevot, \& Prevot-Burnichon}]{Bouchet85}
Bouchet, P., Lequeux, J., Maurice, E., Prevot, L., \& Prevot-Burnichon, M. 1985, Astronomy and Astrophysics, 149, 330

\bibitem[{{Bruzual} \& {Charlot}(2003)}]{Bruzual2003}
{Bruzual}, G., \& {Charlot}, S. 2003, \mnras, 344, 1000, \dodoi{10.1046/j.1365-8711.2003.06897.x}

\bibitem[{Calzetti {et~al.}(2000)Calzetti, Armus, Bohlin, Kinney, Koornneef, \& Storchi-Bergmann}]{Calzetti00}
Calzetti, D., Armus, L., Bohlin, R.~C., {et~al.} 2000, The Astrophysical Journal, 533, 682

\bibitem[{{Camps} \& {Baes}(2015)}]{Camps2015}
{Camps}, P., \& {Baes}, M. 2015, Astronomy and Computing, 9, 20, \dodoi{10.1016/j.ascom.2014.10.004}

\bibitem[{{Camps} \& {Baes}(2020)}]{CampsBaes2020}
---. 2020, Astronomy and Computing, 31, 100381, \dodoi{10.1016/j.ascom.2020.100381}

\bibitem[{Cardelli {et~al.}(1989)Cardelli, Clayton, \& Mathis}]{Cardelli89}
Cardelli, J.~A., Clayton, G.~C., \& Mathis, J.~S. 1989, The Astrophysical Journal, 345, 245

\bibitem[{{Charlot} \& {Fall}(2000)}]{CharlotFall2000}
{Charlot}, S., \& {Fall}, S.~M. 2000, \apj, 539, 718, \dodoi{10.1086/309250}

\bibitem[{{Chevallard} {et~al.}(2013){Chevallard}, {Charlot}, {Wandelt}, \& {Wild}}]{Chevallard13}
{Chevallard}, J., {Charlot}, S., {Wandelt}, B., \& {Wild}, V. 2013, \mnras, 432, 2061, \dodoi{10.1093/mnras/stt523}

\bibitem[{{Choban} {et~al.}(2022){Choban}, {Kere{\v{s}}}, {Hopkins}, {Sandstrom}, {Hayward}, \& {Faucher-Gigu{\`e}re}}]{Choban2022}
{Choban}, C.~R., {Kere{\v{s}}}, D., {Hopkins}, P.~F., {et~al.} 2022, \mnras, 514, 4506, \dodoi{10.1093/mnras/stac1542}

\bibitem[{{Choban} {et~al.}(2024{\natexlab{a}}){Choban}, {Kere{\v{s}}}, {Sandstrom}, {Hopkins}, {Hayward}, \& {Faucher-Gigu{\`e}re}}]{Choban2024}
{Choban}, C.~R., {Kere{\v{s}}}, D., {Sandstrom}, K.~M., {et~al.} 2024{\natexlab{a}}, \mnras, 529, 2356, \dodoi{10.1093/mnras/stae716}

\bibitem[{{Choban} {et~al.}(2024{\natexlab{b}}){Choban}, {Salim}, {Kere{\v{s}}}, {Hayward}, \& {Sandstrom}}]{Choban2024b}
{Choban}, C.~R., {Salim}, S., {Kere{\v{s}}}, D., {Hayward}, C.~C., \& {Sandstrom}, K.~M. 2024{\natexlab{b}}, arXiv e-prints, arXiv:2408.08962, \dodoi{10.48550/arXiv.2408.08962}

\bibitem[{{Cochrane} {et~al.}(2024){Cochrane}, {Angl{\'e}s-Alc{\'a}zar}, {Cullen}, \& {Hayward}}]{Cochrane24}
{Cochrane}, R.~K., {Angl{\'e}s-Alc{\'a}zar}, D., {Cullen}, F., \& {Hayward}, C.~C. 2024, \apj, 961, 37, \dodoi{10.3847/1538-4357/ad02f8}

\bibitem[{{Cochrane} {et~al.}(2023{\natexlab{a}}){Cochrane}, {Hayward}, {Angl{\'e}s-Alc{\'a}zar}, \& {Somerville}}]{2023MNRAS.518.5522C}
{Cochrane}, R.~K., {Hayward}, C.~C., {Angl{\'e}s-Alc{\'a}zar}, D., \& {Somerville}, R.~S. 2023{\natexlab{a}}, \mnras, 518, 5522, \dodoi{10.1093/mnras/stac3451}

\bibitem[{Cochrane {et~al.}(2025)Cochrane, Katz, Begley, Hayward, \& Best}]{Cochrane_2025}
Cochrane, R.~K., Katz, H., Begley, R., Hayward, C.~C., \& Best, P.~N. 2025, The Astrophysical Journal Letters, 978, L42, \dodoi{10.3847/2041-8213/ad9a4d}

\bibitem[{{Cochrane} {et~al.}(2019){Cochrane}, {Hayward}, {Angl{\'e}s-Alc{\'a}zar}, {Lotz}, {Parsotan}, {Ma}, {Kere{\v{s}}}, {Feldmann}, {Faucher-Gigu{\`e}re}, \& {Hopkins}}]{Cochrane19}
{Cochrane}, R.~K., {Hayward}, C.~C., {Angl{\'e}s-Alc{\'a}zar}, D., {et~al.} 2019, \mnras, 488, 1779, \dodoi{10.1093/mnras/stz1736}

\bibitem[{{Cochrane} {et~al.}(2023{\natexlab{b}}){Cochrane}, {Angl{\'e}s-Alc{\'a}zar}, {Mercedes-Feliz}, {Hayward}, {Faucher-Gigu{\`e}re}, {Wellons}, {Terrazas}, {Wetzel}, {Hopkins}, {Moreno}, {Su}, \& {Somerville}}]{Cochrane23}
{Cochrane}, R.~K., {Angl{\'e}s-Alc{\'a}zar}, D., {Mercedes-Feliz}, J., {et~al.} 2023{\natexlab{b}}, \mnras, 523, 2409, \dodoi{10.1093/mnras/stad1528}

\bibitem[{{Cousin} {et~al.}(2019){Cousin}, {Buat}, {Lagache}, \& {Bethermin}}]{Cousin19}
{Cousin}, M., {Buat}, V., {Lagache}, G., \& {Bethermin}, M. 2019, \aap, 627, A132, \dodoi{10.1051/0004-6361/201834674}

\bibitem[{{Dayal} {et~al.}(2022){Dayal}, {Ferrara}, {Sommovigo}, {Bouwens}, {Oesch}, {Smit}, {Gonzalez}, {Schouws}, {Stefanon}, {Kobayashi}, {Bremer}, {Algera}, {Aravena}, {Bowler}, {da Cunha}, {Fudamoto}, {Graziani}, {Hodge}, {Inami}, {De Looze}, {Pallottini}, {Riechers}, {Schneider}, {Stark}, \& {Endsley}}]{Dayal22}
{Dayal}, P., {Ferrara}, A., {Sommovigo}, L., {et~al.} 2022, \mnras, 512, 989, \dodoi{10.1093/mnras/stac537}

\bibitem[{{De Lucia} \& {Blaizot}(2007)}]{deLucia2007}
{De Lucia}, G., \& {Blaizot}, J. 2007, \mnras, 375, 2, \dodoi{10.1111/j.1365-2966.2006.11287.x}

\bibitem[{{Devriendt} {et~al.}(1999){Devriendt}, {Guiderdoni}, \& {Sadat}}]{Devriendt1999}
{Devriendt}, J.~E.~G., {Guiderdoni}, B., \& {Sadat}, R. 1999, \aap, 350, 381, \dodoi{10.48550/arXiv.astro-ph/9906332}

\bibitem[{{Di Mascia} {et~al.}(2024){Di Mascia}, {Pallottini}, {Sommovigo}, \& {Decataldo}}]{DiMascia24}
{Di Mascia}, F., {Pallottini}, A., {Sommovigo}, L., \& {Decataldo}, D. 2024, arXiv e-prints, arXiv:2407.01662, \dodoi{10.48550/arXiv.2407.01662}

\bibitem[{Di Cesare {et~al.}(2022)Di Cesare, Graziani, Schneider, Ginolfi, Venditti, Santini, \& Hunt}]{diCesare23}
Di Cesare, C., Graziani, L., Schneider, R., {et~al.} 2022, Monthly Notices of the Royal Astronomical Society, 519, 4632, \dodoi{10.1093/mnras/stac3702}

\bibitem[{Di Mascia {et~al.}(2021)Di Mascia, Gallerani, Ferrara, Pallottini, Maiolino, Carniani, \& D’Odorico}]{DiMascia21}
Di Mascia, F., Gallerani, S., Ferrara, A., {et~al.} 2021, Monthly Notices of the Royal Astronomical Society, 506, 3946, \dodoi{10.1093/mnras/stab1876}

\bibitem[{{Donnari} {et~al.}(2019){Donnari}, {Pillepich}, {Nelson}, {Vogelsberger}, {Genel}, {Weinberger}, {Marinacci}, {Springel}, \& {Hernquist}}]{2019MNRAS.485.4817D}
{Donnari}, M., {Pillepich}, A., {Nelson}, D., {et~al.} 2019, \mnras, 485, 4817, \dodoi{10.1093/mnras/stz712}

\bibitem[{Draine(1989)}]{Draine89}
Draine, B. 1989, in Infrared spectroscopy in astronomy, Vol. 290

\bibitem[{Draine(2003)}]{Draine03}
Draine, B. 2003, Annual Review of Astronomy and Astrophysics, 41, 241, \dodoi{10.1146/annurev.astro.41.011802.094840}

\bibitem[{Draine \& Salpeter(1979)}]{Draine79}
Draine, B., \& Salpeter, E. 1979, Astrophysical Journal, Part 1, vol. 231, July 15, 1979, p. 438-455., 231, 438

\bibitem[{{Driver} {et~al.}(2011){Driver}, {Hill}, {Kelvin}, {Robotham}, {Liske}, {Norberg}, {Baldry}, {Bamford}, {Hopkins}, {Loveday}, {Peacock}, {Andrae}, {Bland-Hawthorn}, {Brough}, {Brown}, {Cameron}, {Ching}, {Colless}, {Conselice}, {Croom}, {Cross}, {de Propris}, {Dye}, {Drinkwater}, {Ellis}, {Graham}, {Grootes}, {Gunawardhana}, {Jones}, {van Kampen}, {Maraston}, {Nichol}, {Parkinson}, {Phillipps}, {Pimbblet}, {Popescu}, {Prescott}, {Roseboom}, {Sadler}, {Sansom}, {Sharp}, {Smith}, {Taylor}, {Thomas}, {Tuffs}, {Wijesinghe}, {Dunne}, {Frenk}, {Jarvis}, {Madore}, {Meyer}, {Seibert}, {Staveley-Smith}, {Sutherland}, \& {Warren}}]{GAMA}
{Driver}, S.~P., {Hill}, D.~T., {Kelvin}, L.~S., {et~al.} 2011, \mnras, 413, 971, \dodoi{10.1111/j.1365-2966.2010.18188.x}

\bibitem[{{Dwek} \& {Scalo}(1980)}]{Dwek80}
{Dwek}, E., \& {Scalo}, J.~M. 1980, \apj, 239, 193, \dodoi{10.1086/158100}

\bibitem[{{Ferrara} {et~al.}(2022){Ferrara}, {Sommovigo}, {Dayal}, {Pallottini}, {Bouwens}, {Gonzalez}, {Inami}, {Smit}, {Bowler}, {Endsley}, {Oesch}, {Schouws}, {Stark}, {Stefanon}, {Aravena}, {da Cunha}, {De Looze}, {Fudamoto}, {Graziani}, {Hodge}, {Riechers}, {Schneider}, {Algera}, {Barrufet}, {Hygate}, {Labb{\'e}}, {Li}, {Nanayakkara}, {Topping}, \& {van der Werf}}]{Ferrara22REB}
{Ferrara}, A., {Sommovigo}, L., {Dayal}, P., {et~al.} 2022, \mnras, 512, 58, \dodoi{10.1093/mnras/stac460}

\bibitem[{{Fisher} {et~al.}(2025){Fisher}, {Bowler}, {Stefanon}, {Rowland}, {Algera}, {Aravena}, {Bouwens}, {Dayal}, {Ferrara}, {Fudamoto}, {Hodge}, {Inami}, {Ormerod}, {Pallottini}, {Phillips}, {Sartorio}, {Smit}, {Sommovigo}, {Stark}, \& {van der Werf}}]{Fisher25}
{Fisher}, R., {Bowler}, R. A.~A., {Stefanon}, M., {et~al.} 2025, arXiv e-prints, arXiv:2501.10541, \dodoi{10.48550/arXiv.2501.10541}

\bibitem[{{Fitzpatrick}(1999)}]{Fitzpatrick99}
{Fitzpatrick}, E.~L. 1999, \pasp, 111, 63, \dodoi{10.1086/316293}

\bibitem[{Fontanot {et~al.}(2008)Fontanot, Somerville, Silva, Monaco, \& Skibba}]{Fontanot09}
Fontanot, F., Somerville, R.~S., Silva, L., Monaco, P., \& Skibba, R. 2008, Monthly Notices of the Royal Astronomical Society, 392, 553, \dodoi{10.1111/j.1365-2966.2008.14126.x}

\bibitem[{{Foreman-Mackey} {et~al.}(2013){Foreman-Mackey}, {Hogg}, {Lang}, \& {Goodman}}]{emcee}
{Foreman-Mackey}, D., {Hogg}, D.~W., {Lang}, D., \& {Goodman}, J. 2013, PASP, 125, 306, \dodoi{10.1086/670067}

\bibitem[{Gardner {et~al.}(2006)Gardner, {Mather}, {Clampin}, {Doyon}, {Greenhouse}, {Hammel}, {Hutchings}, {Jakobsen}, {Lilly}, {Long}, {Lunine}, {McCaughrean}, {Mountain}, {Nella}, {Rieke}, {Rieke}, {Rix}, {Smith}, {Sonneborn}, {Stiavelli}, {Stockman}, {Windhorst}, \& {Wright}}]{JWST_06}
Gardner, J.~P., {Mather}, J.~C., {Clampin}, M., {et~al.} 2006, \ssr, 123, 485, \dodoi{10.1007/s11214-006-8315-7}

\bibitem[{Gardner {et~al.}(2023)Gardner, Mather, Abbott, Abell, Abernathy, Abney, Abraham, Abraham, Abul-Huda, Acton, Adams, Adams, Adler, Adriaensen, Aguilar, Ahmed, Ahmed, Ahmed, Albat, Albert, Alberts, Aldridge, Allen, Allen, Altenburg, Altunc, Alvarez, Álvarez Márquez, de~Oliveira, Ambrose, Anandakrishnan, Andersen, Anderson, Anderson, Anderson, Anderson, Aprea, Archer, Arenberg, Argyriou, Arribas, Artigau, Arvai, Atcheson, Atkinson, Averbukh, Aymergen, Bacinski, Baggett, Bagnasco, Baker, Balzano, Banks, Baran, Barker, Barrett, Barringer, Barto, Bast, Baudoz, Baum, Beatty, Beaulieu, Bechtold, Beck, Beddard, Beichman, Bellagama, Bely, Berger, Bergeron, Bernier, Bertch, Beskow, Betz, Biagetti, Birkmann, Bjorklund, Blackwood, Blazek, Blossfeld, Bluth, Boccaletti, Boegner~Jr, Bohlin, Boia, Böker, Bonaventura, Bond, Bosley, Boucarut, Bouchet, Bouwman, Bower, Bowers, Bowers, Boyce, Boyer, Boyer, Boyer, Boyer, Bradley, Brady, Brandl, Brannen, Breda, Bremmer, Brennan, Bresnahan, Bright, Broiles,
  Bromenschenkel, Brooks, Brooks, Brown, Brown, Brown, Bruce, Bryson, Bujanda, Bullock, Bunker, Bureo, Burt, Bush, Bushouse, Bussman, Cabaud, Cale, Calhoon, Calvani, Canipe, Caputo, Cara, Carey, Case, Cesari, Cetorelli, Chance, Chandler, Chaney, Chapman, Charlot, Chayer, Cheezum, Chen, Chen, Cherinka, Chichester, Chilton, Chittiraibalan, Clampin, Clark, Clark, Clark, Claybrooks, Cleveland, Cohen, Cohen, Colón, Coleman, Colina, Comber, Comeau, Comer, Reis, Connolly, Conroy, Contos, Contreras, Cook, Cooper, Cooper, Correia, Correnti, Cossou, Costanza, Coulais, Cox, Coyle, Cracraft, Crew, Curtis, Cusveller, Maciel, Dailey, Daugeron, Davidson, Davies, Davis, Davis, Day, de~Chambure, de~Jong, De~Marchi, Dean, Decker, Delisa, Dell, Dellagatta, Dembinska, Demosthenes, Dencheva, Deneu, DePriest, Deschenes, Dethienne, Detre, Diaz, Dicken, DiFelice, Dillman, Disharoon, Dixon, Doggett, Dominguez, Donaldson, Doria-Warner, Santos, Doty, Douglas, Doyon, Dressler, Driggers, Driggers, Dunn, DuPrie, Dupuis, Durning, Dutta,
  Earl, Eccleston, Ecobichon, Egami, Ehrenwinkler, Eisenhamer, Eisenhower, Eisenstein, El~Hamel, Elie, Elliott, Elliott, Engesser, Espinoza, Etienne, Etxaluze, Evans, Fabreguettes, Falcolini, Falini, Fatig, Feeney, Feinberg, Fels, Ferdous, Ferguson, Ferrarese, Ferreira, Ferruit, Ferry, Filippazzo, Firre, Fix, Flagey, Flanagan, Fleming, Florian, Flynn, Foiadelli, Fontaine, Fontanella, Forshay, Fortner, Fox, Framarini, Francisco, Franck, Franx, Franz, Friedman, Friend, Frost, Fu, Fullerton, Gaillard, Galkin, Gallagher, Galyer, García~Marín, Gardner, Garland, Garrett, Gasman, Gáspár, Gastaud, Gaudreau, Gauthier, Geers, Geithner, Gennaro, Gerber, Gereau, Giampaoli, Giardino, Gibbons, Gilbert, Gilman, Girard, Giuliano, Gkountis, Glasse, Glassmire, Glauser, Glazer, Goldberg, Golimowski, Gonzaga, Gordon, Gordon, Goudfrooij, Gough, Graham, Grau, Green, Greene, Greene, Greenfield, Greenhouse, Greve, Greville, Grimaldi, Groe, Groebner, Grumm, Grundy, Güdel, Guillard, Guldalian, Gunn, Gurule, Gutman, Guy, Guyot,
  Hack, Haderlein, Hagan, Hagedorn, Hainline, Haley, Hami, Hamilton, Hammann, Hammel, Hanley, Hansen, Hardy, Harnisch, Harr, Harris, Hart, Hartig, Hasan, Hashim, Hashimoto, Haskins, Hawkins, Hayden, Hayden, Healy, Hecht, Heeg, Hejal, Helm, Hengemihle, Henning, Henry, Henry, Henshaw, Hernandez, Herrington, Heske, Hesman, Hickey, Hilbert, Hines, Hinz, Hirsch, Hitcho, Hodapp, Hodge, Hoffman, Holfeltz, Holler, Hoppa, Horner, Howard, Howard, Huber, Hunkeler, Hunter, Hunter, Hurd, Hurst, Hutchings, Hylan, Ignat, Illingworth, Irish, Isaacs~III, Jackson~Jr, Jaffe, Jahic, Jahromi, Jakobsen, James, James, James, Jamieson, Jandra, Jayawardhana, Jedrzejewski, Jeffers, Jensen, Joanne, Johns, Johnson, Johnson, Johnson, Johnson, Johnson, Johnson, Johnstone, Jollet, Jones, Jones, Jones, Jones, Jones, Jordan, Jordan, Jue, Jurkowski, Justis, Justtanont, Kaleida, Kalirai, Kalmanson, Kaltenegger, Kammerer, Kan, Kanarek, Kao, Karakla, Karl, Kassin, Kauffman, Kavanagh, Kelley, Kelly, Kendrew, Kennedy, Kenny, Keski-Kuha, Keyes,
  Khan, Kidwell, Kimble, King, King, Kinzel, Kirk, Kirkpatrick, Klaassen, Klingemann, Klintworth, Knapp, Knight, Knollenberg, Knutsen, Koehler, Koekemoer, Kofler, Kontson, Kovacs, Kozhurina-Platais, Krause, Kriss, Krist, Kristoffersen, Krogel, Krueger, Kulp, Kumari, Kwan, Kyprianou, Labador, Labiano, Lafrenière, Lagage, Laidler, Laine, Laird, Lajoie, Lallo, Lam, LaMassa, Lambros, Lampenfield, Lander, Langston, Larson, Larson, LaVerghetta, Law, Lawrence, Lee, Lee, Lee, Leisenring, Leveille, Levenson, Levi, Levine, Lewis, Lewis, Lewis, Libralato, Lidon, Liebrecht, Lightsey, Lilly, Lim, Lim, Ling, Link, Link, Lipinski, Liu, Lo, Lobmeyer, Logue, Long, Long, Long, Long, López-Caniego, Lotz, Love-Pruitt, Lubskiy, Luers, Luetgens, Luevano, G.~Flores~Lui, Lund~III, Lundquist, Lunine, Lützgendorf, Lynch, MacDonald, MacDonald, Macias, Macklis, Maghami, Maharaja, Maiolino, Makrygiannis, Malla, Malumuth, Manjavacas, Marini, Marrione, Marston, Martel, Martin, Martin, Martinez, Maschmann, Masci, Masetti, Maszkiewicz,
  Matthews, Matuskey, McBrayer, McCarthy, McCaughrean, McClare, McClare, McCloskey, McClurg, McCoy, McElwain, McGregor, McGuffey, McKay, McKenzie, McLean, McMaster, McNeil, De~Meester, Mehalick, Meixner, Meléndez, Menzel, Menzel, Merz, Mesterharm, Meyer, Meyett, Meza, Midwinter, Milam, Miller, Miller, Miskey, Misselt, Mitchell, Mohan, Montoya, Moran, Morishita, Moro-Martín, Morrison, Morrison, Morse, Moschos, Moseley, Mosier, Mosner, Mountain, Muckenthaler, Mueller, Mueller, Muhiem, Mühlmann, Mullally, Mullen, Munger, Murphy, Murray, Muzerolle, Mycroft, Myers, Myers, R.~Myers, Myers, Myrick, Nagle, Nayak, Naylor, Neff, Nelan, Nella, Nguyen, Nguyen, Nickson, Nidhiry, Niedner, Nieto-Santisteban, Nikolov, Nishisaka, Noriega-Crespo, Nota, O’Mara, Oboryshko, O’Brien, Ochs, Offenberg, Ogle, Ohl, Olmsted, Osborne, O’Shaughnessy, Östlin, O’Sullivan, Otor, Ottens, Ouellette, Outlaw, Owens, Pacifici, Page, Paranilam, Park, Parrish, Paschal, Patapis, Patel, Patrick, Pattishall~Jr, Paul, Paul, Pauly,
  Pavlovsky, Peña-Guerrero, Pedder, Peek, Pelham, Penanen, Perriello, Perrin, Perrine, Perrygo, Peslier, Petach, Peterson, Pfarr, Pierson, Pietraszkiewicz, Pilchen, Pipher, Pirzkal, Pitman, Player, Plesha, Plitzke, Pohner, Poletis, Pollizzi, Polster, Pontius, Pontoppidan, Porges, Potter, Prescott, Proffitt, Pueyo, Quispe~Neira, Radich, Rager, Rameau, Ramey, Alarcon, Rampini, Rapp, Rashford, Rauscher, Ravindranath, Rawle, Rawlings, Ray, Regan, Rehm, Rehm, Reid, Reis, Renk, Reoch, Ressler, Rest, Reynolds, Richon, Richon, Ridgaway, Riedel, Rieke, Rieke, Rifelli, Rigby, Riggs, Ringel, Ritchie, Rix, Robberto, Robinson, Robinson, Robinson, Rock, Rodriguez, del Pino, Roellig, Rohrbach, Roman, Romelfanger, Romo~Jr, Rosales, Rose, Roteliuk, Roth, Rothwell, Rouzaud, Rowe, Rowlands, Roy, Royer, Rui, Rumler, Rumpl, Russ, Ryan, Ryan, Saad, Sabata, Sabatino, Sabbi, Sabelhaus, Sabia, Sahu, Saif, Salvignol, Samara-Ratna, Samuelson, Sanders, Sappington, Sargent, Sauer, Savadkin, Sawicki, Schappell, Scheffer, Scheithauer,
  Scherer, Schiff, Schlawin, Schmeitzky, Schmitz, Schmude, Schneider, Schreiber, Schroeven-Deceuninck, Schultz, Schwab, Schwartz, Scoccimarro, Scott, Scott, Seaton, Seely, Seery, Seidleck, Sembach, Shanahan, Shaughnessy, Shaw, Shay, Sheehan, Sheth, Shih, Shivaei, Siegel, Sienkiewicz, Simmons, Simon, Sirianni, Sivaramakrishnan, Slade, Sloan, Slocum, Slowinski, Smith, Smith, Smith, Smith, Smith, Smith, Smolik, Soderblom, Sohn, Sokol, Sonneborn, Sontag, Sooy, Soummer, Southwood, Spain, Sparmo, Speer, Spencer, Sprofera, Stallcup, Stanley, Stansberry, Stark, Starr, Stassi, Steck, Steeley, Stephens, Stephenson, Stewart, Stiavelli, Jr, Strada, Straughn, Streetman, Strickland, Strobele, Stuhlinger, Stys, Such, Sukhatme, Sullivan, Sullivan, Sumner, Sun, Sunnquist, Swade, Swam, Swenton, Swoish, Tam~Litten, Tamas, Tao, Taylor, Taylor, Plate, Van~Tea, Teague, Telfer, Temim, Texter, Thatte, Thompson, Thompson, Thomson, Thronson, Tierney, Tikkanen, Tinnin, Tippet, Todd, Tran, Trauger, Trejo, Vinh~Truong, Tsukamoto, Tufail,
  Tumlinson, Tustain, Tyra, Ubeda, Underwood, Uzzo, Vaclavik, Valenduc, Valenti, Van~Campen, van~de Wetering, Van Der~Marel, van Haarlem, Vandenbussche, van Dishoeck, Vanterpool, Vernoy, Vila~Costas, Volk, Voorzaat, Voyton, Vydra, Waddy, Waelkens, Wahlgren, Walker~Jr, Wander, Warfield, Warner, Wasiak, Wasiak, Wehner, Weiler, Weilert, Weiss, Wells, Welty, Wheate, Wheeler, White, Whitehouse, Whiteleather, Whitman, Williams, Willmer, Willott, Willoughby, Wilson, Wilson, Wilson, Windhorst, Wislowski, Wolfe, Wolfe, Wolff, Wondel, Woo, Woods, Worden, Workman, Wright, Wu, Wu, Wun, Wymer, Yadetie, Yan, Yang, Yates, Yeager, Yerger, Young, Young, Yu, Yu, Zak, Zeidler, Zepp, Zhou, Zincke, Zonak, \& Zondag}]{JWST_23}
Gardner, J.~P., Mather, J.~C., Abbott, R., {et~al.} 2023, Publications of the Astronomical Society of the Pacific, 135, 068001, \dodoi{10.1088/1538-3873/acd1b5}

\bibitem[{{Garn} \& {Best}(2010)}]{Garn2010}
{Garn}, T., \& {Best}, P.~N. 2010, \mnras, 409, 421, \dodoi{10.1111/j.1365-2966.2010.17321.x}

\bibitem[{Gehrz(1989)}]{Gehrz89}
Gehrz, R.~D. 1989, in Symposium-International astronomical union, Vol. 135, Cambridge University Press, 445--453

\bibitem[{{Gibson} {et~al.}(2022){Gibson}, {Lehner}, {Oppenheimer}, {Howk}, {Cooksey}, \& {Fox}}]{Gibson22}
{Gibson}, J.~L., {Lehner}, N., {Oppenheimer}, B.~D., {et~al.} 2022, \aj, 164, 9, \dodoi{10.3847/1538-3881/ac69d0}

\bibitem[{{Goodman} \& {Weare}(2010)}]{Goodman10}
{Goodman}, J., \& {Weare}, J. 2010, Communications in Applied Mathematics and Computational Science, 5, 65, \dodoi{10.2140/camcos.2010.5.65}

\bibitem[{{Hahn} {et~al.}(2022){Hahn}, {Starkenburg}, {Angl{\'e}s-Alc{\'a}zar}, {Choi}, {Dav{\'e}}, {Dickey}, {Iyer}, {Maller}, {Somerville}, {Tinker}, \& {Yung}}]{Hahn2022}
{Hahn}, C., {Starkenburg}, T.~K., {Angl{\'e}s-Alc{\'a}zar}, D., {et~al.} 2022, \apj, 926, 122, \dodoi{10.3847/1538-4357/ac4253}

\bibitem[{{Hamed} {et~al.}(2023){Hamed}, {Ma{\l}ek}, {Buat}, {Junais}, {Ciesla}, {Donevski}, {Riccio}, \& {Figueira}}]{Hamed2023}
{Hamed}, M., {Ma{\l}ek}, K., {Buat}, V., {et~al.} 2023, \aap, 674, A99, \dodoi{10.1051/0004-6361/202245818}

\bibitem[{{Hirashita} \& {Murga}(2020)}]{Hirashita20}
{Hirashita}, H., \& {Murga}, M.~S. 2020, \mnras, 492, 3779, \dodoi{10.1093/mnras/stz3640}

\bibitem[{Hu {et~al.}(2023)Hu, Sternberg, \& van Dishoeck}]{Hu_2023}
Hu, C.-Y., Sternberg, A., \& van Dishoeck, E.~F. 2023, The Astrophysical Journal, 952, 140, \dodoi{10.3847/1538-4357/acdcfa}

\bibitem[{{Hu} {et~al.}(2019){Hu}, {Zhukovska}, {Somerville}, \& {Naab}}]{Hu_2019}
{Hu}, C.-Y., {Zhukovska}, S., {Somerville}, R.~S., \& {Naab}, T. 2019, \mnras, 487, 3252, \dodoi{10.1093/mnras/stz1481}

\bibitem[{Hunter(2007)}]{matplotlib}
Hunter, J.~D. 2007, Computing in Science Engineering, 9, 90, \dodoi{10.1109/MCSE.2007.55}

\bibitem[{{Inami} {et~al.}(2022){Inami}, {Algera}, {Schouws}, {Sommovigo}, {Bouwens}, {Smit}, {Stefanon}, {Bowler}, {Endsley}, {Ferrara}, {Oesch}, {Stark}, {Aravena}, {Barrufet}, {da Cunha}, {Dayal}, {De Looze}, {Fudamoto}, {Gonzalez}, {Graziani}, {Hodge}, {Hygate}, {Nanayakkara}, {Pallottini}, {Riechers}, {Schneider}, {Topping}, \& {van der Werf}}]{Inami22}
{Inami}, H., {Algera}, H. S.~B., {Schouws}, S., {et~al.} 2022, \mnras, 515, 3126, \dodoi{10.1093/mnras/stac1779}

\bibitem[{{Inoue}(2005)}]{Inoue05}
{Inoue}, A.~K. 2005, \mnras, 359, 171, \dodoi{10.1111/j.1365-2966.2005.08890.x}

\bibitem[{Iyer \& Gawiser(2017)}]{Iyer17}
Iyer, K., \& Gawiser, E. 2017, The Astrophysical Journal, 838, 127, \dodoi{10.3847/1538-4357/aa63f0}

\bibitem[{{Jones} {et~al.}(1996){Jones}, {Tielens}, \& {Hollenbach}}]{Jones1996ApJ}
{Jones}, A.~P., {Tielens}, A.~G.~G.~M., \& {Hollenbach}, D.~J. 1996, \apj, 469, 740, \dodoi{10.1086/177823}

\bibitem[{{Jones} {et~al.}(2022){Jones}, {Stanway}, \& {Carnall}}]{Jones2022}
{Jones}, G.~T., {Stanway}, E.~R., \& {Carnall}, A.~C. 2022, \mnras, 514, 5706, \dodoi{10.1093/mnras/stac1667}

\bibitem[{{Jonsson} {et~al.}(2006){Jonsson}, {Cox}, {Primack}, \& {Somerville}}]{Jonsson06}
{Jonsson}, P., {Cox}, T.~J., {Primack}, J.~R., \& {Somerville}, R.~S. 2006, \apj, 637, 255, \dodoi{10.1086/497567}

\bibitem[{{Kennicutt}(1998)}]{Kennicutt98}
{Kennicutt}, Robert~C., J. 1998, \apj, 498, 541, \dodoi{10.1086/305588}

\bibitem[{{Koornneef} \& {Code}(1981{\natexlab{a}})}]{Koornneef81}
{Koornneef}, J., \& {Code}, A.~D. 1981{\natexlab{a}}, \apj, 247, 860, \dodoi{10.1086/159096}

\bibitem[{{Koornneef} \& {Code}(1981{\natexlab{b}})}]{1981ApJ...247..860K}
---. 1981{\natexlab{b}}, \apj, 247, 860, \dodoi{10.1086/159096}

\bibitem[{{Lacey} {et~al.}(2011){Lacey}, {Baugh}, {Frenk}, \& {Benson}}]{Lacey2011}
{Lacey}, C.~G., {Baugh}, C.~M., {Frenk}, C.~S., \& {Benson}, A.~J. 2011, \mnras, 412, 1828, \dodoi{10.1111/j.1365-2966.2010.18021.x}

\bibitem[{{Lacey} {et~al.}(2016){Lacey}, {Baugh}, {Frenk}, {Benson}, {Bower}, {Cole}, {Gonzalez-Perez}, {Helly}, {Lagos}, \& {Mitchell}}]{Lacey2016}
{Lacey}, C.~G., {Baugh}, C.~M., {Frenk}, C.~S., {et~al.} 2016, \mnras, 462, 3854, \dodoi{10.1093/mnras/stw1888}

\bibitem[{Lara-López {et~al.}(2013)Lara-López, Hopkins, López-Sánchez, Brough, Gunawardhana, Colless, Robotham, Bauer, Bland-Hawthorn, Cluver, Driver, Foster, Kelvin, Liske, Loveday, Owers, Ponman, Sharp, Steele, Taylor, \& Thomas}]{Lopez13}
Lara-López, M.~A., Hopkins, A.~M., López-Sánchez, A.~R., {et~al.} 2013, Monthly Notices of the Royal Astronomical Society, 434, 451, \dodoi{10.1093/mnras/stt1031}

\bibitem[{{Leja} {et~al.}(2019){Leja}, {Carnall}, {Johnson}, {Conroy}, \& {Speagle}}]{Leja2019}
{Leja}, J., {Carnall}, A.~C., {Johnson}, B.~D., {Conroy}, C., \& {Speagle}, J.~S. 2019, \apj, 876, 3, \dodoi{10.3847/1538-4357/ab133c}

\bibitem[{{Lewis} {et~al.}(2023){Lewis}, {Ocvirk}, {Dubois}, {Aubert}, {Chardin}, {Gillet}, \& {Th{\'e}lie}}]{2023MNRAS.519.5987L}
{Lewis}, J. S.~W., {Ocvirk}, P., {Dubois}, Y., {et~al.} 2023, \mnras, 519, 5987, \dodoi{10.1093/mnras/stad081}

\bibitem[{{Li} \& {Greenberg}(2003)}]{LiGreenberg03dust}
{Li}, A., \& {Greenberg}, J.~M. 2003, in Solid State Astrochemistry, ed. V.~{Pirronello}, J.~{Krelowski}, \& G.~{Manic{\`o}}, Vol. 120, 37--84, \dodoi{10.48550/arXiv.astro-ph/0204392}

\bibitem[{{Li} {et~al.}(2008){Li}, {Liang}, {Kann}, {Wei}, {Klose}, \& {Wang}}]{Li08}
{Li}, A., {Liang}, S.~L., {Kann}, D.~A., {et~al.} 2008, \apj, 685, 1046, \dodoi{10.1086/591049}

\bibitem[{Lin {et~al.}(2025)Lin, Yang, Li, \& Witstok}]{Lin25}
Lin, Q., Yang, X., Li, A., \& Witstok, J. 2025, \dodoi{10.48550/arXiv.2502.08113}

\bibitem[{{Lin} {et~al.}(2021){Lin}, {Hirashita}, {Camps}, \& {Baes}}]{Lin21}
{Lin}, Y.-H., {Hirashita}, H., {Camps}, P., \& {Baes}, M. 2021, \mnras, 507, 2755, \dodoi{10.1093/mnras/stab2242}

\bibitem[{{Lorenz} {et~al.}(2023){Lorenz}, {Kriek}, {Shapley}, {Reddy}, {Sanders}, {Barro}, {Coil}, {Mobasher}, {Price}, {Runco}, {Shivaei}, {Siana}, \& {Weisz}}]{Lorenz2023}
{Lorenz}, B., {Kriek}, M., {Shapley}, A.~E., {et~al.} 2023, \apj, 951, 29, \dodoi{10.3847/1538-4357/accdd1}

\bibitem[{{Lovell} {et~al.}(2024){Lovell}, {Starkenburg}, {Ho}, {Angl{\'e}s-Alc{\'a}zar}, {Dav{\'e}}, {Gabrielpillai}, {Iyer}, {Matthews}, {Roper}, {Somerville}, {Sommovigo}, \& {Villaescusa-Navarro}}]{Lovell24}
{Lovell}, C.~C., {Starkenburg}, T., {Ho}, M., {et~al.} 2024, arXiv e-prints, arXiv:2411.13960, \dodoi{10.48550/arXiv.2411.13960}

\bibitem[{{Lower} {et~al.}(2022){Lower}, {Narayanan}, {Leja}, {Johnson}, {Conroy}, \& {Dav{\'e}}}]{Lower22}
{Lower}, S., {Narayanan}, D., {Leja}, J., {et~al.} 2022, \apj, 931, 14, \dodoi{10.3847/1538-4357/ac6959}

\bibitem[{{Maraston} {et~al.}(2001){Maraston}, {Greggio}, \& {Thomas}}]{Maraston01}
{Maraston}, C., {Greggio}, L., \& {Thomas}, D. 2001, \apss, 276, 893, \dodoi{10.1023/A:1017577207926}

\bibitem[{{Markov} {et~al.}(2024){Markov}, {Gallerani}, {Ferrara}, {Pallottini}, {Parlanti}, {Di Mascia}, {Sommovigo}, \& {Kohandel}}]{Markov24}
{Markov}, V., {Gallerani}, S., {Ferrara}, A., {et~al.} 2024, arXiv e-prints, arXiv:2402.05996, \dodoi{10.48550/arXiv.2402.05996}

\bibitem[{{Markov} {et~al.}(2023){Markov}, {Gallerani}, {Pallottini}, {Sommovigo}, {Carniani}, {Ferrara}, {Parlanti}, \& {Di Mascia}}]{Markov23}
{Markov}, V., {Gallerani}, S., {Pallottini}, A., {et~al.} 2023, \aap, 679, A12, \dodoi{10.1051/0004-6361/202346723}

\bibitem[{Mauerhofer \& Dayal(2023)}]{Mauerhofer23}
Mauerhofer, V., \& Dayal, P. 2023, Monthly Notices of the Royal Astronomical Society, 526, 2196, \dodoi{10.1093/mnras/stad2734}

\bibitem[{Meert {et~al.}(2014)Meert, Vikram, \& Bernardi}]{Meert15}
Meert, A., Vikram, V., \& Bernardi, M. 2014, Monthly Notices of the Royal Astronomical Society, 446, 3943, \dodoi{10.1093/mnras/stu2333}

\bibitem[{{Meldorf} {et~al.}(2024){Meldorf}, {Palmese}, \& {Salim}}]{Meldorf2024}
{Meldorf}, C., {Palmese}, A., \& {Salim}, S. 2024, \mnras, 531, 3242, \dodoi{10.1093/mnras/stae1373}

\bibitem[{{Meurer} {et~al.}(1999){Meurer}, {Heckman}, \& {Calzetti}}]{Meurer99}
{Meurer}, G.~R., {Heckman}, T.~M., \& {Calzetti}, D. 1999, \apj, 521, 64, \dodoi{10.1086/307523}

\bibitem[{{Mushtaq} {et~al.}(2023){Mushtaq}, {Ceverino}, {Klessen}, {Reissl}, \& {Puttasiddappa}}]{Mushtaq2023}
{Mushtaq}, M., {Ceverino}, D., {Klessen}, R.~S., {Reissl}, S., \& {Puttasiddappa}, P.~H. 2023, \mnras, 525, 4976, \dodoi{10.1093/mnras/stad2602}

\bibitem[{{Nagaraj} {et~al.}(2022){Nagaraj}, {Forbes}, {Leja}, {Foreman-Mackey}, \& {Hayward}}]{DustE22}
{Nagaraj}, G., {Forbes}, J.~C., {Leja}, J., {Foreman-Mackey}, D., \& {Hayward}, C.~C. 2022, \apj, 932, 54, \dodoi{10.3847/1538-4357/ac6c80}

\bibitem[{{Nakajima} {et~al.}(2022){Nakajima}, {Ouchi}, {Xu}, {Rauch}, {Harikane}, {Nishigaki}, {Isobe}, {Kusakabe}, {Nagao}, {Ono}, {Onodera}, {Sugahara}, {Kim}, {Komiyama}, {Lee}, \& {Zahedy}}]{Nakajima22}
{Nakajima}, K., {Ouchi}, M., {Xu}, Y., {et~al.} 2022, \apjs, 262, 3, \dodoi{10.3847/1538-4365/ac7710}

\bibitem[{{Nandy} {et~al.}(1981{\natexlab{a}}){Nandy}, {Morgan}, {Willis}, {Wilson}, \& {Gondhalekar}}]{Nandy81}
{Nandy}, K., {Morgan}, D.~H., {Willis}, A.~J., {Wilson}, R., \& {Gondhalekar}, P.~M. 1981{\natexlab{a}}, \mnras, 196, 955, \dodoi{10.1093/mnras/196.4.955}

\bibitem[{{Nandy} {et~al.}(1981{\natexlab{b}}){Nandy}, {Morgan}, {Willis}, {Wilson}, \& {Gondhalekar}}]{1981MNRAS.196..955N}
---. 1981{\natexlab{b}}, \mnras, 196, 955, \dodoi{10.1093/mnras/196.4.955}

\bibitem[{{Narayanan} {et~al.}(2018){Narayanan}, {Conroy}, {Dav{\'e}}, {Johnson}, \& {Popping}}]{Narayanan18}
{Narayanan}, D., {Conroy}, C., {Dav{\'e}}, R., {Johnson}, B.~D., \& {Popping}, G. 2018, \apj, 869, 70, \dodoi{10.3847/1538-4357/aaed25}

\bibitem[{{Narayanan} {et~al.}(2021){Narayanan}, {Turk}, {Robitaille}, {Kelly}, {McClellan}, {Sharma}, {Garg}, {Abruzzo}, {Choi}, {Conroy}, {Johnson}, {Kimock}, {Li}, {Lovell}, {Lower}, {Privon}, {Roberts}, {Sethuram}, {Snyder}, {Thompson}, \& {Wise}}]{Narayanan21}
{Narayanan}, D., {Turk}, M.~J., {Robitaille}, T., {et~al.} 2021, \apjs, 252, 12, \dodoi{10.3847/1538-4365/abc487}

\bibitem[{Narayanan {et~al.}(2024)Narayanan, Lower, Torrey, Brammer, Cui, Davé, Iyer, Li, Lovell, Sales, Stark, Marinacci, \& Vogelsberger}]{Narayanan_2024}
Narayanan, D., Lower, S., Torrey, P., {et~al.} 2024, The Astrophysical Journal, 961, 73, \dodoi{10.3847/1538-4357/ad0966}

\bibitem[{{Nelson} {et~al.}(2018){Nelson}, {Pillepich}, {Springel}, {Weinberger}, {Hernquist}, {Pakmor}, {Genel}, {Torrey}, {Vogelsberger}, {Kauffmann}, {Marinacci}, \& {Naiman}}]{Nelson19}
{Nelson}, D., {Pillepich}, A., {Springel}, V., {et~al.} 2018, \mnras, 475, 624, \dodoi{10.1093/mnras/stx3040}

\bibitem[{{Newman} {et~al.}(2025){Newman}, {Lovell}, {Maraston}, {Giavalisco}, {Roper}, {Saxena}, {Vijayan}, \& {Wilkins}}]{Newman25}
{Newman}, S.~L., {Lovell}, C.~C., {Maraston}, C., {et~al.} 2025, arXiv e-prints, arXiv:2501.03133, \dodoi{10.48550/arXiv.2501.03133}

\bibitem[{{Noll} {et~al.}(2009){Noll}, {Burgarella}, {Giovannoli}, {Buat}, {Marcillac}, \& {Mu{\~n}oz-Mateos}}]{Noll2009}
{Noll}, S., {Burgarella}, D., {Giovannoli}, E., {et~al.} 2009, \aap, 507, 1793, \dodoi{10.1051/0004-6361/200912497}

\bibitem[{{Osborne} \& {Salim}(2024)}]{Osborne2024}
{Osborne}, C., \& {Salim}, S. 2024, \apj, 962, 59, \dodoi{10.3847/1538-4357/ad17c8}

\bibitem[{Pacifici {et~al.}(2023)Pacifici, Iyer, Mobasher, da~Cunha, Acquaviva, Burgarella, Calistro~Rivera, Carnall, Chang, Chartab, Cooke, Fairhurst, Kartaltepe, Leja, Małek, Salmon, Torelli, Vidal-García, Boquien, Brammer, Brown, Capak, Chevallard, Circosta, Croton, Davidzon, Dickinson, Duncan, Faber, Ferguson, Fontana, Guo, Haeussler, Hemmati, Jafariyazani, Kassin, Larson, Lee, Mantha, Marchi, Nayyeri, Newman, Pandya, Pforr, Reddy, Sanders, Shah, Shahidi, Stevans, Triani, Tyler, Vanderhoof, de~la Vega, Wang, \& Weston}]{Pacifici_2023}
Pacifici, C., Iyer, K.~G., Mobasher, B., {et~al.} 2023, The Astrophysical Journal, 944, 141, \dodoi{10.3847/1538-4357/acacff}

\bibitem[{{Pallottini} {et~al.}(2022){Pallottini}, {Ferrara}, {Gallerani}, {Behrens}, {Kohandel}, {Carniani}, {Vallini}, {Salvadori}, {Gelli}, {Sommovigo}, {D'Odorico}, {Di Mascia}, \& {Pizzati}}]{Pallottini22}
{Pallottini}, A., {Ferrara}, A., {Gallerani}, S., {et~al.} 2022, \mnras, 513, 5621, \dodoi{10.1093/mnras/stac1281}

\bibitem[{{Parsotan} {et~al.}(2021){Parsotan}, {Cochrane}, {Hayward}, {Angl{\'e}s-Alc{\'a}zar}, {Feldmann}, {Faucher-Gigu{\`e}re}, {Wellons}, \& {Hopkins}}]{Parsotan2021}
{Parsotan}, T., {Cochrane}, R.~K., {Hayward}, C.~C., {et~al.} 2021, \mnras, 501, 1591, \dodoi{10.1093/mnras/staa3765}

\bibitem[{{P{\'e}roux} \& {Howk}(2020)}]{Peroux20}
{P{\'e}roux}, C., \& {Howk}, J.~C. 2020, \araa, 58, 363, \dodoi{10.1146/annurev-astro-021820-120014}

\bibitem[{{Pillepich} {et~al.}(2018){Pillepich}, {Nelson}, {Hernquist}, {Springel}, {Pakmor}, {Torrey}, {Weinberger}, {Genel}, {Naiman}, {Marinacci}, \& {Vogelsberger}}]{Pillepich18}
{Pillepich}, A., {Nelson}, D., {Hernquist}, L., {et~al.} 2018, \mnras, 475, 648, \dodoi{10.1093/mnras/stx3112}

\bibitem[{{Pillepich} {et~al.}(2019){Pillepich}, {Nelson}, {Springel}, {Pakmor}, {Torrey}, {Weinberger}, {Vogelsberger}, {Marinacci}, {Genel}, {van der Wel}, \& {Hernquist}}]{Pillepich19}
{Pillepich}, A., {Nelson}, D., {Springel}, V., {et~al.} 2019, \mnras, 490, 3196, \dodoi{10.1093/mnras/stz2338}

\bibitem[{{Poetrodjojo} {et~al.}(2021){Poetrodjojo}, {Groves}, {Kewley}, {Sweet}, {Sanchez}, {Medling}, {L{\'o}pez-S{\'a}nchez}, {Brough}, {Cortese}, {van de Sande}, {Vaughan}, {Richards}, {Bryant}, {Croom}, {Bland-Hawthorn}, {Goodwin}, {Lawrence}, {Owers}, \& {Scott}}]{Poetrodjojo21}
{Poetrodjojo}, H., {Groves}, B., {Kewley}, L.~J., {et~al.} 2021, \mnras, 502, 3357, \dodoi{10.1093/mnras/stab205}

\bibitem[{Popping \& Péroux(2022)}]{Popping22}
Popping, G., \& Péroux, C. 2022, Monthly Notices of the Royal Astronomical Society, 513, 1531, \dodoi{10.1093/mnras/stac695}

\bibitem[{{Popping} {et~al.}(2017){Popping}, {Somerville}, \& {Galametz}}]{Popping17}
{Popping}, G., {Somerville}, R.~S., \& {Galametz}, M. 2017, \mnras, 471, 3152, \dodoi{10.1093/mnras/stx1545}

\bibitem[{{Popping} {et~al.}(2022){Popping}, {Pillepich}, {Calistro Rivera}, {Schulz}, {Hernquist}, {Kaasinen}, {Marinacci}, {Nelson}, \& {Vogelsberger}}]{Popping2022_tng}
{Popping}, G., {Pillepich}, A., {Calistro Rivera}, G., {et~al.} 2022, \mnras, 510, 3321, \dodoi{10.1093/mnras/stab3312}

\bibitem[{Prevot {et~al.}(1984)Prevot, Lequeux, Maurice, Pr{\'e}vot, \& Rocca-Volmerange}]{Prevot84}
Prevot, M., Lequeux, J., Maurice, E., Pr{\'e}vot, L., \& Rocca-Volmerange, B. 1984, Astronomy and Astrophysics, 132, 389

\bibitem[{{Qin} {et~al.}(2022){Qin}, {Zheng}, {Fang}, {Pan}, {Wuyts}, {Shi}, {Peng}, {Gonzalez}, {Bian}, {Huang}, {Gu}, {Liu}, {Tan}, {Shi}, {Ren}, {Zhang}, {Qiao}, {Wen}, \& {Liu}}]{Qin2022}
{Qin}, J., {Zheng}, X.~Z., {Fang}, M., {et~al.} 2022, \mnras, 511, 765, \dodoi{10.1093/mnras/stac132}

\bibitem[{Reddy {et~al.}(2015)Reddy, Kriek, Shapley, Freeman, Siana, Coil, Mobasher, Price, Sanders, \& Shivaei}]{Reddy15}
Reddy, N.~A., Kriek, M., Shapley, A.~E., {et~al.} 2015, The Astrophysical Journal, 806, 259, \dodoi{10.1088/0004-637X/806/2/259}

\bibitem[{{Reissl} {et~al.}(2016){Reissl}, {Wolf}, \& {Brauer}}]{2016A&A...593A..87R}
{Reissl}, S., {Wolf}, S., \& {Brauer}, R. 2016, \aap, 593, A87, \dodoi{10.1051/0004-6361/201424930}

\bibitem[{{R{\'e}my-Ruyer} {et~al.}(2014){R{\'e}my-Ruyer}, {Madden}, {Galliano}, {Galametz}, {Takeuchi}, {Asano}, {Zhukovska}, {Lebouteiller}, {Cormier}, {Jones}, {Bocchio}, {Baes}, {Bendo}, {Boquien}, {Boselli}, {DeLooze}, {Doublier-Pritchard}, {Hughes}, {Karczewski}, \& {Spinoglio}}]{RemyRuyer14}
{R{\'e}my-Ruyer}, A., {Madden}, S.~C., {Galliano}, F., {et~al.} 2014, \aap, 563, A31, \dodoi{10.1051/0004-6361/201322803}

\bibitem[{{Salim} {et~al.}(2018){Salim}, {Boquien}, \& {Lee}}]{Salim18}
{Salim}, S., {Boquien}, M., \& {Lee}, J.~C. 2018, \apj, 859, 11, \dodoi{10.3847/1538-4357/aabf3c}

\bibitem[{{Salim} \& {Narayanan}(2020)}]{Salim20}
{Salim}, S., \& {Narayanan}, D. 2020, \araa, 58, 529, \dodoi{10.1146/annurev-astro-032620-021933}

\bibitem[{Salim {et~al.}(2023)Salim, Tacchella, Osborne, Faber, Lee, \& Ellison}]{Salim23}
Salim, S., Tacchella, S., Osborne, C., {et~al.} 2023, The Astrophysical Journal, 958, 183, \dodoi{10.3847/1538-4357/ad04db}

\bibitem[{Salim {et~al.}(2016)Salim, Lee, Janowiecki, Cunha, Dickinson, Boquien, Burgarella, Salzer, \& Charlot}]{Salim16}
Salim, S., Lee, J.~C., Janowiecki, S., {et~al.} 2016, The Astrophysical Journal Supplement Series, 227, 2, \dodoi{10.3847/0067-0049/227/1/2}

\bibitem[{Salmon {et~al.}(2016)Salmon, Papovich, Long, Willner, Finkelstein, Ferguson, Dickinson, Duncan, Faber, Hathi, Koekemoer, Kurczynski, Newman, Pacifici, Pérez-González, \& Pforr}]{Salmon16}
Salmon, B., Papovich, C., Long, J., {et~al.} 2016, The Astrophysical Journal, 827, 20, \dodoi{10.3847/0004-637X/827/1/20}

\bibitem[{{Schneider} \& {Maiolino}(2023)}]{Schneider23}
{Schneider}, R., \& {Maiolino}, R. 2023, arXiv e-prints, arXiv:2310.00053, \dodoi{10.48550/arXiv.2310.00053}

\bibitem[{{Schulz} {et~al.}(2020){Schulz}, {Popping}, {Pillepich}, {Nelson}, {Vogelsberger}, {Marinacci}, \& {Hernquist}}]{Schulz20}
{Schulz}, S., {Popping}, G., {Pillepich}, A., {et~al.} 2020, \mnras, 497, 4773, \dodoi{10.1093/mnras/staa1900}

\bibitem[{Scoville {et~al.}(2015)Scoville, Faisst, Capak, Kakazu, Li, \& Steinhardt}]{Scoville15}
Scoville, N., Faisst, A., Capak, P., {et~al.} 2015, The Astrophysical Journal, 800, 108, \dodoi{10.1088/0004-637X/800/2/108}

\bibitem[{{Seon} \& {Draine}(2016)}]{SeonDraine16}
{Seon}, K.-I., \& {Draine}, B.~T. 2016, \apj, 833, 201, \dodoi{10.3847/1538-4357/833/2/201}

\bibitem[{{Shapley} {et~al.}(2020){Shapley}, {Cullen}, {Dunlop}, {McLure}, {Kriek}, {Reddy}, \& {Sanders}}]{Shapley20}
{Shapley}, A.~E., {Cullen}, F., {Dunlop}, J.~S., {et~al.} 2020, \apjl, 903, L16, \dodoi{10.3847/2041-8213/abc006}

\bibitem[{{Shen} {et~al.}(2020){Shen}, {Vogelsberger}, {Nelson}, {Pillepich}, {Tacchella}, {Marinacci}, {Torrey}, {Hernquist}, \& {Springel}}]{2020MNRAS.495.4747S}
{Shen}, X., {Vogelsberger}, M., {Nelson}, D., {et~al.} 2020, \mnras, 495, 4747, \dodoi{10.1093/mnras/staa1423}

\bibitem[{{Shivaei} {et~al.}(2020{\natexlab{a}}){Shivaei}, {Reddy}, {Rieke}, {Shapley}, {Kriek}, {Battisti}, {Mobasher}, {Sanders}, {Fetherolf}, {Azadi}, {Coil}, {Freeman}, {de Groot}, {Leung}, {Price}, {Siana}, \& {Zick}}]{Shivaei20}
{Shivaei}, I., {Reddy}, N., {Rieke}, G., {et~al.} 2020{\natexlab{a}}, \apj, 899, 117, \dodoi{10.3847/1538-4357/aba35e}

\bibitem[{{Shivaei} {et~al.}(2020{\natexlab{b}}){Shivaei}, {Reddy}, {Rieke}, {Shapley}, {Kriek}, {Battisti}, {Mobasher}, {Sanders}, {Fetherolf}, {Azadi}, {Coil}, {Freeman}, {de Groot}, {Leung}, {Price}, {Siana}, \& {Zick}}]{Shivaei2020}
---. 2020{\natexlab{b}}, \apj, 899, 117, \dodoi{10.3847/1538-4357/aba35e}

\bibitem[{{Shivaei} {et~al.}(2022){Shivaei}, {Popping}, {Rieke}, {Reddy}, {Pope}, {Kennicutt}, {Mobasher}, {Coil}, {Fudamoto}, {Kriek}, {Lyu}, {Oesch}, {Sanders}, {Shapley}, \& {Siana}}]{Shivaei22}
{Shivaei}, I., {Popping}, G., {Rieke}, G., {et~al.} 2022, \apj, 928, 68, \dodoi{10.3847/1538-4357/ac54a9}

\bibitem[{{Soliman} {et~al.}(2024){Soliman}, {Hopkins}, \& {Grudi{\'c}}}]{2024ApJ...974..136S}
{Soliman}, N.~H., {Hopkins}, P.~F., \& {Grudi{\'c}}, M.~Y. 2024, \apj, 974, 136, \dodoi{10.3847/1538-4357/ad6ddd}

\bibitem[{{Somerville} {et~al.}(2012){Somerville}, {Gilmore}, {Primack}, \& {Dom{\'\i}nguez}}]{Somerville2012}
{Somerville}, R.~S., {Gilmore}, R.~C., {Primack}, J.~R., \& {Dom{\'\i}nguez}, A. 2012, \mnras, 423, 1992, \dodoi{10.1111/j.1365-2966.2012.20490.x}

\bibitem[{{Sommovigo} {et~al.}(2020){Sommovigo}, {Ferrara}, {Pallottini}, {Carniani}, {Gallerani}, \& {Decataldo}}]{Sommovigo20}
{Sommovigo}, L., {Ferrara}, A., {Pallottini}, A., {et~al.} 2020, \mnras, 497, 956, \dodoi{10.1093/mnras/staa1959}

\bibitem[{Suess {et~al.}(2022)Suess, Leja, Johnson, Bezanson, Greene, Kriek, Lower, Narayanan, Setton, \& Spilker}]{Suess_2022}
Suess, K.~A., Leja, J., Johnson, B.~D., {et~al.} 2022, The Astrophysical Journal, 935, 146, \dodoi{10.3847/1538-4357/ac82b0}

\bibitem[{{Tielens}(1999)}]{Tielens99}
{Tielens}, A.~G.~G.~M. 1999, in NATO Advanced Study Institute (ASI) Series C, Vol. 523, Formation and Evolution of Solids in Space, ed. J.~M. {Greenberg} \& A.~{Li}, 331.
\newblock \url{https://ui.adsabs.harvard.edu/abs/1999ASIC..523..331T}

\bibitem[{Todini \& Ferrara(2001)}]{Todini00}
Todini, P., \& Ferrara, A. 2001, Monthly Notices of the Royal Astronomical Society, 325, 726, \dodoi{10.1046/j.1365-8711.2001.04486.x}

\bibitem[{{Torrey} {et~al.}(2015){Torrey}, {Snyder}, {Vogelsberger}, {Hayward}, {Genel}, {Sijacki}, {Springel}, {Hernquist}, {Nelson}, {Kriek}, {Pillepich}, {Sales}, \& {McBride}}]{Torrey15}
{Torrey}, P., {Snyder}, G.~F., {Vogelsberger}, M., {et~al.} 2015, \mnras, 447, 2753, \dodoi{10.1093/mnras/stu2592}

\bibitem[{{Torrey} {et~al.}(2019){Torrey}, {Vogelsberger}, {Marinacci}, {Pakmor}, {Springel}, {Nelson}, {Naiman}, {Pillepich}, {Genel}, {Weinberger}, \& {Hernquist}}]{Torrey19}
{Torrey}, P., {Vogelsberger}, M., {Marinacci}, F., {et~al.} 2019, \mnras, 484, 5587, \dodoi{10.1093/mnras/stz243}

\bibitem[{{Trayford} {et~al.}(2020){Trayford}, {Lagos}, {Robotham}, \& {Obreschkow}}]{Trayford20}
{Trayford}, J.~W., {Lagos}, C. d.~P., {Robotham}, A. S.~G., \& {Obreschkow}, D. 2020, \mnras, 491, 3937, \dodoi{10.1093/mnras/stz3234}

\bibitem[{{Trayford} {et~al.}(2015){Trayford}, {Theuns}, {Bower}, {Schaye}, {Furlong}, {Schaller}, {Frenk}, {Crain}, {Dalla Vecchia}, \& {McCarthy}}]{Trayford15}
{Trayford}, J.~W., {Theuns}, T., {Bower}, R.~G., {et~al.} 2015, \mnras, 452, 2879, \dodoi{10.1093/mnras/stv1461}

\bibitem[{Tress {et~al.}(2018)Tress, Mármol-Queraltó, Ferreras, Pérez-González, Barro, Pampliega, Cava, Domínguez-Sánchez, Eliche-Moral, Espino-Briones, Esquej, Hernán-Caballero, Rodighiero, \& Rodriguez-Muñoz}]{Tress18}
Tress, M., Mármol-Queraltó, E., Ferreras, I., {et~al.} 2018, Monthly Notices of the Royal Astronomical Society, 475, 2363, \dodoi{10.1093/mnras/stx3334}

\bibitem[{{Triani} {et~al.}(2020){Triani}, {Sinha}, {Croton}, {Pacifici}, \& {Dwek}}]{Triani20}
{Triani}, D.~P., {Sinha}, M., {Croton}, D.~J., {Pacifici}, C., \& {Dwek}, E. 2020, \mnras, 493, 2490, \dodoi{10.1093/mnras/staa446}

\bibitem[{van~der Walt {et~al.}(2011)van~der Walt, Colbert, \& Varoquaux}]{numpy}
van~der Walt, S., Colbert, S.~C., \& Varoquaux, G. 2011, Computing in Science Engineering, 13, 22, \dodoi{10.1109/MCSE.2011.37}

\bibitem[{Van~Rossum \& de~Boer(1991)}]{python2}
Van~Rossum, G., \& de~Boer, J. 1991, CWI Quarterly, 4, 283

\bibitem[{Van~Rossum \& Drake(2009)}]{python3}
Van~Rossum, G., \& Drake, F.~L. 2009, Python 3 Reference Manual (Scotts Valley, CA: CreateSpace)

\bibitem[{{Vijayan} {et~al.}(2019){Vijayan}, {Clay}, {Thomas}, {Yates}, {Wilkins}, \& {Henriques}}]{Vijayan19}
{Vijayan}, A.~P., {Clay}, S.~J., {Thomas}, P.~A., {et~al.} 2019, \mnras, 489, 4072, \dodoi{10.1093/mnras/stz1948}

\bibitem[{{Vijayan} {et~al.}(2024){Vijayan}, {Thomas}, {Lovell}, {Wilkins}, {Greve}, {Irodotou}, {Roper}, \& {Seeyave}}]{Vijayan24}
{Vijayan}, A.~P., {Thomas}, P.~A., {Lovell}, C.~C., {et~al.} 2024, \mnras, 527, 7337, \dodoi{10.1093/mnras/stad3594}

\bibitem[{{Virtanen} {et~al.}(2020){Virtanen}, {Gommers}, {Oliphant}, {Haberland}, {Reddy}, {Cournapeau}, {Burovski}, {Peterson}, {Weckesser}, {Bright}, {van der Walt}, {Brett}, {Wilson}, {Millman}, {Mayorov}, {Nelson}, {Jones}, {Kern}, {Larson}, {Carey}, {Polat}, {Feng}, {Moore}, {VanderPlas}, {Laxalde}, {Perktold}, {Cimrman}, {Henriksen}, {Quintero}, {Harris}, {Archibald}, {Ribeiro}, {Pedregosa}, {van Mulbregt}, \& {SciPy 1. 0 Contributors}}]{scipy}
{Virtanen}, P., {Gommers}, R., {Oliphant}, T.~E., {et~al.} 2020, Nature Methods, 17, 261, \dodoi{10.1038/s41592-019-0686-2}

\bibitem[{{Weingartner} \& {Draine}(2001)}]{Weingartner01}
{Weingartner}, J.~C., \& {Draine}, B.~T. 2001, \apj, 548, 296, \dodoi{10.1086/318651}

\bibitem[{{Witstok} {et~al.}(2023){Witstok}, {Shivaei}, {Smit}, {Maiolino}, {Carniani}, {Curtis-Lake}, {Ferruit}, {Arribas}, {Bunker}, {Cameron}, {Charlot}, {Chevallard}, {Curti}, {de Graaff}, {D'Eugenio}, {Giardino}, {Looser}, {Rawle}, {Rodr{\'\i}guez del Pino}, {Willott}, {Alberts}, {Baker}, {Boyett}, {Egami}, {Eisenstein}, {Endsley}, {Hainline}, {Ji}, {Johnson}, {Kumari}, {Lyu}, {Nelson}, {Perna}, {Rieke}, {Robertson}, {Sandles}, {Saxena}, {Scholtz}, {Sun}, {Tacchella}, {Williams}, \& {Willmer}}]{Witstok23}
{Witstok}, J., {Shivaei}, I., {Smit}, R., {et~al.} 2023, arXiv e-prints, arXiv:2302.05468, \dodoi{10.48550/arXiv.2302.05468}

\bibitem[{{Yan} {et~al.}(2004){Yan}, {Lazarian}, \& {Draine}}]{Yan2004}
{Yan}, H., {Lazarian}, A., \& {Draine}, B.~T. 2004, \apj, 616, 895, \dodoi{10.1086/425111}

\bibitem[{Zeimann {et~al.}(2015)Zeimann, Ciardullo, Gronwall, Bridge, Brooks, Fox, Gawiser, Gebhardt, Hagen, Schneider, \& Trump}]{Zeimann15}
Zeimann, G.~R., Ciardullo, R., Gronwall, C., {et~al.} 2015, The Astrophysical Journal, 814, 162, \dodoi{10.1088/0004-637X/814/2/162}

\bibitem[{{Zhao} \& {Furlanetto}(2024)}]{Zhao2024}
{Zhao}, R.~J., \& {Furlanetto}, S.~R. 2024, \jcap, 2024, 018, \dodoi{10.1088/1475-7516/2024/09/018}

\bibitem[{Zucker {et~al.}(2021)Zucker, Goodman, Alves, Bialy, Koch, Speagle, Foley, Finkbeiner, Leike, Enßlin, Peek, \& Edenhofer}]{Zucker_2021}
Zucker, C., Goodman, A., Alves, J., {et~al.} 2021, The Astrophysical Journal, 919, 35, \dodoi{10.3847/1538-4357/ac1f96}

\end{thebibliography}
